      [1999/12/01 v1.4c Il Nuovo Cimento]
\documentclass[varenna]{cimento}

\usepackage{graphics}
\usepackage{epsfig}
\usepackage{amsfonts}
\usepackage{amsmath}
\usepackage{amssymb}
\usepackage{bm}

\usepackage{todonotes}
\newcommand*{\beqs}{\begin{equation*}}
\newcommand*{\eeqs}{\end{equation*}}
\newcommand*{\ve}[1]{\mathbf{#1}}

\newcommand{\beq}{\begin{equation}}
\newcommand{\eeq}{\end{equation}}

\newcommand{\beqa}{\begin{eqnarray}}
\newcommand{\eeqa}{\end{eqnarray}}

\newcommand{\ket}[1]{| #1 \rangle}                   
\newcommand{\bra}[1]{\langle #1 |}                   

\newcommand{\nOneD}{n_{1d}}

\DeclareMathSymbol{\R}{\mathalpha}{AMSb}{"52}        
\DeclareMathSymbol{\Q}{\mathalpha}{AMSb}{"51}        
\DeclareMathSymbol{\N}{\mathalpha}{AMSb}{"4E}        
\DeclareMathSymbol{\Z}{\mathalpha}{AMSb}{"5A}        
\DeclareMathSymbol{\C}{\mathalpha}{AMSb}{"43}        

\begin{document}

\title{Interferometry with Atoms}
\author{J.-F. Schaff, T. Langen \atque J. Schmiedmayer}
\institute{Vienna Center for Quantum Science and Technology, Atominstitut, Technische Unversit\"at Wien, Stadionallee 2, 1020 Vienna, Austria}

\PACSes{
\PACSit{03.65.-w}{Quantum mechanics}
\PACSit{37.25.+k}{Atom interferometry}
\PACSit{03.75.-b}{Matter waves}
\PACSit{05.30.-d}{Quantum statistical mechanics}
}


\maketitle

\begin{abstract}
Optics and interferometry with matter waves is the art of coherently manipulating the translational motion of particles like neutrons, atoms and molecules. Coherent atom optics is an extension of techniques that were developed for manipulating \emph{internal} quantum states. Applying these ideas to \emph{translational} motion required the development of techniques to localize atoms and transfer population coherently between distant localities. In this view position and momentum are (continuouse) quantum mechanical degree of freedom analogous to discrete internal quantum states. In our contribution we start with an introduction into matter-wave optics in section~\ref{sec:intro}, discuss coherent atom optics and atom interferometry techniques for molecular beams in section~\ref{sec:OIA_grating} and for trapped atoms in section~\ref{sec:IFMTrappedBEC}. In section~\ref{sec:ProbingManyBody} we then describe tools and experiments that allow us to probe the evolution of quantum states of many-body systems by atom interference.
\end{abstract}

\newpage
\tableofcontents
\newpage
\section{Optics and interferometry with atoms: an introduction}  \label{sec:intro}

Interference is one of the hallmark features of all wave theories. Atom interferometry \cite{Cronin2009} is the art of coherently manipulating the translational and internal states of atoms (and molecules), and with it one of the key experimental techniques to exploit matter waves. From the first experiments demonstrating the wave-like nature of light~\cite{Young1845,Bornwolf97} to the ground-breaking achievements of matter-wave interferometry with electrons~\cite{Davisson27}, neutrons~\cite{Rauch74}, atoms~\cite{Keith91} and even large molecules~\cite{Arndt99}, interference has led to new insights into the laws of nature and served as a sensitive tool for metrology. 

Interference with atomic and molecular matter-waves now forms a rich branch of atomic physics and quantum optics. It started with atom diffraction of He from a LiF crystal surface~\cite{ESS30} and the separated oscillatory fields technique ~\cite{Ramsey49} now used in atomic clocks. Broadly speaking, at the start of the 20th century atomic beams were developed to isolate atoms from their environment; this is a requirement for maintaining quantum coherence of any sort. In 1924 Hanle studied coherent superpositions of atomic internal states that lasted for tens of ns in atomic vapors~\cite{Hanle24}. But with atomic beams, Stern-Gerlach magnets were used to select and preserve atoms in specific quantum states for several ms. A big step forward was the ability to change internal quantum states using RF resonance, as demonstrated by Rabi et al. in 1938~\cite{Rabi38}. Subsequently, long-lived coherent superpositions of internal quantum states were created and detected by Ramsey in 1949~\cite{Ramsey49} which is the basis of modern atomic clocks and most of quantum metrology. 

Applying these ideas to spatial degrees fo freedom required the development of techniques to transfer atoms coherently between different locations. The simplest way is to create coherent superpositions of states with differetn momenta. Here, coherently means with respect to the phase of the de~Broglie wave that represents this motion. 

We give an introduction ot the coherent atom optics techniques in sections~\ref{sec:intro} and \ref{sec:OIA_grating} (A much more complete overview is given in the review by Cronin et al. \cite{Cronin2009}). Although some experiments with Bose-Einstein condensates are included, the focus of these two first sections is on linear matter wave optics where each single atom interferes with itself. Techniques for trapped atoms are then discussed in section~\ref{sec:IFMTrappedBEC} and in section~\ref{sec:ProbingManyBody} we describe recent tools and experiments, where atomic interference is used to probe the complex quantum states of interacting many-body systems.

\subsection{Basics of matter-wave optics} 
In this first section we discuss the basics of {\em matter-wave optics} and illustrate the similarities and differences to the more familiar {\em light optics}. This is by no way a detailed and in depth theoretical discussion, but should merely highlight the differences and similarities between {\em matter-waves} and {\em light}. For a detailed  theoretical discussion we refer the reader to the lectures of Ch. Bord\'e in these proceedings. 

\subsubsection{The wave equations} \label{s:WaveEquations}

A first approach to comparing {\em matter-wave} optics to {\em light} optics is to study the underlying wave equations. The differences and similarities between light optics and matter-wave optics can then be nicely illustrated in the following way: Light optics is described by Maxwell's equations.  They can be transformed and rewritten as the d'Alembert equation for the vector potential $\ve{A}$
\begin{equation} \label{e:d'Alambert_eq}
        \left[ \nabla^2  - \frac{n^2}{c^2}
        \frac{\partial^2}{\partial t^2} \right] \ve{A}(\ve{r},t) = 0 .
\end{equation}
Here for simplicity, we have assumed an isotropic and homogeneous propagation medium with refractive index $n$, and $c$ denotes the speed of light in vacuum.

If we now consider a monochromatic wave oscillating at an angular frequency $\omega$, i.e.\ we go to Fourier space with respect to time, $\ve{A}(\ve{r},t) \rightarrow \ve{A}(\ve{r},\omega) \exp(-i \omega t)$, we obtain the Helmholtz equation for $\ve{A}(\ve{r},\omega)$
\begin{equation}
  \left[ \nabla^2 + \frac{n^2 \omega^2}{c^2} \right] \ve{A}(\ve{r},\omega) = 0 . \label{e:Helmholz_eq}
\end{equation}

The propagation of matter-waves for a non-relativistic particle is governed by the time-dependent Schr\"{o}dinger equation. For non-interacting particles, or sufficiently dilute beams, a single particle approach is sufficient:
\begin{equation}
\left[-\frac{\hbar^2}{2m} \nabla^2 + V(\ve{r},t)\right] \psi(\ve{r},t) =
i \hbar \frac{\partial\psi(\ve{r},t)}{\partial t} , \label{e:Schr_eq1}
\end{equation}
where $m$ is the mass of the atom and $V(\ve{r})$ is a scalar potential.

Comparing equations~\eqref{e:d'Alambert_eq} and \eqref{e:Schr_eq1} one finds a hyperbolic differential equation for light optics (equation~\eqref{e:d'Alambert_eq}), whereas the Schr\"{o}dinger equation~\eqref{e:Schr_eq1} is of parabolic form.  This difference in the fundamental wave equations would suggest significantly different behavior for matter-waves and light waves.

If one only considers time independent problems, like, for example, propagation of a plane wave in a time independent potential $V(\ve{r})$, we can eliminate the explicit time dependence in equation~\eqref{e:Schr_eq1} by substituting $\psi(\ve{r},t) \rightarrow \psi(\ve{r}) \, \exp(-i\frac{E t}{\hbar})$ where $E$ is the total energy, which is a constant of the motion for time independent interactions. The propagation of the de~Broglie waves is then described by the time independent Schr\"{o}dinger equation\begin{equation}
        \label{e:Schr_eq_tindep}
        \left[ \nabla^2 + \frac{2m}{\hbar^2} \left(E - V(\ve{r})\right) \right] \psi(\ve{r}) = 0 .
\end{equation}
We can define the local $k$-vector for a particle with mass $m$ in a potential $V(\ve{r})$ as
\begin{equation}
        \label{e:kvector}
        k(\ve{r})= \frac{1}{\hbar} \sqrt{2m(E-V(\ve{r}))} ,
\end{equation}
and equation (\ref{e:Schr_eq_tindep}) becomes equivalent to the Helmholtz equation \eqref{e:Helmholz_eq} for the propagation of electromagnetic fields.
Therefore, at the level of equations~\eqref{e:Helmholz_eq} and \eqref{e:Schr_eq_tindep}, for a {\em monochromatic} and {\em time-independent} wave, matter waves and classical electromagnetic waves behave similarly. If identical boundary conditions can be realized the solutions for the wave function in matter-wave optics and the electric field in light optics will be the same. Many of the familiar phenomena of light optics, like refraction (see section \ref{s:RefInd}) and diffraction (see section \ref{ssec:diff}), will also appear in matter-wave optics.

\subsubsection{Dispersion Relations} \label{s:DispRel}

The dispersion relation of a wave relates its energy to its $k$-vector. Dispersion relations become apparent in the Helmholtz equations \eqref{e:Helmholz_eq} and \eqref{e:Schr_eq_tindep} by applying a Fourier transform to the spacial coordinates of the fields, i.e.\ by substituting $\psi(\ve{r}) \rightarrow \psi(\ve{k})\exp(i \ve{k} \cdot \ve{r})$.
%
%
Writing $k_0$ as the wave-vector in vacuum, we obtain vacuum dispersion relations which are {\em linear} for {\em light}
\begin{equation}
        \label{e:DispRel_Light}
        \omega = \frac{c k}{n} = c k_0 ,
\end{equation}
and {\em quadratic} for {\em matter-waves}\footnote{In a relativistic description the dispersion relation is given by
$\omega^2 = \frac{m^2 c^4}{\hbar^2} + \frac{k^2 c^2}{n^2}$ which reduces to $\omega \simeq \frac{m c^2}{\hbar} +
\frac{\hbar k^2}{2m \, n^2}$ in the non-relativistic limit . The difference with equation~\eqref{e:DispRel_matter} is caused by the energy associated with the rest mass of the particle.}
\begin{equation} \label{e:DispRel_matter}
        \omega = \frac{\hbar k_0^2}{2m} = \frac{\hbar k^2}{2m n^2} = \frac{E-V}{\hbar} .
\end{equation}
An important fact we observe in equation~\eqref{e:DispRel_matter} is that the mass enters the dispersion relation for matter-waves.  Moreover, the quadratic dispersion relation for matter-waves causes even the vacuum to be dispersive. A consequence of this dispersion is, for example, the {\em spreading of a wave packet}, which happens even in the longitudinal direction. For example, a Gaussian minimum uncertainty wave packet, prepared a time $t=0$ as
\begin{equation} \label{e:gwp}
        \psi(\ve{r},t) \propto \int A(\ve{k},\ve{k_0}) e^{i(\ve{k}\cdot\ve{r} - \omega t)} \, d\ve{k},
\end{equation}
with a momentum distribution
\begin{equation}  \label{e:gwp_Akt0}
        A(\ve{k},\ve{k_0}) \propto e^{-\frac{(\ve{k}-\ve{k_0})^2}{2 \sigma_k^2}},
\end{equation}
spreads even when propagating in vacuum, that is in the {\em absence} of a refractive medium. Here, $\sigma_x(0) = \hbar / \sigma_k$ is the width of the wave packet in position space. For our Gaussian minimum uncertainty wave packet the spreading results in a time dependent width in real space which can be written as
\begin{equation} \label{e:gwp_spread}
        \sigma_x (t) = \sqrt{\sigma_x^2 (0) +
        \left(\frac{\hbar t}{m} \right)^2 \sigma_k^2 (0)}
\end{equation}
This spreading of the wave packet is nothing else than the wave
mechanical equivalence of the dependence of the propagation velocity on the kinetic energy of a massive particle in classical mechanics. In wave mechanics, the wave packet spreads due to its different $k$-space components moving at different velocities.

\subsubsection{Phase and Group Velocity} \label{s:PhGrVel}

Another consequence of the different dispersion relations is that the phase velocity $v_{ph}$ and group velocity $v_{g}$ for de~Broglie waves are different from those of light.  In a medium with refractive index $n$ (section \ref{s:RefInd}) one finds for matter-waves
\begin{eqnarray} \label{e:vp_m}
        v_{ph} \,&=&\,\frac{\omega}{k} \,=\,
                \frac{\hbar k_0}{2n \, m} \,=\,
                \frac{1}{n} \sqrt{\frac{E}{2m} }  \,=\,
                \frac{1}{2 n} v_0 \,=\, \frac{v}{2}, \\
        \label{e:vg_m}
        v_{g} \,&=&\,\frac{\text{d}\omega}{\text{d}k} \,=\,
                \frac{\hbar k_0}{n \, m} \,=\,
                \frac{1}{n} \sqrt{\frac{2E}{m}} \,=\, \frac{1}{n} v_0 \,=\,  v.
\end{eqnarray}
The group velocity $v_{g}$, as given in (\ref{e:vg_m}), corresponds to the classical velocity $v$ of the particle.  For a wave packet this corresponds to the velocity of the wave packet envelope.  For matter-waves the vacuum is dispersive, that is $v_{ph} \neq v_{g}$ for $n=1$.  Furthermore, $v_{ph}$ and $v_{g}$ are both inversely proportional to the refractive index\footnote{In a relativistic description the phase and group velocity for matter-waves can be obtained as follows ($E$ is now the total energy including the rest mass):
\begin{eqnarray} 
        v_{ph} \,&=&\,\frac{\omega}{k} \,=\,
                \frac{1}{n} \frac{E}{\hbar k_0} \, \simeq \,
                \frac{ m c^2 }{n \hbar k_0} + \frac{\hbar k_0}{2n \, m} \, = \,
                \frac{c^2 }{n^2 \, v} + \frac{v}{2}, \nonumber \\
        v_{g} \,&=&\,\frac{d\omega}{dk} \,=\,
                \frac{\hbar k c^2}{n^2 \, E} \,=\,
                \frac{1}{n} \frac{\hbar k_0 c^2}{E} \,=\,
                \frac{1}{n} \frac{\hbar k_0}{m} \,=
                \, \frac{1}{n} v_0 \, = \,  v,  \nonumber
\end{eqnarray}
where $v_{ph}$ and $v_{g}$ are again both inversely proportional to the refractive index, but the product $v_{ph} v_{g}$ is now given by
 $v_{ph} v_{g} = c^2 / n^2$, which is the same as for light.}, with 
\begin{equation}
v_{ph}v_{g} = \frac{v_0^2}{2 n^2} = v^2/2.
\end{equation}
 
It is interesting to note that similar phenomena, like non-linear dispersion relations, the spreading of wave packets and the non-equivalence of phase and group velocity can also be found in the propagation of electromagnetic waves in refractive media, or in wave guides. The details of this correspondence depend on the detailed dispersion characteristics in the refractive index $n(k)$ of the medium, or the wave guide.

\subsection{Path Integral Formulation} \label{s:PathInteg}

The wave function description of the propagation of light or matter waves is very illustrating and powerful. Nevertheless many problems can be solved more easily by an equivalent approach developed by Feynman, where the amplitude and phase of the propagating wave at a position in space and time are expressed as the sum over all possible paths between the source and the observation point\footnote{A good and easily readable summary, adapted for atom optics is given by P.~Storey and C.~Cohen-Tannoudji in ref.~\cite{Storey94}.}.  This method can, for most cases in matter-wave optics, be simplified and gives straightforward, easily interpretable results.

In the regime of classical dynamics the path a particle takes is determined by the equation of motion. The actual path taken can be found by the principle of {\em least action} from the Lagrangian 
\begin{equation}
    L = \frac{1}{2} m \dot{z}^2 (t) - V(z)  = p \, \dot{z} - H.
\end{equation}
Here, $z$ is the spatial coordinate, $H$ the Hamiltonian, and the momentum $p$ is defined as $p = \frac{\partial L}{\partial \dot{z}}$.  The classical action is defined as the integral of the Lagrangian over the path $\Gamma$
\begin{equation}
    S{_\Gamma} = \int_{t_1}^{t_2} L(z(t),\dot{z}(t)) \, \mathrm{d}t.
\end{equation}
In general, the dynamics of the system is described by the Lagrangian equations of motion
\begin{equation}
        \label{e:ClEqMotion}
    \frac{\partial L}{\partial z} \,-\, \frac{\mathrm{d}}{\mathrm{d}t}
    \frac{\partial L}{\partial \dot{z}}\, = \, 0,
\end{equation}
which are the differential form of the principle of least action and completely equivalent to Newtons equations.  Note that the principle of least action which defines the classical paths for particles is equivalent to Fermat's principle for rays in classical light optics.

For a quantum description one has to calculate the phase and amplitude of the wave function.  As it was pointed out by Feynman \cite{FEN48, FEN65}, the wave function at point $a$ can be calculated by superposing all possible paths that lead to $a$.  In general the state of the quantum system at time $t_b$ is connected to its state at an earlier time $t_a$ by the time-evolution operator $U(t_b,t_a) = \exp(-H(t_b-t_a)/\hbar)$ via
\begin{equation}
    |\psi(t_b) \rangle \, = \, U(t_b,t_a) |\psi(t_a) \rangle.
\end{equation}
The wave function $\psi(x_b,t_b)$ at point $x_b$ is given by the projection onto position
\begin{equation}
        \label{QuantProp_eq}
        \psi(x_b,t_b) = \langle x_b |\psi(t_b) \rangle =
        \int \, \mathrm{d}x_a \, K(x_b,t_b; x_a,t_a) \psi(x_a,t_a),
\end{equation}
where the quantum propagator $K(x_b,t_b; x_a,t_a)$ is defined as
\begin{equation}
        \label{QuantProp_eq1}
        K(x_b,t_b; x_a,t_a) \equiv \langle x_b |U(t_b,t_a)|x_a \rangle.
\end{equation}
Equation (\ref{QuantProp_eq}) is a direct manifestation of the quantum mechanical superposition principle and shows the similarities of quantum mechanical wave propagation to the Fresnel-Huygens principle in optics: The value of the wave function at point $(x_b,t_b)$ is a superposition of all wavelets emitted by all point sources $(x_a,t_a)$.

Furthermore, the quantum propagator has some properties which are very useful for real calculations.  One such property comes from the fact that the evolution of a quantum system from time $t_a$ to time $t_b$ can always be broken up into two pieces at a time $t_c$ with $t_a < t_c < t_b$.  The calculation can be done in two steps from $t_a$ to time $t_c$ and then from $t_c$ to $t_b$ using the identity $U(t_b,t_a)=U(t_b,t_c)U(t_c,t_a)$.  Therefore, the composition property of the quantum propagator is given by
\begin{eqnarray} \label{Coomp_QuantProp_eq}
        K(x_b,t_b;x_a,t_a)  &=& \langle x_b |U(t_b,t_c)U(t_c,t_a)|x_a \rangle \\ 
                &=& \int \, \mathrm{d}x_c \, K(x_b,t_b;x_c,t_c) K(x_c,t_c;x_a,t_a). \nonumber
\end{eqnarray}
This shows that the propagation may be interpreted as summation over all possible intermediate states.  It is also interesting to note that this composition property applies to the amplitudes and not the probabilities. This is a distinct feature of the quantum evolution, which is equivalent to the superposition of the electric fields in optics.

Based on this composition property of the quantum propagator we can give Feynman's formulation of $K(x_b,t_b;x_a,t_a)$ as a sum over all contributions from all possible paths connecting $(x_a,t_a)$ to $(x_b,t_b)$ \cite{FEN48, FEN65}
\begin{equation} \label{Feyn_QuantProp_eq} 
	K(x_b, t_b;x_a, t_a) = N \, \sum_{\Gamma} \, e^{i S_\Gamma / \hbar} ,
\end{equation} 
where $N$ is a normalization and $\sum_{\Gamma}$ is the sum (integral) over all possible paths connecting $(x_a, t_a)$ to $(x_b, t_b)$. Each path contributes with the same modulus but with a phase factor determined by $S_\Gamma / \hbar $ where $S_\Gamma$ is the classical action along the path $\Gamma$.  Feynman's formulation is completely equivalent to the formulation of equation~\eqref{QuantProp_eq1}.

In the quasi-classical limit, where $S_\Gamma \,\gg \hbar $, the phase varies very rapidly along the path and most of the interference will be destructive, except where the classical action has an extremum.  Only paths close to the classical path described by equation~\ref{e:ClEqMotion} will then contribute significantly to the sum in equation~\eqref{Feyn_QuantProp_eq}.

The method of path integrals is a very powerful method to solve the problem of propagating matter waves. However it is, even for very simple geometries, very hard to implement in its most general form.  In most cases of matter-wave optics we can use approximations to the full Feynman path integral formulation.  The possible approximations follow from the observation that the largest contribution to the path integral comes from the paths close to a path with an extremum in the classical action $S_{\Gamma}$.

\subsubsection{JWKB approximation}

The first approximation that can be done is the JWKB approximation\footnote{This method was first introduced by Lord Rayleigh for the solution of wave propagation problems.  It was then applied to quantum mechanics by H.  Jeffreys (1923) and further developed by G. Wentzel, H.A.  Kramers and L.  Brillouin (1926).}, often also called the quasi-classical approximation: one uses the {\em classical} path to calculate phase and amplitude of the wave function at a \mbox{specific location, i.e.}
\beq \label{e:psi_JWKB} 
\psi = \frac{C}{\sqrt{p}} e^{\frac{i}{\hbar}S_{cl}} = \frac{C}{\sqrt{p}} e^{\frac{i}{\hbar} \int |p| dx}.
\eeq
In the JWKB approximation one easily sees how the wavefronts and the classical trajectories correspond to each other.  For a fixed energy $E$, a wavefront is given by the relation $S(\ve{x}) = S_0 = \text{constant}$.  One can show\footnote{see for example chapter 6 in Messiah's book on quantum mechanics~\cite{Messiah}} that for a scalar potential the wavefronts are orthogonal to the classical trajectories.

In the case of a vector potential $\ve{A}$ one has to replace the classical momentum by the canonical momentum and finds the relation $\ve{p} = \nabla S - e\ve{A}/c$ between the propagation and the wavefronts.  In this case the wavefronts are no longer orthogonal to the classical trajectories.  This is analogous to geometric optics in an anisotropic medium.

The JWKB approximation is, in general, applicable when the change in the amplitude of the wave function is small at a scale of one wavelength, i.e. for a slowly varying amplitude of the wave function.  This is usually not the case for reflections, or at classical turning points where $k \rightarrow 0$.  However, for most of these cases, methods were developed to calculate the additional phase shifts that are neglected when using the JWKB approximation.  Very good results are typically obtained by adding these additional phase shifts to the phase found by the JWKB approximation, even in cases where $k \rightarrow 0$.

\subsubsection{Eikonal approximation} 

For most experiments in matter-wave optics the even simpler {\em eikonal approximation} of classical optics is sufficient.  There, the phase of a wave function is calculated along the straight and unperturbed path between the starting point (source) and the observation point.

\subsection{Coherence}
\label{s:Coherence}
Many of the phenomena in wave optics are concerned with the superposition of many waves. Therefore, one of the central questions is concerned with the coherence properties of this superposition. Naturally, this is also an important question in matter-wave optics. In general, one can define the coherence of matter waves analogously to the coherence in light optics, by using correlation functions.

The first order correlation function for matter-waves with respect to coordinate $\eta$ and a displacement $\delta\eta$ is defined by
\begin{equation} \label{e:coherence}
        g^{(1)}(\delta \eta) = \langle \psi | {\textstyle T}_\eta (\delta \eta) | \psi \rangle,
\end{equation}
where T$_\eta(\delta \eta)$ is the translation operator with respect to a displacement of $\delta \eta$.  The width of this function with respect to the displacement $\eta$ is called the amount of coherence with respect to $\eta$.


\subsubsection{Spatial coherence}
In the case of spatial coherence, T$_{\ve{x}}(\delta \ve{x})=\exp( i \ve{p} \delta \ve{x}/\hbar)$ is the spatial translation operator and the correlation function takes the familiar form
\beq \label{e:spat_coherence}
        g^{(1)}(\delta x)  =  \langle \psi(x) | \psi(x + \delta x) \rangle
                 =  \int{ \psi^{*}(x) \psi(x + \delta x) \, d x}.
\eeq
In analogy to optics with light, in a beam of matter-waves one distinguishes between {\em longitudinal} and {\em transverse} coherence\footnote{For massive particles the distinction between longitudinal and transversal coherence is not always as clear as for light.  This can be easily seen if one notices that for a nonrelativistic particle longitudinal and transversal motion can be transformed into each other by a simple Galilean transformation.  The distinction breaks down especially if the particles are brought to rest.  Therefore, in the following discussion we will assume a particle beam with mean $k$-vector $\ve{k}$ much larger than the momentum distribution ($\ve{k} \gg \sigma_{\ve{k}}$).}.

\paragraph{Longitudinal coherence} 
An interesting and common example is the longitudinal coherence length $l_c$ of a particle beam with a Gaussian distribution of $k$-vectors, which propagates along the $x$-direction. An example of such a beam is given in equation~(\ref{e:gwp_Akt0}). Its first order longitudinal correlation function is given by
\begin{equation}
        g^{(1)}(\delta x) = \langle \psi(x) | \psi(x+\delta x) \rangle =
                e^{- \frac{\delta x^2 \sigma_k^2}{2}},
\end{equation}
and the longitudinal coherence length $\ell_c$ is related to the momentum distribution in the beam by
\beq \label{e:lcoh_long}
     \ell_c = \frac{1}{\sigma_k} = \frac{1}{k} \frac{\langle v \rangle}{\sigma_v} = \frac{\langle \lambda_\text{dB} \rangle}{2 \pi} \frac{\langle v \rangle}{\sigma_v}
\eeq
where $\sigma_v$ is the rms velocity spread connected to the momentum distribution and $\langle \lambda_\text{dB} \rangle$ is the mean deBroglie wavelength.

\begin{figure}[t]
\begin{center}
\includegraphics[width = 13.5cm]{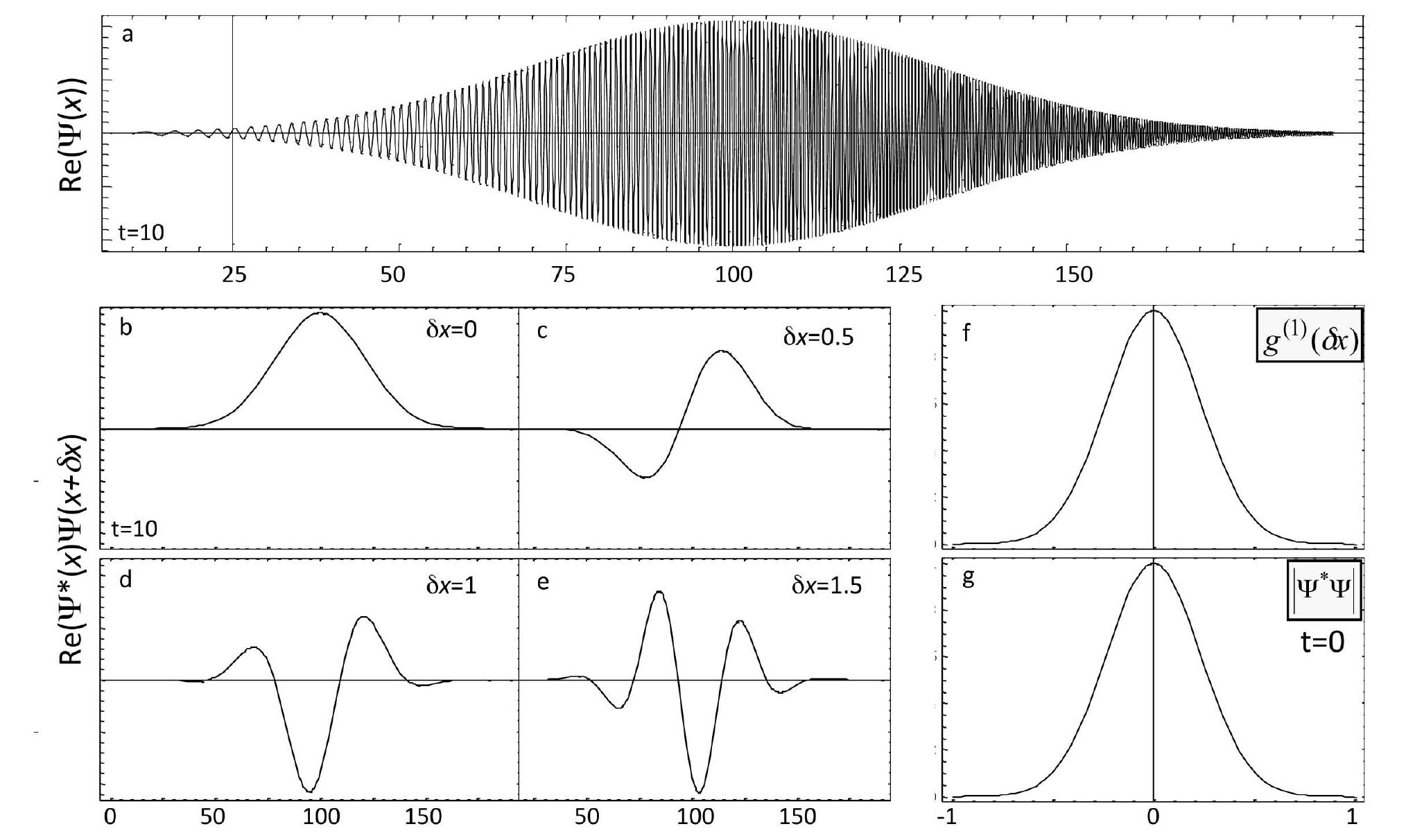} 
\caption{Gaussian Wavepacket with $k_0=10$ and $\sigma_k=3$ after a propagation time of $t=10$ (in natural units $\hbar=1$, $m=1$).  At $t=0$ the wavepacket is a minimum uncertainty wavepacket with a width in $x$ of $\sigma_x=0.3$ (\emph{g}).  \emph{(a)} Real part of the wave function Re($\psi(x)$). \emph{(b-e)}   Re($\psi^*(x) \psi(x + \delta x)$) for 4 different values of $\delta x$.  \emph{(b)}:  $\delta x=0$. \emph{(c-e)} for increasing $\delta x$ the oscillating contribution of Re($\psi^*(x) \psi(x + \delta x)$) leads to a vanishing contribution to the coherence function $g^{(1)}(\delta x)$ displayed in \emph{(f)}. \emph{(g)} Initial size of the wavepacket.} \label{fig:WavePacket}
\end{center}
\end{figure}

Because of the different dispersion relations, light and matter-waves exhibit differences in their correlation functions. For the linear dispersion relation of light propagating in vacuum, the coherence length can pictorially be associated with the size of a Fourier transform limited pulse with the same frequency width.  For matter waves, the quadratic dispersion relation leads to a spreading of the wave packet, and the size of a wave packet can not be related to the coherence length, as illustrated in figure \ref{fig:WavePacket}. This difference led to various discussions in the early matter-wave experiments with neutrons \cite{Kais83a, Klei83, Cosm83, Kais83b, Bern87}.

It is interesting to note that the coherence length (transversal and longitudinal) has nothing to do with the size of the particles. For example for the matter-wave interferometer experiments with Na$_2$ molecules at MIT the coherence length was about a factor 4 smaller than the size of the molecule (size of the Na$_2$ molecule $\sim 400$ pm, its de~Broglie wavelength: $\lambda_\textrm{dB} \approx 10$ pm and its coherence length $l_\textrm{coh} \approx 100$ pm). Nevertheless, the same interference contrast as in the experiments with $Na$ atoms was observed \cite{CEH95}. Similar conclusion at a more extreme scale can then be drawn from the later experiments on large molecules by the Vienna group (see the lectures of M. Arndt in these proceedings).

Consequently a measurement of the longitudinal coherence length in a time-independent experiment using an interferometer \cite{Kais83a, Klei83, Schm93} generally tells us {\em nothing} about the length or even the existence of a wave packet \cite{Cosm83, Kais83b, Bern87}. One can easily show that the longitudinal correlation functions for the following examples are identical: a minimum uncertainty Gaussian wave packet, the same wave packet after spreading for an arbitrary time, and even a superposition of plane waves with the same k-vector distribution but random phases $\varphi_r$.  The latter is the correct description for an thermal atomic source, as it is used in most atom optics experiments. Only in experiments when the beam is chopped at a timescale comparable to the inverse energy spread, can one hope to prepare an atomic beam in a not completely chaotic state. 

\paragraph{Transverse coherence} The transverse coherence of a matter-wave is obtained similar to the longitudinal coherence length, but by translation in the transverse direction $\delta z$ $g^{(1)}_\text{trans}(\delta z) = \langle \psi(z) | \psi(z+\delta z) \rangle$.  In analogy to light optics it is related to the transverse momentum distribution by the von Cittert-Zernike theorem \cite{Bornwolf97}.  For an atomic beam the transverse coherence of a beam can be defined by preparation (selection) in space (e.g. collimation by slits), transverse to the propagation direction. For waves emitted with an angular spead $\alpha$, the longitudinal coherence length is 
\beq 
\ell_{tcoh} \sim \frac{1}{2}\frac{\lambda_{dB}}{\alpha} = \frac{\lambda_{dB}}{2}\frac{z}{w} 
\eeq 
where $z$ is the distance from the source aperture of width $w$. For a BEC, transverse coherence is related to fluctuations, and the resulting multimode structure of the quantum gas, similar to a laser that emits in multiple transverse modes.

\subsubsection{Coherence in momentum space}

To address the problems arising in the interpretation of the experiments on the longitudinal coherence length for matter-waves a more useful concept is to study coherence in {\em momentum space}.  In analogy to coherence in real space the correlation function in momentum space is given by
\beq
        g^{(1)}(\delta k)  =  \langle \psi(k) | \psi(k + \delta k) \rangle  =  \int{ \psi^{*}(k) \psi(k + \delta k) d k}.
\eeq
This correlation function is, in principle, sensitive to the phase relations between different $k$-components of an atomic beam, and can therefore distinguish between the different interpretations of the longitudinal coherence length.

This correlation function in momentum space can only be measured in a time-dependent experiment.  One possibility the measure $g^{(1)}(\delta k)$ for matter-waves is the sideband interferometer as described by B.  Golub and S. Lamoreaux \cite{sidebandIFM}. There, $g^{(1)}(\delta k)$ is measured by superposing two paths, where the energy of the propagating wave is shifted in one of the two\footnote{Experiments describing interference of neutron paths with different energies are described in ref.~\cite{Badu86}.}. The result of $g^{(1)}(\delta k)$ is the given as the relative contrast of the {\em time-averaged} interference fringes in the interferometer.  Similarly coherence in momentum space can be probed by a differentially tuned separated oscillatory field experiment as demonstrated at MIT \cite{SDK98}

\subsubsection{Higher-order coherence}
\label{s:HghO_Coherence}

The higher-order correlation functions for matter-waves can be defined similarly to light optics.  The {\em second-order correlation function} is given by the joint detection probability of two particles at two locations:
\beq \label{e:g2_matterwaves_2}
	g^{(2)} (x_1,t_1 ;x_2,t_2) = \langle \Psi | b^\dag(x_1,t_1) b^\dag(x_2,t_2) \;
	b(x_1,t_1) b(x_2,t_2) |\Psi \rangle
\eeq
where $b^\dag(x,t)$ and $b(x,t)$ are the creation and annihilation operators for an atom at time $t$ and location $x$, and $\Psi$ is a multi-particle wave function.

The second-order correlation function is trivially zero for one particle experiments.  Since the form is similar to light optics one would expect bunching behavior for bosons and antibunching behavior for fermions from a chaotic source.  High phase-space density per propagating mode is needed for realistic experiments to observe these higher-order coherences.

Second- and higher-order correlation functions for identical particles emitted from highly-excited nuclei were investigated in refs.~\cite{Kopylov72, Koonin89} as a tool to measure the coherence properties of nuclear states. Second-order correlation functions for fermions were indeed observed in nuclear physics experiments by studying neutron correlations in the decay of highly-excited $^{44}$Ca nuclei \cite{Duennweber90}. Higher-order correlation functions in pions emitted in heavy-ion collisions were widely used to investigate the state of this form of matter~\cite{Goldhaber60, Fung78}. In atom optics, the bunching of cold atoms in a slow atomic beam, analogous to the Hanbury-Brown-Twiss experiment in light optics~\cite{HBT,BROWN56}, was first observed for Ne~\cite{Yasu96} and Li \cite{Kasevich96} and recently using ultracold $^4$He~\cite{He4-HBT}. It was measured for atom lasers~\cite{AtomLaser-HBT} or to study the BEC phase transition~\cite{LRO-HBT,Perrin12,RuGwayWu13} or higher-order correlation functions~\cite{Dall13}. Fermi antibunching in an atomic beam of $^3$He was observed in ref.~\cite{He3-HBT}.

\subsection{Index of refraction for matter waves} \label{s:RefInd}

We define the index of refraction for matter waves in the same way as for light: as the ratio of the free propagation $k$-vector $k_0$ to the local $k$-vector $k(\ve{r})$
\begin{equation}
        \label{e:refIndex_define}
        n(\ve{r}) = \frac{k(\ve{r})}{k_0}.
\end{equation}
There are two different phenomena that can give rise to a refractive index for matter-waves:
\begin{itemize}
        \item  One can describe the action of a potential as being equivalent to a refractive index.
        \item  One finds a refractive index from the scattering of the propagating particles off a medium. This is equivalent to the refractive index in light optics.
\end{itemize}

\subsubsection{Index of refraction caused by a classical potential}

If we compare the local $k$-vector in equation~(\ref{e:kvector}) to our above definition of a refractive index (equation~\eqref{e:refIndex_define}), one sees that in matter-wave optics the action of a {\em scalar} potential $V(\ve{r})$ can be described as a position dependent refractive index $n(\ve{r})$.  In most cases the potential $V(\ve{r})$ will be much smaller than the kinetic energy $E_\text{kin}$ ($V(\ve{r}) \ll E_\text{kin}$) of the atom, and one finds
\begin{equation}
        \label{e:refIndex_pot}
        n(\ve{r}) = \sqrt{1-\frac{V(\ve{r})}{E_\text{kin}}} \simeq 1-\frac{V(\ve{r})}{2E_{kin}}.
\end{equation}
Therefore, the refractive index will be larger than unity ($n(\ve{r}) > 1$) in regions with an attractive potential ($V(\ve{r}) < 0$) and smaller than unity ($n(\ve{r}) < 1$) in regions with a repulsive potential.  It is interesting to note that the refractive index caused by a classical potential has a strong dispersion, as $n$ changes with $\lambda_\text{dB}^2$ ($1/k_0^2$).

The above relation in equation (\ref{e:refIndex_pot}) can be extended to a vector potential $\ve{A}$. In this case, the canonical momentum $\ve{p}_c$ (wave vector $\ve{k}_c$) and the kinetic momentum $\ve{p}_k$ (wave vector $\ve{k}_k$) are not necessary parallel.  In a wave description, the wavefront (orthogonal to $\ve{k}_c$) and the propagation direction (parallel to $\ve{k}_k$) are not orthogonal to each other.  This is similar to propagation of light in an anisotropic medium.  The refractive index is then direction dependent.

\subsubsection{Index of refraction from scattering}
A second phenomena that gives rise to a refractive index for matter-waves is the interaction of the matter-wave with a medium.  This is analogous to the refractive index for light, which results from the coherent forward scattering of the light in the medium.  Similarly, the scattering processes of massive particles inside a medium result in a phase-shift of the forward scattered wave and define a refractive index for matter-waves. Here we give a schematic introduction. For a full treatment, see one of the standard books on scattering theory.

From the perspective of wave optics the evolution of the wave function $\psi$ while propagating through a medium is given by
\begin{equation}
        \label{e:scatteringindex}
        \psi(x)=\psi(0)\, e^{-i\,k_\text{lab}\,x} \; e^{i\,\frac{2\pi}{k_\text{c}}\,Nx \,{\rm Re}(f(k_c,0))}
        \; e^{-\frac{2\pi}{k_c}\,Nx \,{\rm Im}(f(k_c,0))}.
\end{equation}
Here $k_\text{lab}$ is the wave vector in the laboratory frame, $k_\text{c}$ the wave vector in the center-of-mass frame of the collision, $N$ is the areal density of scatterers in the medium and $f(k_c,0)$ is the center-of-mass, forward scattering amplitude.  The amplitude of propagating wave function $\psi$ is {\em attenuated} in proportion to the {\em imaginary} part of the forward scattering amplitude, which is related to the total scattering cross section by the optical theorem
\begin{equation}
        \sigma _{tot} = \frac{4\pi}{k_c} \,{\rm Im}(f(k_c,0)).
\end{equation}
In addition, there is a {\em phase shift} $\phi$ proportional to the {\em real} part of the forward scattering amplitude
\begin{equation}
        \phi(x) = \frac{2\pi}{k_c} \, Nx \,{\rm Re}(f(k_c,0))).
\end{equation}
In analogy to light optics one can define the {\it complex} index of refraction
\begin{equation}
        n=1+\frac{2\pi}{k_\text{lab}\,k_\text{c}}\,N \, f(k_c,0) .
\end{equation}

The refractive index of matter for de Broglie waves has been extensively studied in neutron optics \cite{SEA90, NeutScatLng}, especially using neutron interferometers. It has also been widely used in electron holography \cite{Lich88}.  In neutron optics, scattering is dominantly $s$-wave and measuring the refractive index gives information about the $s$-wave scattering length $a$ defined as\begin{equation} \label{e:scatlng_def}
        a = - \lim_{k\rightarrow 0}f_0,
\end{equation}
where $f_0$ is the $s$-wave scattering amplitude.

In atom optics with thermal beams, usually many partial waves, typically a few hundred, contribute to scattering of thermal atoms. The number of contributing partial waves $l$ can be estimated by $ l \sim x_r k_\text{c}$ where $x_r$ is the range of the inter atomic potential and $k_\text{c}$ is the center-of-mass wave vector of the collision. The refractive index will depend on the forward scattering amplitude and therefore on details of the scattering process. Measuring $n$ will lead to new information about atomic and molecular scattering \cite{SCE95}, especially the real part of the scattering amplitude, not directly accessible in standard scattering experiments. 

For ultracold atoms and an ultracold media the scattering is predominately $s$-wave and can be described by a scattering length $a$ very similar to neutron optics.  This regime can be reached for scattering inside a sample of ultracold atoms like a BEC or by scattering between two samples of ultracold atoms. We would like to note that for scattering processes between identical atoms at ultra low energies, quantum statistic becomes important. For example scattering between identical Fermions vanishes because symmetry leads to suppression of $s$-wave scattering.  On the other hand the dominance of $s$-wave scattering at low energies is only valid if the interaction potential decays {\em faster} than $1/r^3$. For scattering of two dipoles, higher partial waves contribute even in the limit of zero collision energy.

We now look closer at the low-energy limit where $s$-wave scattering is predominant.  The scattering can in first approximation be described by only one parameter, the scattering length $a$.  Here we can derive simple relations for the dispersion of the refractive index $n$ starting from a low energy expansion of the $s$-wave scattering amplitude
\begin{equation}
        f(k,0) = \frac{1}{2ik} (e^{-2i \delta_0} - 1) \simeq - a (1 - i k a)
\end{equation}
where $\delta_0 \simeq - k a$ is the $s$-wave scattering phase-shift\footnote{To be more precise for larger  $k$ one can use the {\em effective range} approximation $k \cot(\delta_0) = -1/a + k^2 r_\text{eff}$ ($r_\text{eff}$ is the effective range of the potential}.  The refractive index is then given by
\begin{equation}
        n(k_\text{lab}) = 1+\frac{2\pi}{k_\text{lab}\,k_\text{c}}\,N \, a (1 - i k a)
\end{equation}
For $k \rightarrow 0$~ $n$ becomes predominantly real and diverges with $1/k^2$.

Consequently we can now reverse the above argument defining a refractive index for a classical potential and define an effective optical potential $U_\text{opt}$ for a particle in a medium with scattering length $a$
\begin{equation} \label{e:nUopt}
        U_\text{opt}(r)      = \frac{2 \pi \hbar^2}{m_c} N(r) a
\end{equation}
where $m_c$ is the reduced mass.  This potential $U_\text{opt}$ is one of the basics ingredients for many neutron optics experiments and neutron optics devices.

An important phenomenon in matter-wave optics, is that matter interacts with itself. matter-wave optics is inherently non-linear, and the non-linearities can define the dominant energy scale.  The local refractive index, and therefore the propagation of matter waves depends on the local density of the propagating particles. A simple description can be found in the limit when the propagating beam can be viewed as weakly interacting, that is if the mean particle spacing is much larger than $a$ ($a \ll \rho^{1/3}$, where $\rho$ is the density). The self-interaction can then be described by the optical potential (equation~\eqref{e:nUopt}).  In its simplest form, it leads to an additional term in the Schr\"{o}dinger equation (equation~\eqref{e:Schr_eq1}) which is then {\em nonlinear} and called the Gross-Pitaevskii equation~\cite{GrPitEq}
\begin{equation} \label{e:GrPit}
        \left[-\frac{\hbar^2}{2m} \nabla^2 + V(r,t) \right] \psi(r,t) +
        \frac{4 \pi \hbar^2 \, a}{m} |\psi(r,t)|^2=
        i \hbar \frac{\partial\psi(r,t)}{\partial t} .
\end{equation}
This self-interaction leads to a new type of nonlinear optics where even the vacuum is nonlinear. This has to be contrasted with the fact that for light, nonlinearities are very small and come into play only in special media.

\newpage

\section{Optics and interferometry using gratings}  \label{sec:OIA_grating}

In this section we will give an overview of optics and interferometry with beams of atoms or molecules using gratings.  We will only discuss the main aspects and phenomena, and refer the reader for details about experiments to the review article on atom interferometry by Cronin et al.~\cite{Cronin2009}.

\subsection{Diffraction}  \label{ssec:diff}

Diffraction of matter waves from phase and amplitude modulating objects is a hallmark example of wave propagation and interference. It arises from the coherent superposition and interference of the propagating matter wave which is modified in amplitude and phase by the diffracting structure. It is described by the solution of the Schr\"{o}dinger equation (equation~\eqref{e:Schr_eq1}) with the appropriate boundary conditions.  An elegant approach to solve this problem is to express the amplitude and phase of the matter wave at a position in space and time as the sum over all possible paths between the source and the observation point (see section \ref{s:PathInteg}). Beamsplitters for atom beam interferometers are often based on diffraction. Comparing the diffraction of matter waves and light, we expect differences arising from the different dispersion relations.  These manifest themselves in time dependent diffraction problems, and will give rise to a new phenomenon: {\em diffraction in time}.

\subsubsection{Diffraction in space}

First we discuss diffraction in space, transverse to the propagation of the beam. A diffraction \emph{grating} is a diffracting region that is periodic. Spatial modulation of the wave by the grating generates multiple momentum components for the scattered waves which interfere. The fundamental relationship between the mean momentum transferred to waves in the $n^\text{th}$ component and the grating period, $d$, is 
\beq 
	\delta p_n = n\frac {h}{d} = n \hbar G \;\; \quad \quad  
	\theta_n \approx \frac { \delta p_n}{p_{beam}} = n\frac{\lambda_{dB}}{d} 
\eeq
where $G=2\pi/d$ is the reciprocal lattice vector of the grating, $h$ is Planck's constant, and $\lambda_\text{dB} = h/p_{beam}$ is the de~Broglie wavelength of the incoming beam. In the far field diffraction is observed with respect to the diffraction angle $\theta_n$.  To resolve the different diffraction orders in the far field the transverse momentum distribution of the incoming beam must be smaller than the transverse momentum given by the diffraction grating $\delta p = \hbar G$. This is equivalent to the requirement that the transverse coherence length must be larger than a few grating periods. This is usually accomplished by collimating the incident beam\footnote{The transverse coherence length is $\ell_{tcoh} \approx \lambda_\textrm{dB} / \vartheta_\textrm{coll}$, where $\lambda_\textrm{dB}$ is the de Broglie wavelength and $\vartheta_\textrm{coll}$ is the (local) collimation angle of the beam (the angle subtended by a collimating slit). Since for thermal atomic beams $\lambda_\textrm{dB} \sim 10$ pm a collimation of $\vartheta_\textrm{coll}<10\mu rad$ is required for a 1 $\mu m$ coherent illumination.}.

The first examples of atom interference were diffraction experiments.  Just three years after the electron diffraction experiment by Davisson and Germer~\cite{Davisson27} Estermann and Stern observed diffraction of He beam off a LiF crystal surface~\cite{ESS30}. 

Classical wave optics recognizes two limiting cases, near- and far-field. In the \emph{far-field} the curvature of the atom wave fronts is negligible and Fraunhofer diffraction is a good description. The diffraction pattern is then given by the Fourier transform of the transmission function, including the imprinted phase shifts.
In the \emph{near-field} limit the curvature of the wave fronts must be considered and the intensity pattern of the beam is characterized by Fresnel diffraction.  Edge diffraction and the Talbot self-imaging of periodic structures are typical examples.  
 

\begin{figure}[t]
\begin{center}
\includegraphics[width = 4cm]{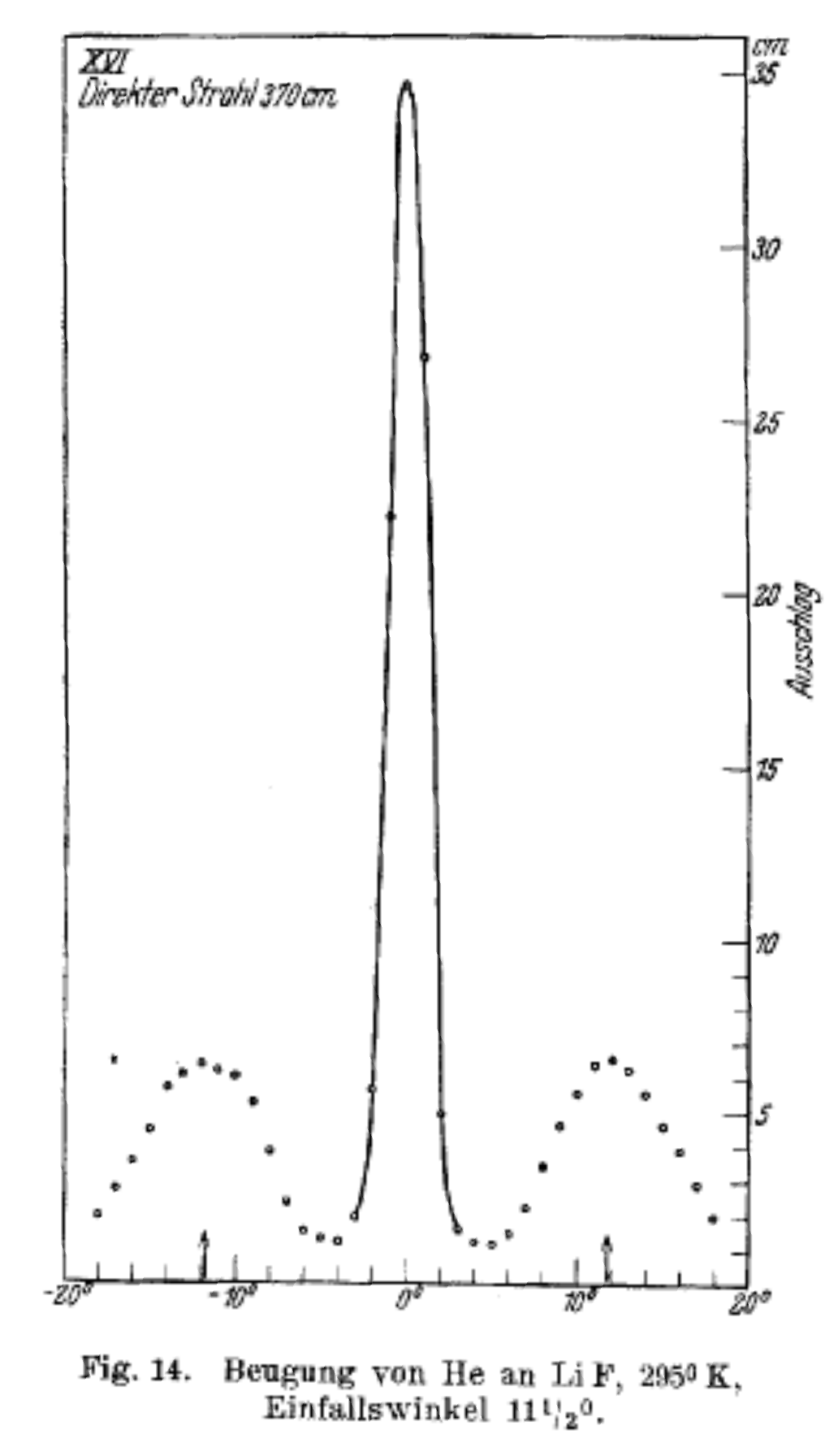} 
\includegraphics[width = 8cm]{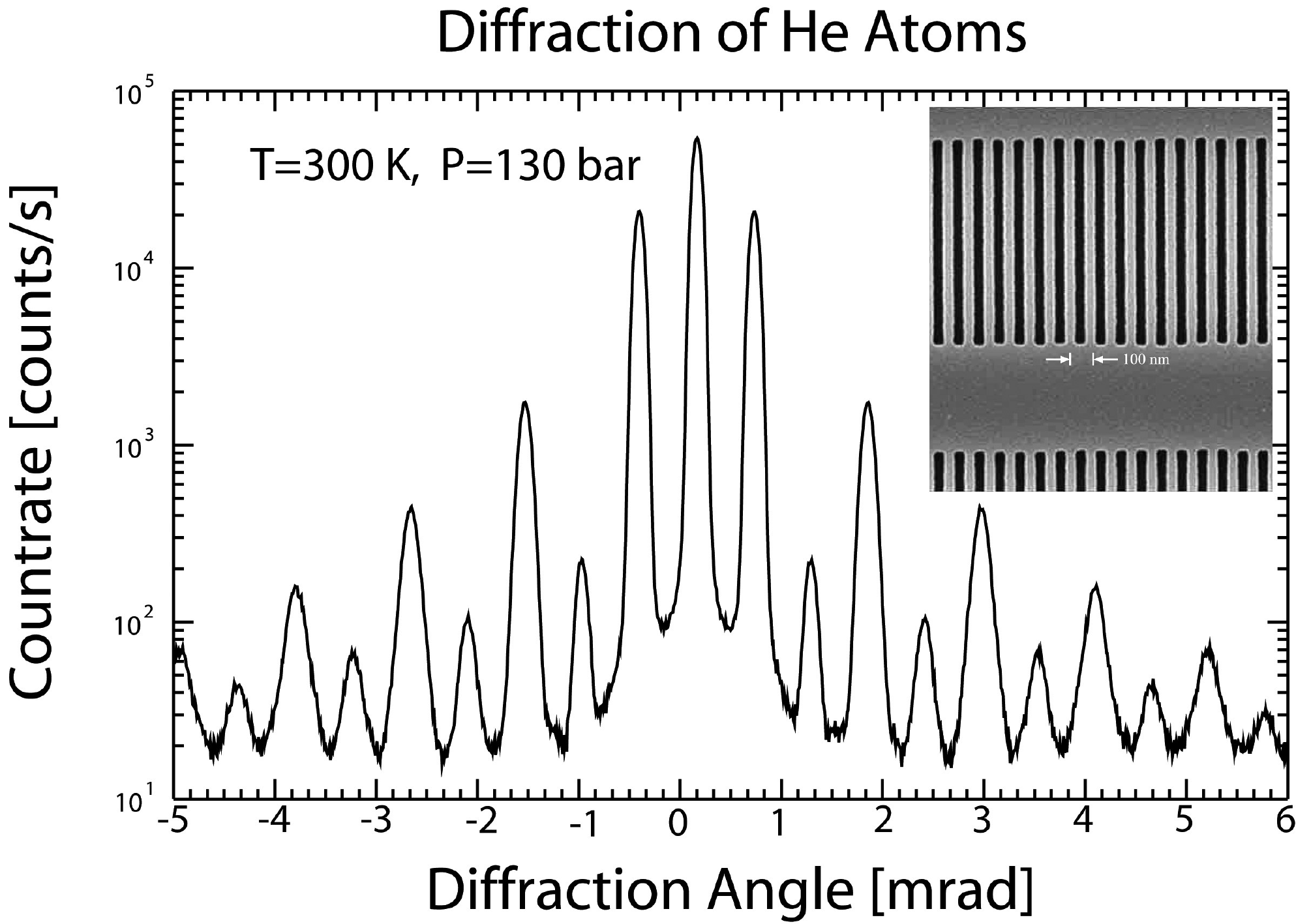}     
\caption{(Left)  Historic data showing diffraction of He atoms from a LiF crystal surface~\cite{ESS30}. (Right) Diffraction of He atoms transmitted through a nanostructure grating. The average velocity, the velocity spread of the beam, the uniformity of the material grating, and the strength of atom-sruface van der Waals forces can all be determined from these data~\cite{GST99}. The insert shows a $100\,$nm period grating for atom waves.  The dark regions are slits, and light regions are free-standing silicon nitride bars. Figure courtesy of T.A. Savas and H.I. Smith at the MIT NanoStructure laboratory~\cite{sava96,SAS90,sss95}.  Figure curtesy of J.P. Toennies, W. Schoellkopfand and O. Kornilov.} \label{fig:He diffn}
\end{center}
\end{figure}

\subsubsection{{Diffraction from nano-fabricated structures}}  \label{sssec:nonofab}

With the advent of modern nanotechnology it became possible to fabricate elaborate arrays of holes and slits with feature sizes well below $50\,$nm in a thin membrane that allow atoms to pass through. Diffraction from a fabricated grating was first observed for neutrons by H. Kurz and H. Rauch in 1969 \cite{Kurz1969} and for atoms by the Pritchard group at MIT~\cite{KSS88}. The latter experiment used a transmission grating with $200\,$nm wide slits. Similar mechanical structures have been used for single slits, double slits, diffraction gratings, zone plates, hologram masks, mirrors, and phase shifting elements for atoms and molecules. 

The benefits of using mechanical structures for atom optics include the possibility to create feature sizes smaller than light wavelengths, arbitrary patterns, rugged designs, and the ability to diffract any atom or molecule. The primary disadvantage is that atoms or molecules can stick to (or bounce back from) surfaces, so that most structures serve as absorptive atom optics with a corresponding loss of transmission. When calculating the diffraction patterns, one has to consider that, first, nanofabrication is never perfect, and that the slits and holes can have variations in their size. Second, the van der Waals interaction between atoms and molecules and the material of the gratings can lead to effectively much smaller slits and holes in the diffracting structures. Moreover, the wave-front emerging from the hole can have additional phase shifts from the van der Waals interaction with the surface. Such effects can be particularly significant for molecules with a large electric polarizability and very small diffraction structures. For a detailed discussion of this topic we refer the reader to the lecture of M.~Arndt.

\begin{figure}[t]
\begin{center}
\includegraphics[width = 7cm]{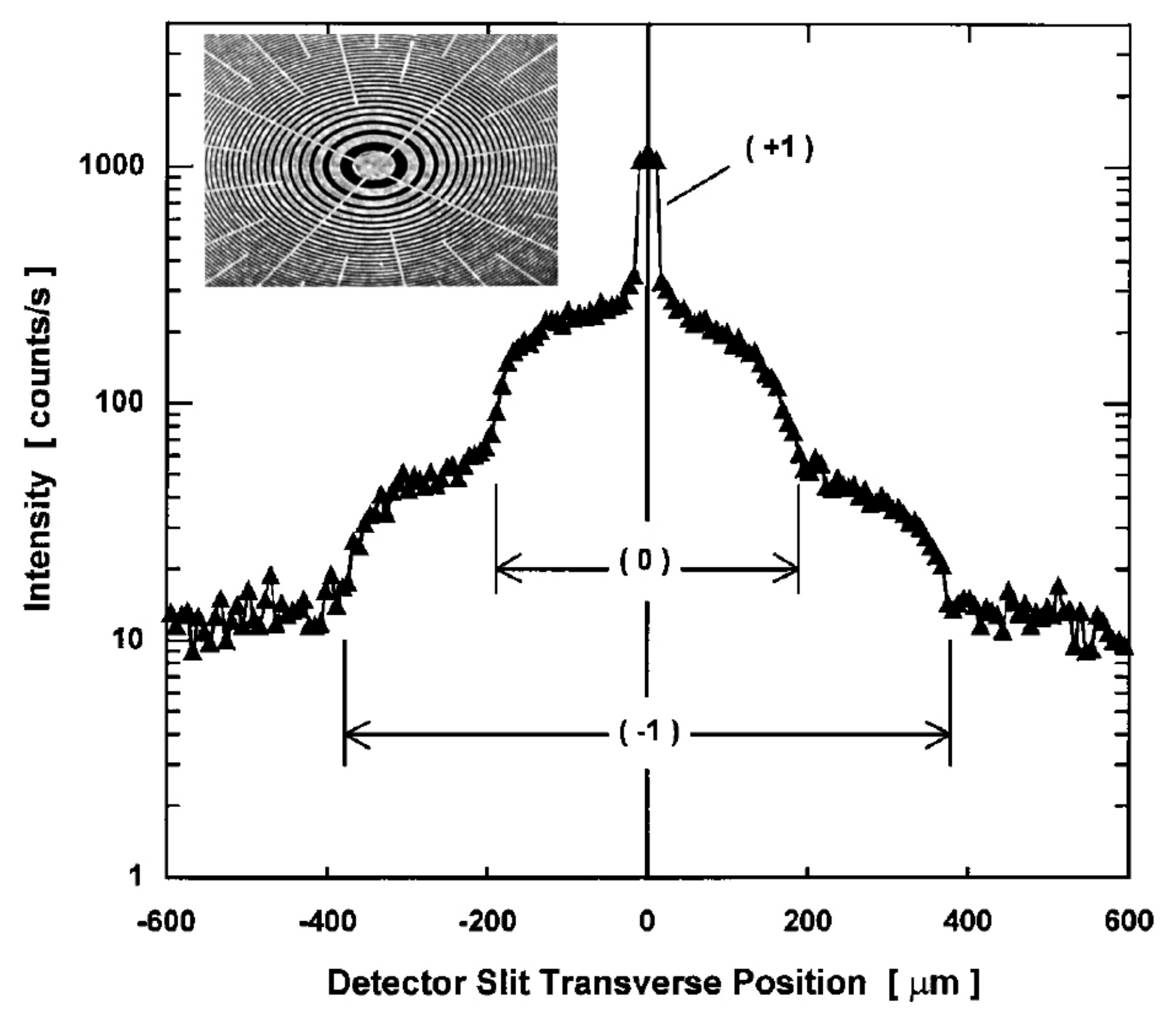} \includegraphics[width = 5.75cm]{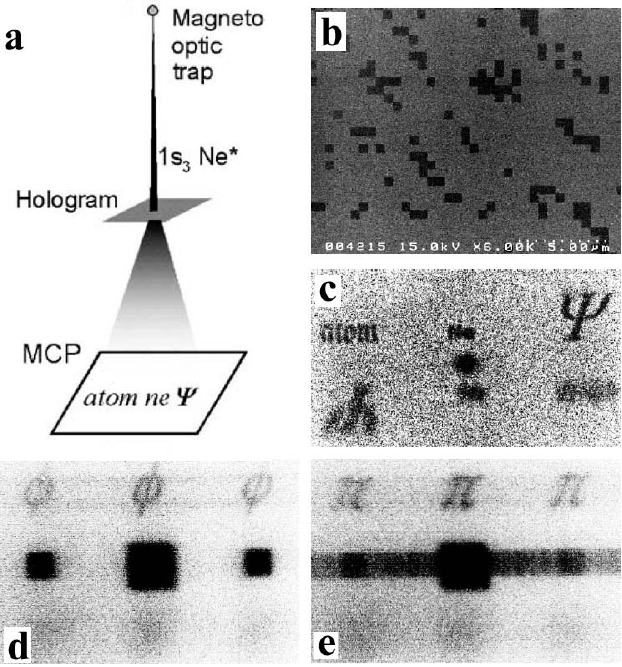}  
\caption{(Left) A zone plate for focusing atom beams.  The plate has free-standing annular rings and radial support struts. (Right) Data showing focused and defocused atom beam components. Figure from ref.~\cite{DGR99}.
(Left) Image reconstruction by atom holography. (Middle Top) The hologram is designed by computer and realized with a SiN membrane with square holes. (Right) data from a switchable hologram.~\cite{MYK96,FMS00b,FMS00}.} \label{fig:hologram}
\end{center}
\end{figure}

Nano-fabricated gratings have been used to diffract atoms and molecules such as $^4$He, $^4$He$_2$,  $^4$He$_3$ and larger $^4$He clusters (Fig. \ref{fig:He diffn}), Na$_2$, C$_{60}$, C$_{60}$F$_{48}$, C$_{44}$H$_{30}$N$_4$ and many more.~\cite{CEH95,ANV99,BAZ03,NAZ03,BHU02,HUH03,SHT96,BGK04}.  
\emph{Fresnel zone plates} have been employed to focus atoms~\cite{CSS91,DGR99} and spot sizes below $2\,\mu$m have been achieved.  
\emph{Atom holography} with nanostructures can make the far-field atom flux resemble arbitrary patterns. Adding electrodes to a structure allows electric and magnetic fields that cause adjustable phase shifts for the transmitted atom waves. With this technique, a two-state atom holographic structure was demonstrated~\cite{FMK96,FKM99,FMS00} that produced images of the letters $\phi$ or $\pi$ as shown in Fig.~\ref{fig:hologram}. The different holographic diffraction patterns are generated depending on the voltages applied to each nanoscale aperture.


\subsubsection{{Light gratings from standig waves}}  \label{sssec:LightGrating}

\begin{figure}[t]
\begin{center}
\includegraphics[width = 10cm]{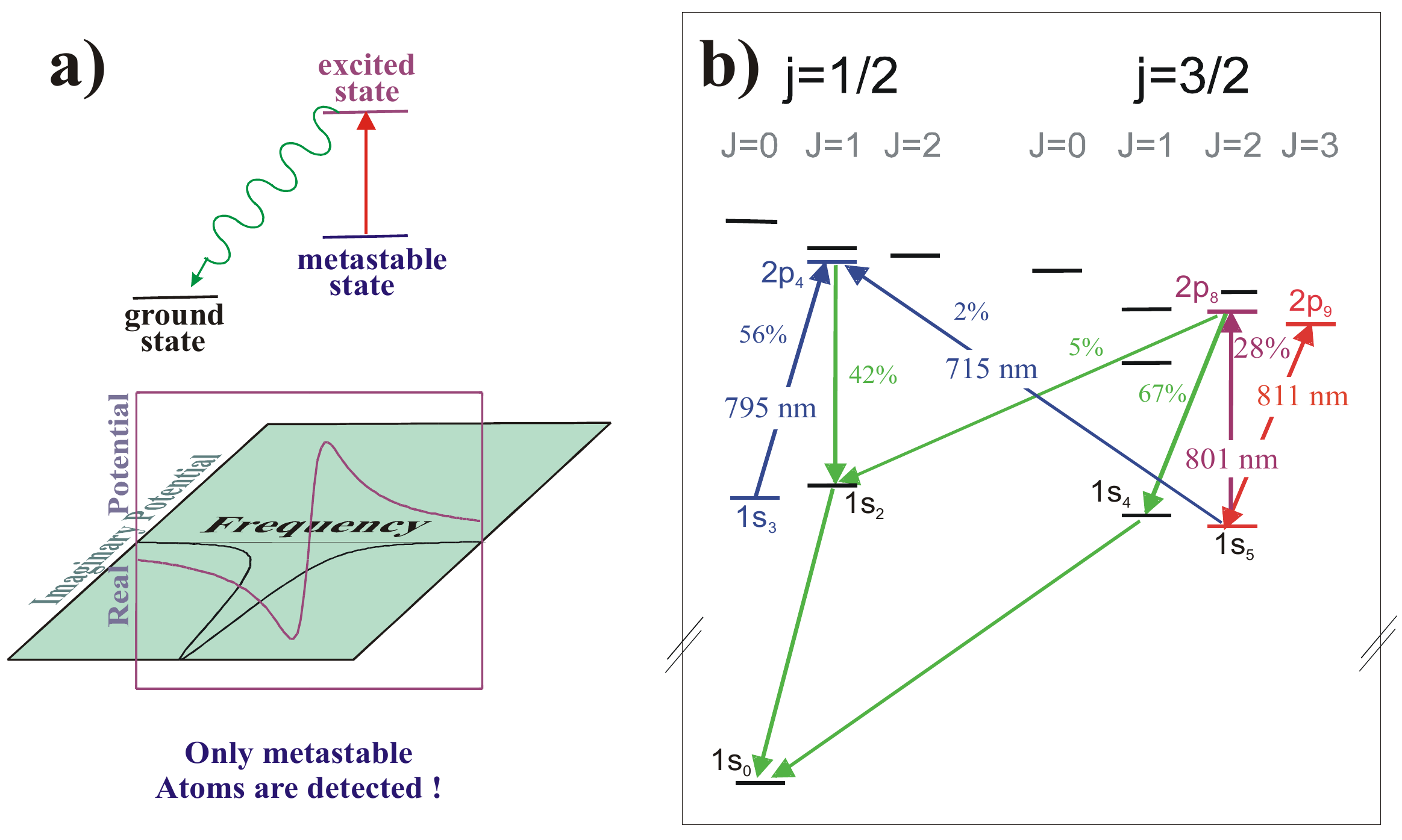} 
\caption{Implementing complex potentials with light: (a) \emph{Top}: Schematic of an open transition from a metastable state.  If the ground state is not detected then emission of a photon corresponds to 'absorption' of the atom, and one can realize also 'absorptive' potentials. \emph{Bottom}: real and imaginary part of the optical potential close to a resonance as depicted above.
(b) Level structure of Ar atom with relevant transitions as an example to implement complex  optical potentials. } \label{fig:LightPotential}
\end{center}
\end{figure}

In an open two-level system the interaction between an atom and the light field (with detuning $\Delta = \omega_\textrm{laser} - \omega_\textrm{atom}$) can be described by an effective optical potential of the form~\cite{OAB96} (figure~\ref{fig:LightPotential}):
 \beq
    U(x) \propto I(x) \frac{1}{2 \Delta +  i \Gamma}
    \label{eq:Uopt}
 \eeq
where $\Gamma$ is the atomic decay rate and $I(x)$ is the light intensity.  The imaginary part of the potential results from the spontaneous scattering processes, the real part from the ac Stark shift. If the spontaneous decay follows a path to a state which is not detected, the imaginary part of the potential in equation~\eqref{eq:Uopt} is equivalent to absorption. On-resonant light can therefore be used to create absorptive structures. Light with large detuning produces a real potential and therefore acts as pure phase object. Near-resonant light can have both roles.

The spatial shape of the potential is given by the local light intensity pattern, $I(x)$, which can be shaped with all the tricks of near and far field optics for light, including holography.  The simplest object is a periodic potential created by two beams of light whose interference forms a standing wave with reciprocal lattice vector $\vec{G} = \vec{k}_1 - \vec{k}_2$.  Such a periodic light field is often called \emph{light crystal} or more recently an \emph{optical lattice} because of the close relation of the periodic potentials in solid state crystals, and thus motivates the use of Bloch states to understand atom diffraction.

\begin{figure}
\center \includegraphics[width = 8cm]{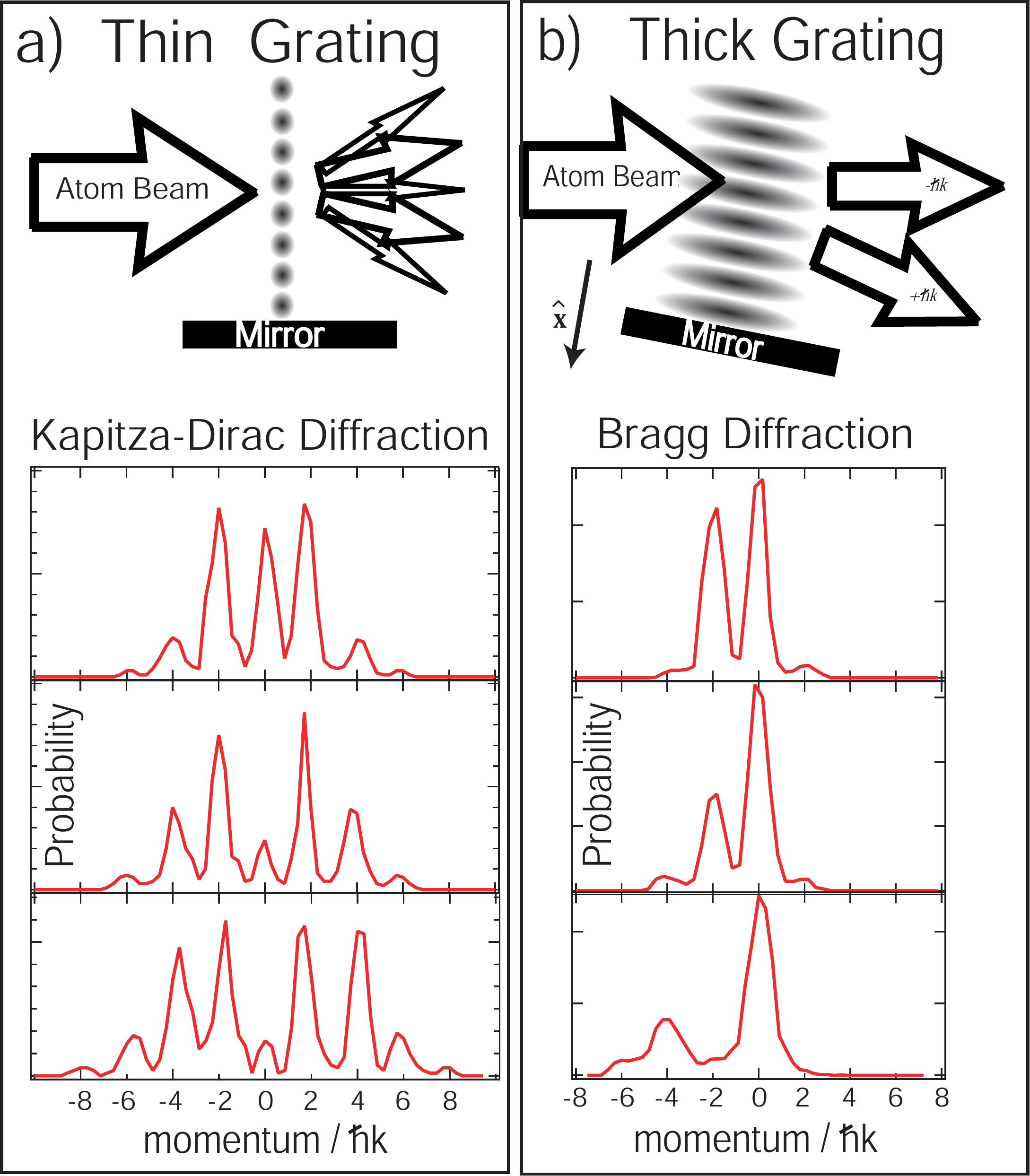}     
\caption{(a) Kapitza Dirac (KD) diffraction (b) Bragg Diffraction. The top row shows the essential difference: \emph{thick} vs. \emph{thin} gratings. The bottom row shows data obtained by the Pritchard group for KD and Bragg diffraction~\cite{GRP86,MOM88}. }
\label{fig:standing}
\end{figure}

Since light gratings can fill space, they can function as either \emph{thin} or \emph{thick} optical elements. In the case of a grating, the relevant scale is the Talbot length $L_\text{Talbot} = d^2 / \lambda_\text{dB} $.  For a spatial extent $D$ along the beam propagation of $D \ll L_\text{Talbot}$  it is considered \emph{thin}, otherwise \emph{thick} (figure~\ref{fig:standing}). 
For a \emph{thin} optical element the extent of the grating along the propagation direction has no influence on the final diffraction (interference).  This limit is also called the Raman-Nath limit.
For a  \emph{thick} optical element the full propagation of the wave throughout the diffracting structure must be considered.  A thick grating acts as a crystal, and the characteristics observed are Bragg scattering or channeling, depending on the height of the potentials.

The second distinction, relevant for \emph{thick} gratings, has to do with the strength of the potential.  One must determine if the potential is only a perturbation, or if the potential modulations are larger then the typical transverse energy scale of the atomic beam or the characteristic energy scale of the grating,
\beq
E_\textrm{G} = \hbar^2 G^2/(2 m) = 4 \hbar \omega_{rec}, \label{eq:EG}
\eeq 
associated with one grating momentum unit $\hbar G$ ($\hbar \omega_{rec}$ is an atom's `recoil energy' due to absorbing or emitting a photon). For weak potentials, $U \ll E_\textrm{G}$, one observes Bragg scattering. The dispersion relation looks like that of a free particle, only with avoided crossings at the edges of the cell boundaries. Strong potentials, with $U \gg E_\textrm{G}$, cause channelling. The dispersion relations are nearly flat, and atoms are tightly bound to the wells.

\paragraph{Diffraction with on-resonant light}

If the spontaneous decay of the excited state proceeds mainly to an internal state which is not detected, then tuning the light frequency of a standing light wave to resonance with an atomic transition ($\Delta = 0$) can make an `absorptive' grating with light. (If the excited state decays back to the ground state, this process produces decoherence.)  For a thin standing wave the atomic transmission is given by
 \beq
 T(x) =\exp\left[-\frac{\kappa}{2} [1+\cos (Gx)]\right], \label{eq:t(x)}
 \eeq
where $\kappa$ is the absorption depth for atoms passing through the antinodes. For sufficiently large absorption only atoms passing near the intensity nodes survive in their original state and the atom density evolves into a comb of narrow peaks. Since the `absorption' involves spontaneous emission, such light structures have been called \emph{measurement induced gratings}. As with all thin gratings, the diffraction pattern is then given by the scaled Fourier transform of the transmission function.

\begin{figure}
\begin{center}
\includegraphics[width = 0.8 \columnwidth]{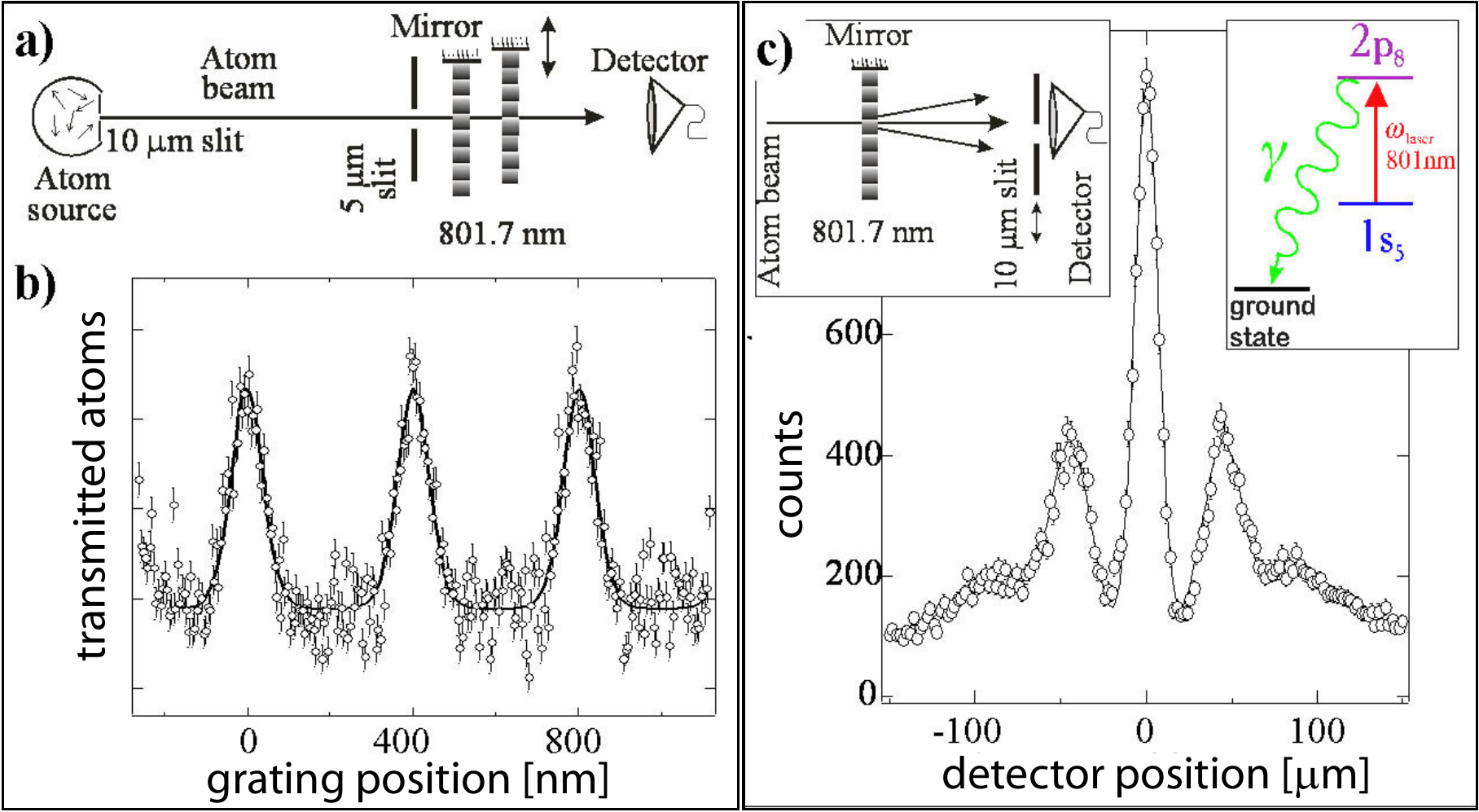}
\end{center}
\caption{(color online) Diffraction from a measurement-induced grating.  (a) Schematic of two on-resonant standing waves of light. The first causes atom diffraction.  The second translates to analyze near-field atomic flux. (b) Periodic structure in the transmitted atomic beam. (c) Far-field atom diffraction from a measurement induced grating. Figure from ref.~\cite{AKB97}.} \label{fig:OnResDiff}
\end{figure}

On-resonant standing waves have been used as gratings for a series of near-field (atom lithography, Talbot effect) and far-field (diffraction, interferometry) experiments~\cite{AKB97,JBB96,JTD98,JGS04,ROB95}. An example is shown in figure~\ref{fig:OnResDiff}. These experiments demonstrate that transmission of atoms through the nodes of the `absorptive' on resonant light masks is a coherent process.

\paragraph{Light crystals}  \label{sssec:LightCryst}

If the standing wave is thick, one must consider the full propagation of the matter wave inside the periodic potential. The physics is characterized by multi-wave (beam) interference. For two limiting cases one can get to simple models. For weak potentials: Bragg scattering; and for strong potentials: coherent channeling.

When an atomic matter wave impinges on a thick but weak light crystal, diffraction occurs only at specific angles, the Bragg angles $\theta_B$ defined by the Bragg condition
\beq
N \lambda_\text{dB} = \lambda_{ph} \sin(\theta_B).
\eeq
Bragg scattering, as shown in figure~\ref{fig:standing} (right column) transfers atoms with momentum $-p_x$ into a state with a single new momentum, $p_x = - p_x + \hbar G$. Momentum states in this case are defined in the frame of the standing wave in direct analogy to neutron scattering from perfect crystals. Bragg scattering of atoms on a standing light wave was first observed at MIT~\cite{MOM88}. Higher order Bragg pulses transfer multiples of $N \hbar G$ of momentum, and this has been demonstrated up to $50^{th}$ order (transter of $\> 100 \hbar k$ and beyond \cite{Chiow11,Muller09,Muller08,Poli11,Ivanov08} (see also cectures by G. Tio, M. Kasevich and H. Mueller) 

Bragg scattering can be described as a multi-beam interference as treated in the dynamical diffraction theory developed for neutron scattering. Inside the crystal one has two waves, the incident `forward' wave ($k_F$) and the diffracted `Bragg' wave ($k_B$). These form a standing atomic wave with a periodicity that is the same as the standing light wave. This is enforced by the diffraction condition \mbox{($k_{B} - k_{F} = G$)}. At any location inside the lattice, the exact location of atomic probability density depends on $k_F$, $k_B$ and the phase difference between these two waves.

For incidence exactly on the Bragg condition the nodal planes of the two wave fields are parallel to the lattice planes. The eigenstates of the atomic wave field in the light crystal are the two Bloch states, one exhibiting maximal ($\psi_{max}$) the other minimal ($\psi_{min}$) interaction:
\begin{eqnarray}
  \psi_{max} &=& \frac{1}{2} \left[ e^{i\frac{G}{2}x} + e^{-i\frac{G}{2}x}\right]
                = \cos \left( \frac{G}{2}x \right), \nonumber \\
  \psi_{min} &=&  \frac{1}{2} \left[ e^{i\frac{G}{2}x} - e^{-i\frac{G}{2}x}\right]
                = i \sin \left(\frac{G}{2}x\right) .
 \end{eqnarray}
For $\psi_{max}$ the anti-nodes of the atomic wave field coincide with the planes of maximal light intensity, for $\psi_{min}$ the anti-nodes of atomic wave fields are at the nodes of the standing light wave. These states are very closely related to the coupled and non-coupled states in velocity selective coherent population trapping (VSCPT).

The total wave function is the superposition of $\psi_{max}$ and $\psi_{min}$ which satisfies the initial boundary condition. The incoming wave is projected onto the two Bloch states which propagate through the crystal accumulating a relative phase shift. At the exit, the final populations in the two beams is determined by interference between $\psi_{max}$ and $\psi_{min}$ and depends on their relative phase.

\begin{figure}
\center\includegraphics[width = 0.8 \columnwidth]{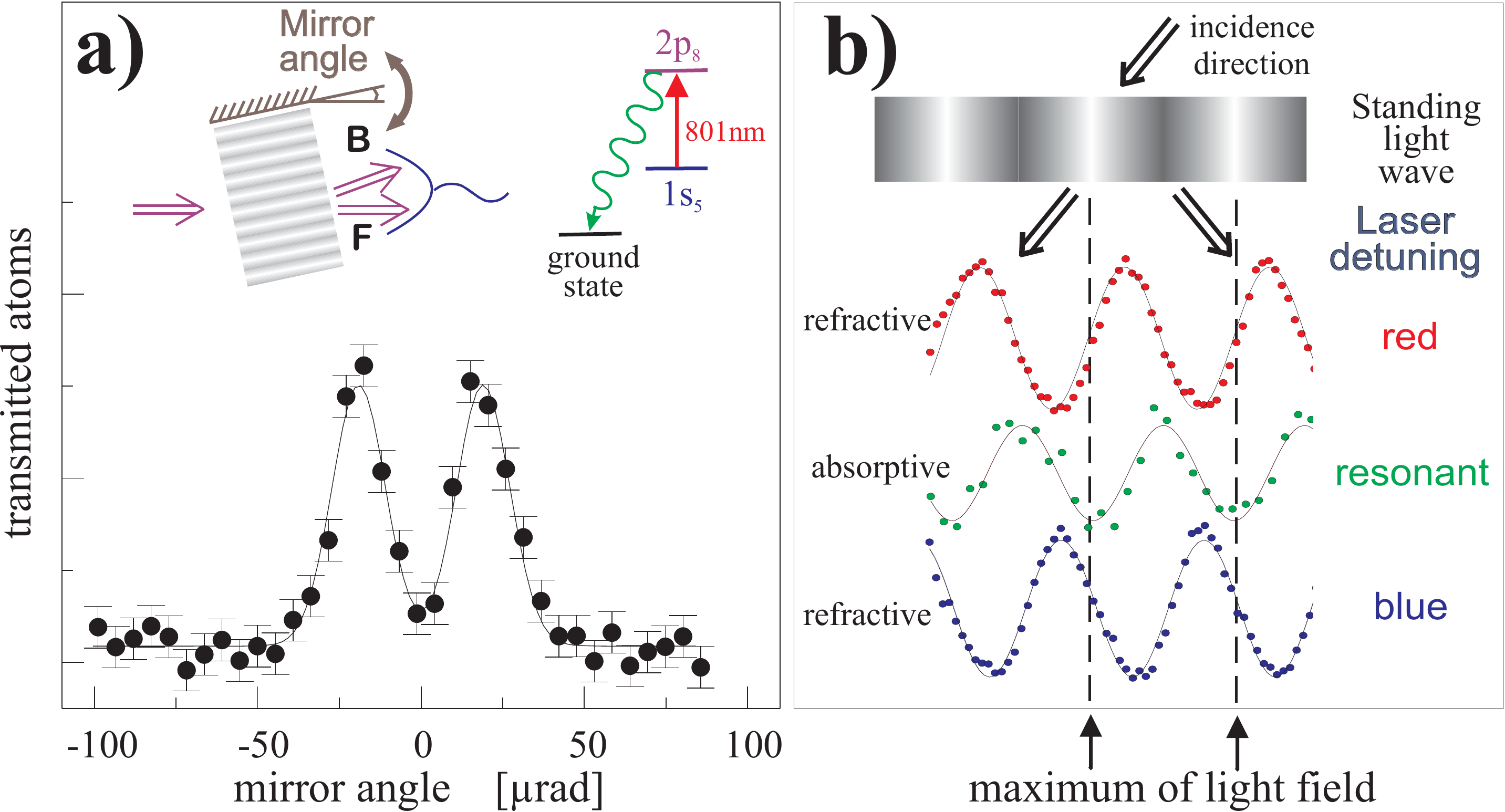}   
\caption{(color online) Bragg diffraction of atoms from resonant standing waves of light.  (a) Anomalous transmission of atoms. (b) A resonant standing wave inside an off-resonant light crystal serves to measure the atom wave fields inside the light crystal. Figure from ref.~\cite{OAB96}.} \label{fig:AnormTrans}
\end{figure}

Bragg scattering can also be observed with absorptive, on-resonant light structures~\cite{OAB96} and combinations of both on- and off-resonant light fields~\cite{KOA97}. One remarkable phenomenon is that one observes the total number of atoms transmitted through a weak on-resonant standing light wave increases if the incident angles fulfill the Bragg condition, as shown in figure~\ref{fig:AnormTrans}. This observation is similar to what Bormann discovered for X-rays and called anomalous transmission~\cite{BOR41}. It can easily be understood in the framework of the two beam approximation outlined above.  The rate of de-population of the atomic state is proportional to coupling between the atom wave field and the standing light field.  The minimally coupled state $\psi_{min}$ will be attenuate much less and propagate much further into the crystal than $\psi_{max}$. For a sufficiently thick light crystal the propagating wave field will be nearly pure $\psi_{min}$ at the exit, and as a consequence one observes two output beams of equal intensity. Inserting an absorptive mask~\cite{AKB97} inside the light crystal one can directly observe the standing matter wave pattern inside the crystal~\cite{OAB96,OAB99} and verify the relative positions between the light field and $\psi_{min}$.

More complex potentials for atoms can be made by superposing different standing light waves~\cite{KOA97}. For example, a superposition of a on- and a off-resonant standing wave with a phase shift $\Delta \varphi = \pm \pi/2$ results in a combined potential of $ U(x)= U_0 e^{\pm i G x}$ which, in contrast to a standing wave, has only \emph{one} momentum component and therefore only diffract in one direction~\cite{KOA97}.

\subsubsection{{Diffraction in time}}

If the optical elements are explicitly time dependent an interesting phenomenon arises for matter waves, which is generally not observed in light optics: {\em diffraction in time}. The physics behind this difference is the different dispersion relations. The time dependent optical element creates different energies which propagates differently according the quadratic dispersion relation for matter waves, which leads to superposition between matter waves emitted at different times at the detector.  

Diffraction in time and the differences between light and matter waves can best seen by looking at the shutter problem as discussed by Moshinsky~\cite{Mosh52}.  We start with a shutter illuminated with monochromatic wave from one side.  The shutter opens at time $t=0$ and we ask the question how does the time dependence of the transmitted radiation behave at a distance $z$ behind the shutter.  For light we expect a sharp increase in the light intensity at time $t=z/c$.  For matter waves the detected intensity will increase at time $t=z/v$, where $v$ is the velocity corresponding to the incident monochromatic wave.  But the increase will be not instantaneous, but will show a typical Fresnell diffraction pattern in time, similar to the diffraction pattern obtained by diffracting from an edge~\cite{Mosh52, Gera76}.  Similar arguments also hold for diffraction from a single slit in time, a double slit in time, or any time structure imprinted on a particle beam.

The experimental difficulty in seeing diffraction in time is that the time scale for switching has to be faster than the inverse frequency (energy) width of the incident matter wave. This condition is the time equivalent to coherent illumination in spatial diffraction. 
The first (explicit) experiments demonstrating diffraction in time used ultra-cold neutrons reflecting from vibrating mirrors~\cite{HKO87,FMG90,HFG98}. Side bands of the momentum components were observed.

\begin{figure}[t]
\begin{center}
\includegraphics[width = 0.8 \columnwidth]{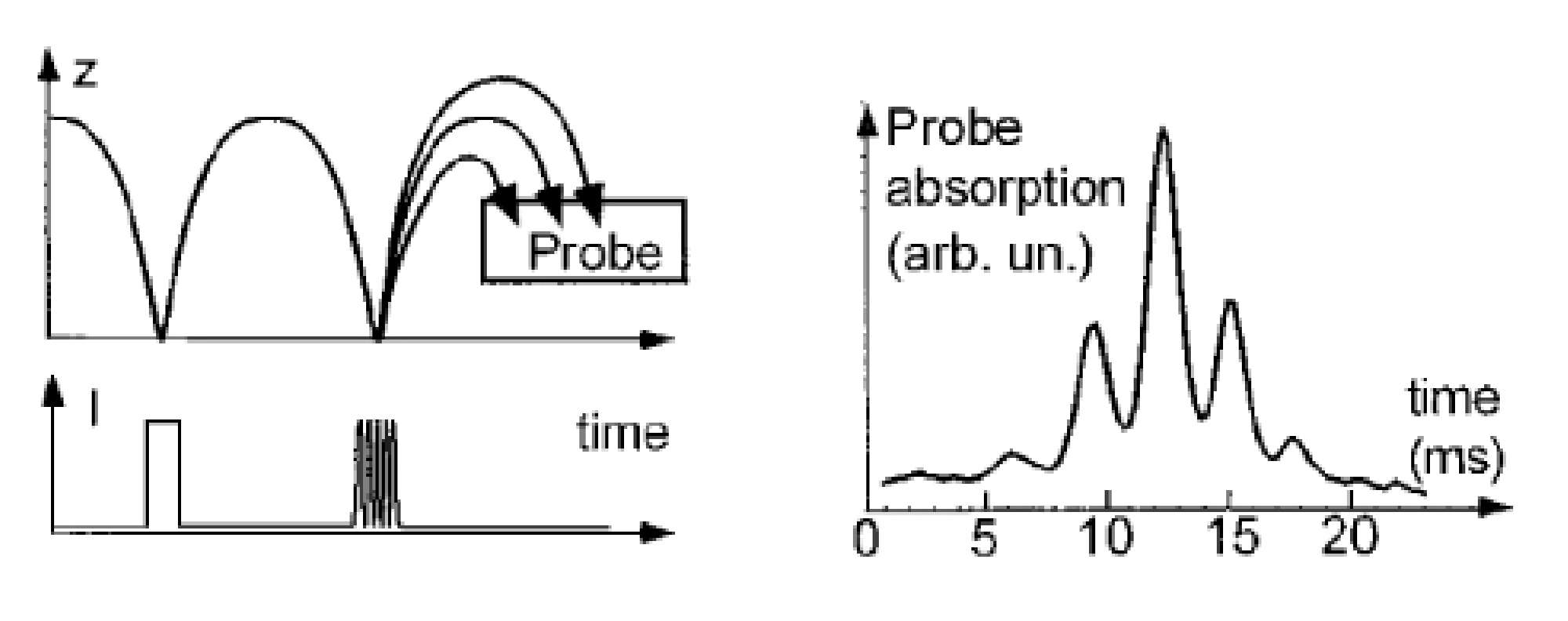}   
\caption{The ENS diffraction in time experiments described in the text. Figure from ref.~\cite{SSD95} and reviewed in ref.~\cite{COH98}.}
\label{diffTimeENS.fig}
\end{center}
\end{figure}

The group of J. Dalibard at the ENS in Paris used ultra cold Cs atoms ($T \sim 3.6$ $\mu$K) released from an optical molasses reflecting from an evanescent wave atom mirror~\cite{SSD95,SGA96,ASD96}.  By pulsing the evanescent light field one can switch the mirror on and off, creating time-dependent apertures that are diffractive structures.  To obtain the necessary temporal coherence a very narrow energy window was first selected by two (0.4 ms) temporal slits separated by 26 ms. If the second slit is very narrow ($ < 10$ $\mu$s) one observes single slit diffraction in time, if the mirror is pulsed on twice within the coherence time of the atomic ensemble one observes double slit interference in time, and many pulses lead to a time-dependent flux analogous to a grating diffraction pattern as shown in figure~\ref{diffTimeENS.fig}. From the measurement of the arrival times of the atoms at the final screen by fluorescence with a light sheet the energy distribution can be reconstructed.

Because the interaction time between the atoms and the mirror potential ($< 1$ $\mu$s) was always much smaller then the modulation time scale ($> 10$ $\mu$s), these experiments are in the `thin grating' (Raman-Nath) regime for diffraction in time.

\paragraph{Modulated light crystals}

The time equivalent of spatial Bragg scattering can be reached if the interaction time between the atoms and the potential is long enough to accommodate many cycles of modulation. When a light crystal is modulated much faster than the transit time, momentum is transferred in reciprocal lattice vector units and energy in sidebands at the modulation frequency. This leads to new resonance conditions and 'Bragg diffraction' at two new incident angles~\cite{BOA96,BAK00}. Consequently Bragg scattering in time can be understood as a transition between two energy and momentum states. The intensity modulation frequency of the standing light wave compensates the detuning of the Bragg angle and frequency of the de~Broglie wave diffracted at the new Bragg angles is shifted by $ \pm \hbar \omega_{mod}$~\cite{BOA96,BAK00}. Thus, an amplitude modulated light crystal realizes a \emph{coherent frequency shifter} for a continuous atomic beam.  It acts on matter waves in an analogous way as an acousto-optic modulator acts on photons, shifting the frequency (kinetic energy) and requiring an accompanying momentum (direction) change. In a complementary point of view the new Bragg angles can be understood from looking at the light crystal itself. The modulation creates side bands $\pm \omega_\textrm{{mod}}$ on the laser light, and creates moving crystals which come from the interference between the carrier and the side bands.  Bragg diffraction off the moving crystals occurs where the Bragg condition is fulfilled in the frame co-moving with the crystal,resulting in diffraction of the incident beam to new incident angles.

\begin{figure}[t]
\center \includegraphics[width = 0.8 \columnwidth]{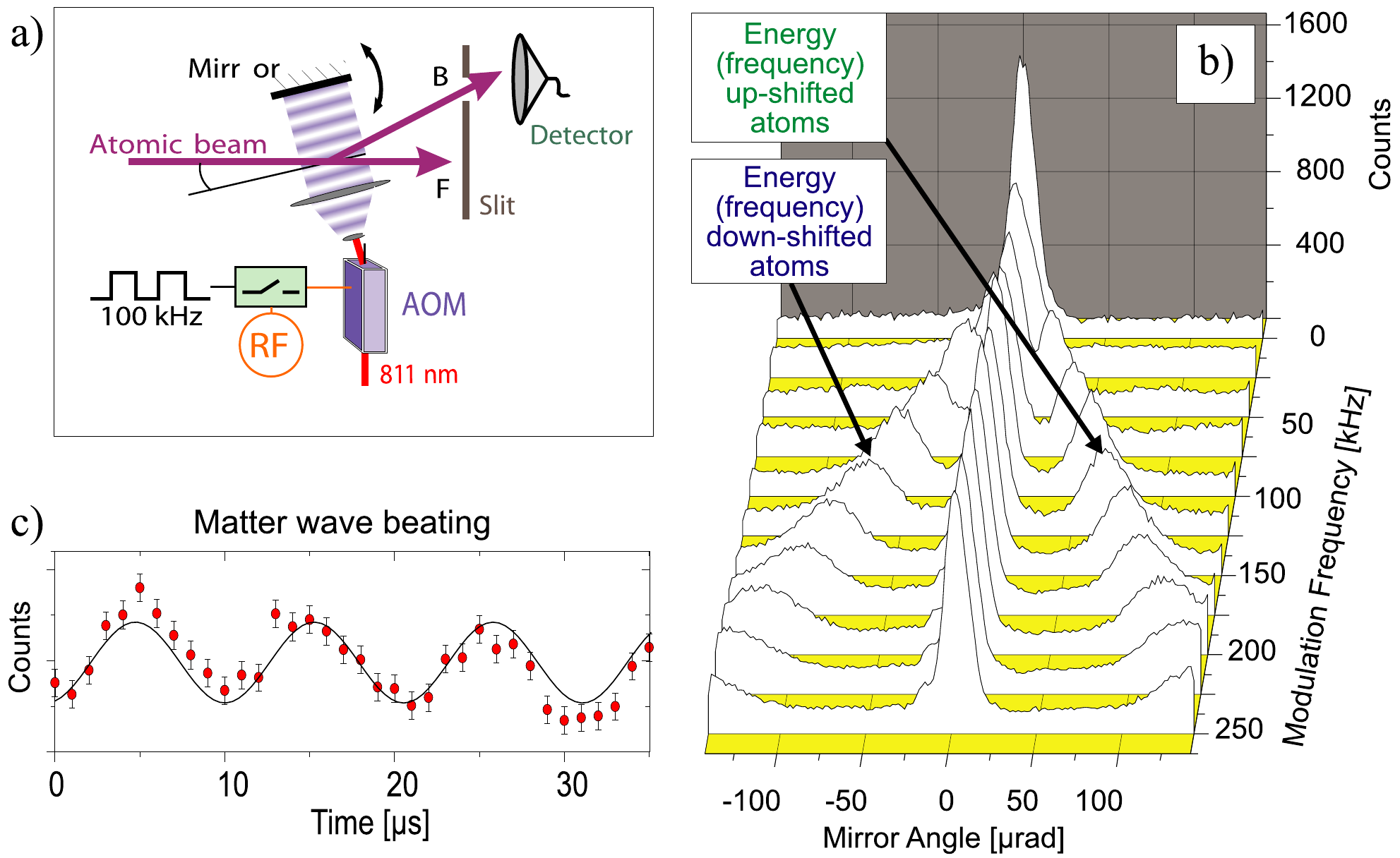}   
\caption{(color online) Frequency shifter for matter waves.  (a) A time-modulated light crystal causes diffraction in time and space. (b) Rocking curves show how the Bragg angle for frequency-shifted matter waves is controlled by the grating modulation frequency. (c) Matter wave interference in time. Figure from ref.~\cite{BOA96}.}
\label{fig:ODBM}
\end{figure}

The coherent frequency shift of the Bragg diffracted atoms can be measured by interferometric superposition with the transmitted beam. Directly behind the light crystal the two outgoing beams form an atomic interference pattern which can be probed by a thin absorptive light grating~\cite{AKB97}.  Since the energy of the diffracted atoms is shifted by $\hbar \omega_\textrm{{mod}}$, the atomic interference pattern continuously moves, This results in a temporally oscillating atomic transmission through the absorption grating (see figure~\ref{fig:ODBM}).

Starting from this basic principle of frequency shifting by diffraction from a time dependent light crystal many other time-dependent interference phenomena were studied for matter waves~\cite{BAK99,BAK00} developing a diffractive matter wave optics in time. For example using light from two different lasers one can create two coinciding light crystals. Combining real and imaginary potentials can produce a driving potential of the form $U(t) \sim e^{ \pm i \omega_m t}$ which contains only positive (negative) frequency components respectively.  Such a modulation can only drive transitions up in energy (or down in energy).

\begin{figure}[t]
\center\includegraphics[width = 0.6 \columnwidth]{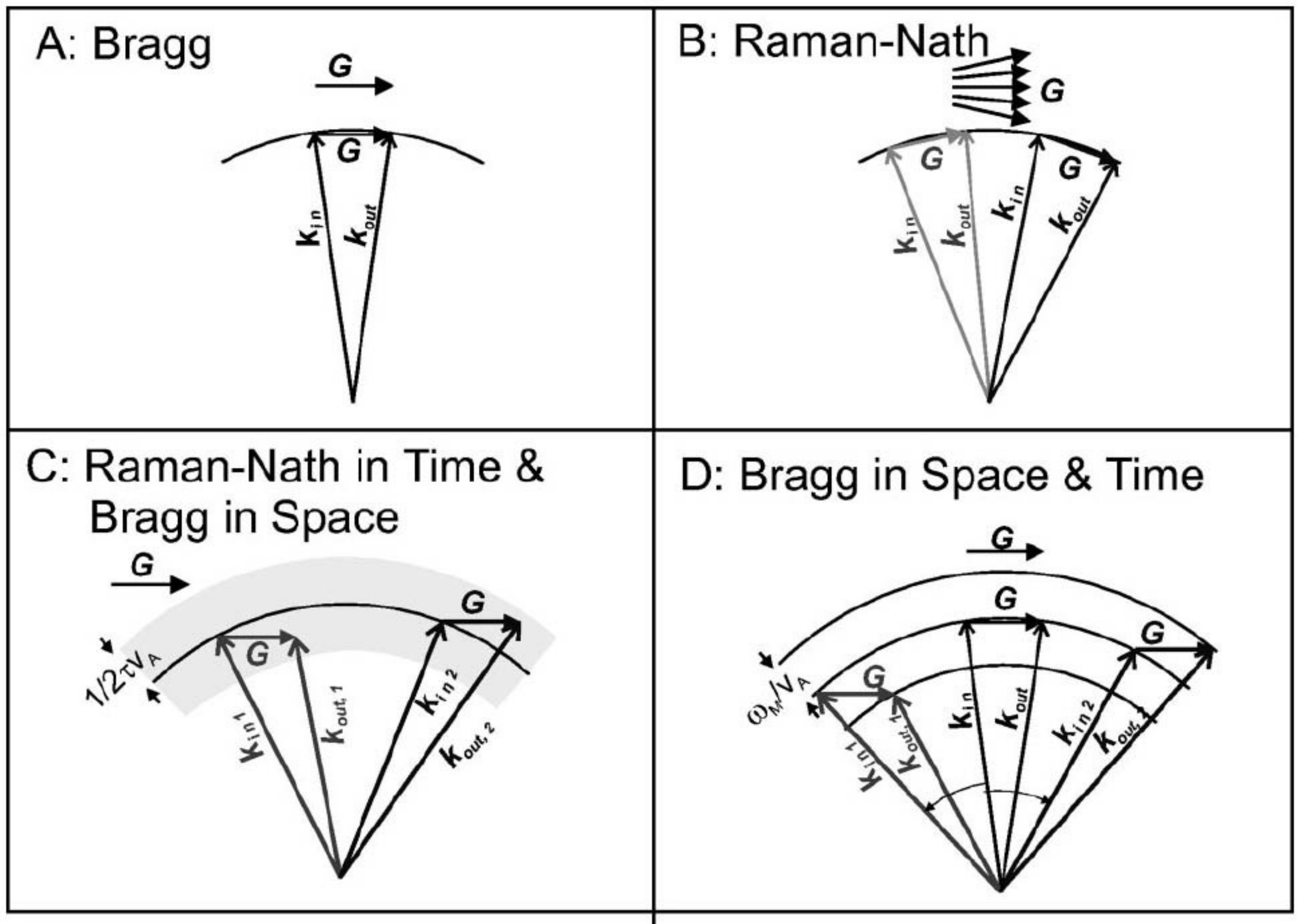}
\caption{Momentum diagrams for cases: (A) a thick grating, (B) a thin grating, (C) a thick pulsed grating (D) a thick harmonically modulated grating. Figure from ref.~\cite{BAK00}. \label{fig:RNB}}
\end{figure}

Figure \ref{fig:RNB} summarizes thick and thin gratings in space and also in time with Ewald constructions to denote energy and momentum of the diffracted and incident atom waves. The diffraction from (modulated) standing waves of light can also be summarized with the Bloch band spectroscopy picture~\cite{BAK00,CBD01}.

\subsection{Interferometers}  \label{ssec:IFM}

Interferometers are, very generally speaking, devices that utilize the superposition principle for waves and allow a measurement through the resulting interference pattern.  A generic interferometer splits an incoming beam $|\psi \rangle$ in (at least two) different components $| \psi_1 \rangle$, $| \psi_2 \rangle$,... which evolve along different paths in configuration space and then recombined to interfere. Interferometer exhibit a closed path and can be viewed, in a topological sense, as a ring.

At the output port of an interferometer the superposition of two interfering waves $\psi_{out} = \psi_1 + \psi_2 $ leads to interference fringes in the detected intensity:
\beqa \label{e_ifm_pattern1}
I=|\psi_{out}|^2&=&|\psi_1 + \psi_2|^2 \nonumber \\
                &=& A_1^2 + A_2^2 + 2 A_1  A_2 \cos(\phi) \\
                &=& \langle I \rangle \left(1+C \cos(\phi) \right) \nonumber
\eeqa
where $A_1=|\psi_1|^2$, $A_2=|\psi_2|^2$ are the amplitudes of the interfering beams and $\phi = \varphi_1 - \varphi_2$ is their phase difference.  The observed interference is then fully characterized by its phase $\Delta\varphi$, and by two of the following: its amplitude $2 A_1 A_2$, its average intensity $\langle I \rangle = A_1^2 + A_2^2$, or its contrast $C$ given by:
\beq \label{e_contrast}
C = \frac{I_{max} - I_{min}}{I_{max} + I_{min}} = \frac{2 A_1  A_2 }{A_1^2 + A_2^2}.
\eeq
If one of the interfering beams is much stronger then the other, for example $|\psi_1 |^2 \gg |\psi_{2} |^2$, then the contrast of the interference pattern scales like
 \beq
    C \sim \frac{2 |\psi_2 |}{|\psi_1 |} = 2 \sqrt{\frac{I_2}{I_1}}.
 \eeq
Consequently one can observe $20 \%$ ($2 \%$) contrast for an intensity ratio of 100:1 ($10^4$:1) in the interfering beams.

While the overall phase of the wave function is not observable, the power of interferometers lies in the possibility of measuring the phase difference $\Delta \varphi$ between waves propagating along two different paths.
\beq \label{e:phaseshift_IFM}
\Delta \varphi =  \varphi_1 - \varphi_2 = \frac{1}{\hbar}(S_{1} - S_{2})
\eeq
where $\varphi_1$ ($S_{1}$) and $\varphi_2$ ($S_{2}$) are the phases (classical action) along path 1 (path 2) of the interferometer. The other parameters of the interference pattern, the amplitude $2 A_1 A_2$, the contrast $C$, and the mean intensity $\langle I \rangle$ also give information about the paths.  Especially the contrast of the interference pattern tells us about the coherence in the interferometer.

From a practical point of view interferometers can be divided in two categories:
\begin{itemize}
\item In {\em internal state interferometers}, the beam splitter produces a superposition of internal states which can be linked to external momentum. Examples are polarisation interferometry with light and Ramsey spectroscopy for internal states of massive particles.

	\item In {\em de~Broglie wave interferometers} the  beam splitter does {\em not} change the internal state but directly creates a superposition of external center of mass states and thus distinctly different paths in {\em real space}.  The spatially distinct interfering paths can be created by \emph{wavefront division} like in{\em Young's double slit} or by \emph{amplitude division} as realizes by a beam splitter in optics or by diffraction.   In matter wave optics amplitude division Mach-Zehnder interferometer can be built with three gratings.

\end{itemize}

When designing and building interferometers for beams of atoms and molecules, one must consider their specifics. 
(1) Beams of atoms and molecules have a wide energy distribution and consequently the coherence lengths for matter waves is in general very short ($\sim$100 pm  for thermal atomic beams, and seldom larger then 10 $\mu m$ even for atom lasers or BEC). This requires that the period and the position of the interference fringes must be independent of the de~Broglie wavelength of the incident atoms.  In optical parlance this is a property of \emph{white light} interferometers like the three grating Mach-Zehnder configuration.
(2) Atoms interact strongly with each other. Therefore the optics with matter waves is non-linear, especially in the cases where the atoms have significant density as in a BEC or atom laser.
(3) Atoms can be trapped which allows a different class of interferometers for confined particles, which will be disused in a later section.

\subsubsection{Three-grating Mach-Zehnder interferometer}  \label{sssec:MZ_IFM}

The challenge of building a white light interferometer for matter waves is most frequently met by the 3-grating Mach Zehnder (MZ) layout. In symmetric setup the fringe spacing is independent of wavelength and the fringe position is independent of the incoming direction \footnote{Diffraction separates the split states by the lattice momentum, then reverses this momentum difference prior to recombination.  Faster atoms will diffract to smaller angles resulting in less transverse separation downstream, but will produce the same size fringes upon recombining with their smaller angle due to their shorter deBroglie wavelength.  For three evenly spaced gratings, the fringe \emph{phase} is independent of incident wavelength.}. This design was used for the first \emph{electron} interferometer~\cite{mar52}, for the first \emph{neutron} interferometer by H. Rauch~\cite{Rauch74}, and for the first atom interferometer that spatially separated the atoms~\cite{Keith91}.  In the 3-grating Mach Zehnder interferometer the role of splitter, recombiner and mirror is taken up by diffraction gratings.  At the position of the third grating (G3) an interference pattern is formed with the phase given by 
\beq \phi = G(x_1 - 2 x_2 + x_3) + \Delta \phi_{int} \label{eq:3g std phase}\eeq 
where $x_1$, $x_2$, and $x_3$ are the relative positions of gratings 1, 2 and 3 with respect to an inertial frame of reference ~\cite{BER97}.

It is interesting to note that many diffraction-based interferometers produce fringes even when illuminated with a source whose transverse coherence length is much less than the (large) physical width of the beam.  The transverse coherence can even be smaller then the grating period.  Under the latter condition, the different diffraction orders will not be separated and the arms of the interferometer it will specially overlap.  Nevertheless, high contrast fringes will still be formed.

The three grating interferometer produces a ``position echo" as discussed by CDK~\cite{CDK85}.  Starting at one grating opening, one arm evolves laterally with $\hbar G$ more momentum for some time, the momenta are reversed, and the other arm evolves with the same momentum excess for the same time, coming back together with the other arm at the third grating.  If the gratings are registered, its trapezoidal pattern starts at a slit on the first grating, is centered on either a middle grating slit or groove, and recombines in a slit at the third grating. Not surprisingly, spin-echo and time-domain echo techniques (discussed below) also offer possibilities for building an interferometer that works even with a distribution of incident transverse atomic momenta.

\paragraph{Interferometer with nano fabricated gratings}

The first 3-grating Mach-Zehnder interferometer for atoms was built by Keith et al.~\cite{Keith91} using three 0.4-$\mu$m period nano fabricated diffraction gratings. Starting from a supersonic Na source with a brightness of $B\approx10^{19}$ s$^{-1}$cm$^{-2}$sr$^{-1}$ the average count rate  $\langle I \rangle$, in the interference pattern was 300 atoms per second. Since then, gratings of 100 nm period have been used to generate fringes with 300000 atoms per seconds.

\begin{figure}[t]
\center \includegraphics[width = \columnwidth]{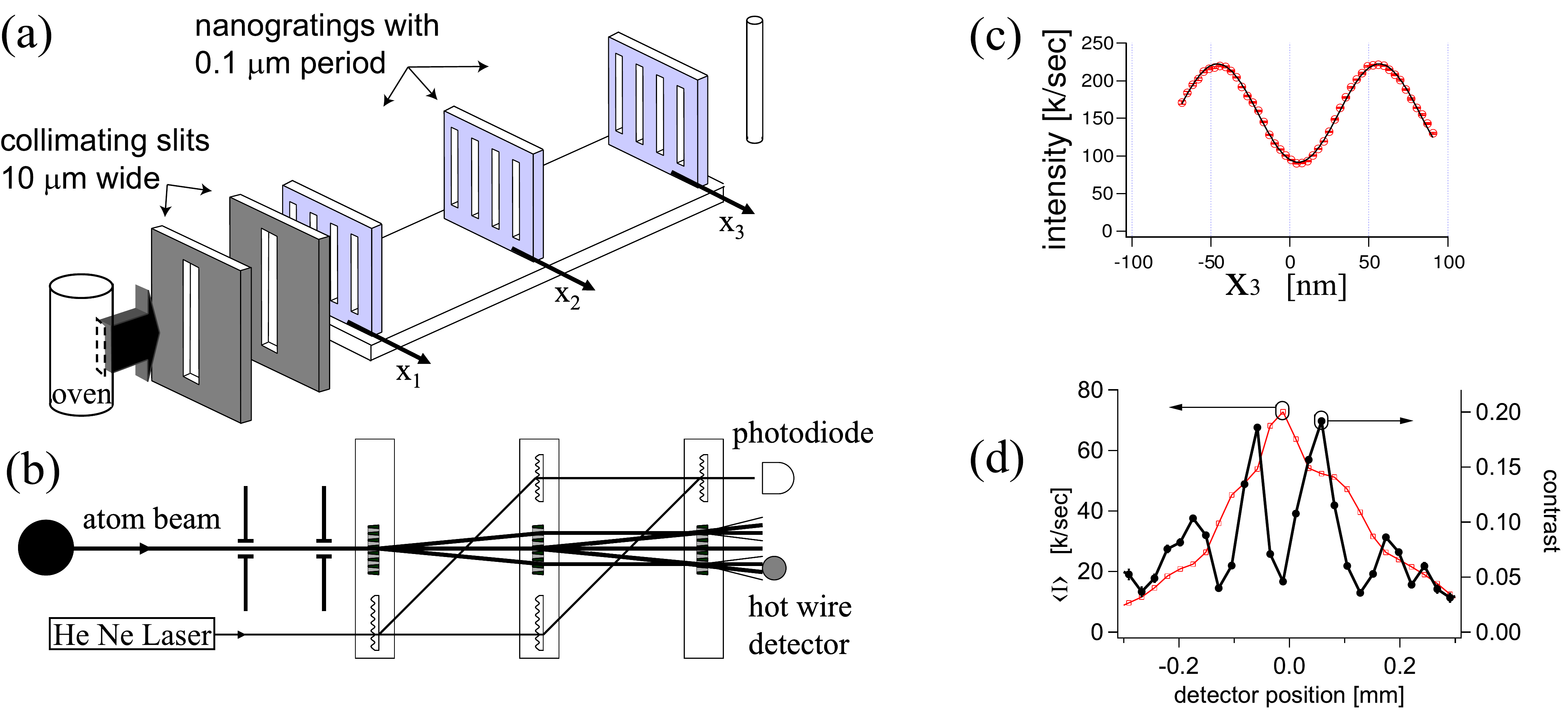}   
\caption{(a) and (b) Atom Interferometer setup used in Keith et al.~\cite{Keith91}. (c) Interference fringe data and best fit with $\langle I \rangle$ = 157,000 counts per second and $C = 0.42$. A total of 5 seconds of data are shown and the uncertainty in phase is $\sigma_{\phi}= 2.7 \times 10^{-3}$ radians. (d) Average intensity $\langle I \rangle$ and contrast $C$ as a function of detector position [(under different conditions than (c)].} \label{fig:marton}
\end{figure}

Following the design shown in Fig.~\ref{fig:marton}) there are two the MZ Interferometers formed starting with a common incident beam.  One by two paths created by $0^\text{th}$, $+1^\text{st}$ and $+1^\text{st}$, $-1^\text{st}$ order diffraction at grating G1, G2 respectively. The second one formed symmetrically by two paths created by $0^\text{th}$, $-1^\text{st}$ and $-1^\text{st}$, $+1^\text{st}$ order diffraction at grating G1, G2 respectively.
In each interferometer loop the difference in momentum is one unit of $\hbar G \hat{x}$ and at the position of the third grating G3 an interference pattern forms in space as a standing matter wave wave with a period of $d=G/(2\pi)$ and a phase that depends on the location $x_1$ and  $x_2$ of the two gratings G1 and G2 as well as the interaction phase $\Delta \phi_{int}$.  These fringes can be read out in different ways. The simplest is to use the third grating as a mask to transmit (or block) the spatially structured matter wave intensity. By translating G3 along $x$ one obtains a moir\'{e} filtered interference pattern which is also sinusoidal and has a mean intensity  and contrast
\beq
    \langle I \rangle = \frac{w_3}{d} \langle \tilde{I} \rangle  \quad C = \frac{\sin(G w_3/2)}{(G w_3/2)} \tilde{C}. \label{eq:CCp}
\eeq
where $\tilde{I}$ and $\tilde{C}$ refer to the intensity and contrast just prior to the mask.

There are in fact many more interferometers formed by the diffraction from absorption  gratings. For example, the 1$^{\mathrm{st}}$ and 2$^{\mathrm{nd}}$ orders can recombine in a skew diamond to produce another interferometer with the white fringe property. The mirror images of these interferometers makes contrast peaks on either side of the original beam axis (figure~\ref{fig:marton}). In the symmetric MZ interferometer all those interferometers have fringes with the same phase, and consequently one can therefore build interferometers with wide uncollimated beams which have high count rate, but lower contrast (the contrast is reduced because additional beam components which do not contribute to the interference patterns such as the zeroth order transmission through each grating will also be detected).

For well-collimated incoming beams, the interfering paths can be separated at the $2^\text{nd}$ grating. For example in the interferometer built at MIT the beams at the $2^\text{nd}$ grating have widths of 30 $\mu$m and can be separated by 100 $\mu$m (using 100-nm period gratings and 1000 m/s sodium atoms ($\lambda_\text{dB} = 16$ pm). Details of this apparatus, including the auxiliary laser interferometer used for alignment and the requirements for vibration isolation, are given in ref.~\cite{BER97}.


\begin{figure}[t]
\includegraphics[width = \columnwidth]{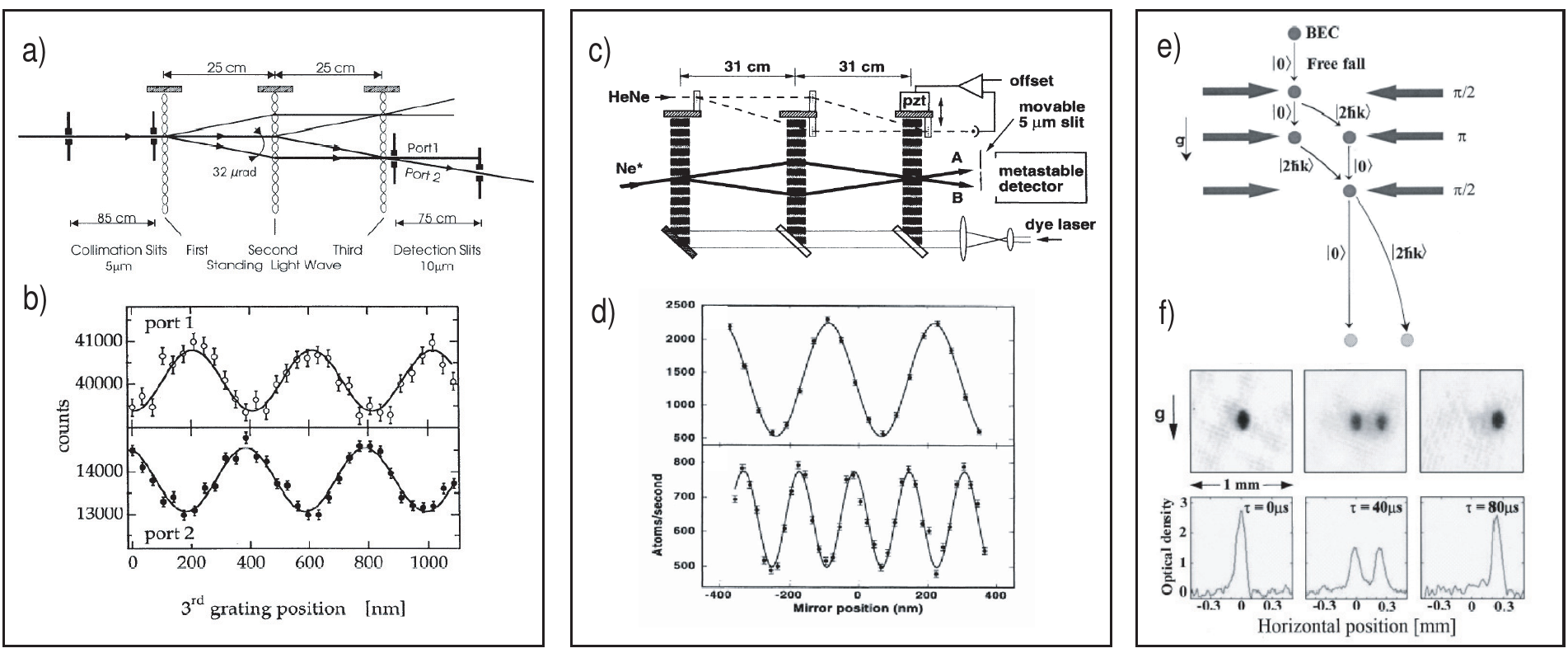}   
\caption{Atom interferometers based on three standing waves of light.  (a) Atom beam and three Kapitza-Dirac gratings.  
(b) Atominterference patterns for both output ports demonstrate complementary intensity variations. This is a consequence of atom number conservation. Figures a and b reproduced from~\cite{ROB95}.
(c) Interferometer based on three Bragg gratings. Dashed line shows the path of auxiliary optical interferometer used for stabilization. 
(d) Intensity fluctuations in beam A vs. position of the Bragg gratings.  For second order Bragg diffraction, fringes of half the period are formed. Figures c and d reproduced from~\cite{GML95}. 
(e) Schematic of the $\pi/2-\pi-\pi/2$ Bragg interferometer for atoms in a BEC falling from a trap. 
(f)Absorption images and density profiles demonstrating different outputs of the interferometer. Figures e and f reproduced from \cite{TSK00}. 
\label{fig:rob1} }
\end{figure}

\paragraph{Interferometers with light gratings}

One can also build MZ interferometers with near-resonant standing light waves which make species-specific phase gratings (Fig.~\ref{fig:rob1}).  The third grating can function to recombine atom waves so their relative phase dictates the probability to find atoms in one output port (beam) or another.  Alternatively, fringes in position space can be detected with fluorescence from a resonant standing wave.  Another detection scheme uses backward Bragg scattering of laser light from the density fringes.  Detecting the direction of exiting beams requires that the incident beams must be collimated well enough to resolve diffraction, and may well ensure that the beams are spatially separated in the interferometer.
Because they transmit all the atoms, light gratings are more efficient than material gratings.

Rasel et al.~\cite{ROB95} used light gratings in the Kapitza-Dirac regime with a $5 \mu$m wide collimated beam. Many different interferometers are formed, due to symmetric KD diffraction into the many orders. Two slits after the interferometer served to select both the specific interferometer, and the momentum of the outgoing beam (ports 1 and 2 in figure~\ref{fig:rob1}a). Fringes show complementary intensity variations, as expected from particle number conservation in a MZ interferometer with phase gratings.

A group in Colorado used Bragg diffraction at various orders to built MZ interferometer for a Ne$^*$ beam ~\cite{GML95} (figure \ref{fig:rob1}b).  A Bragg scattering interferometer for Li atoms with high contrast and a count rate of 17 kc/s was used to measure the polarizability of Li atoms~\cite{DCB02,MJB06,MJB06b}.

\paragraph{Talbot-Lau (near-field) interferometer}

\begin{figure}[t]
\begin{center}
\includegraphics[width = 8cm]{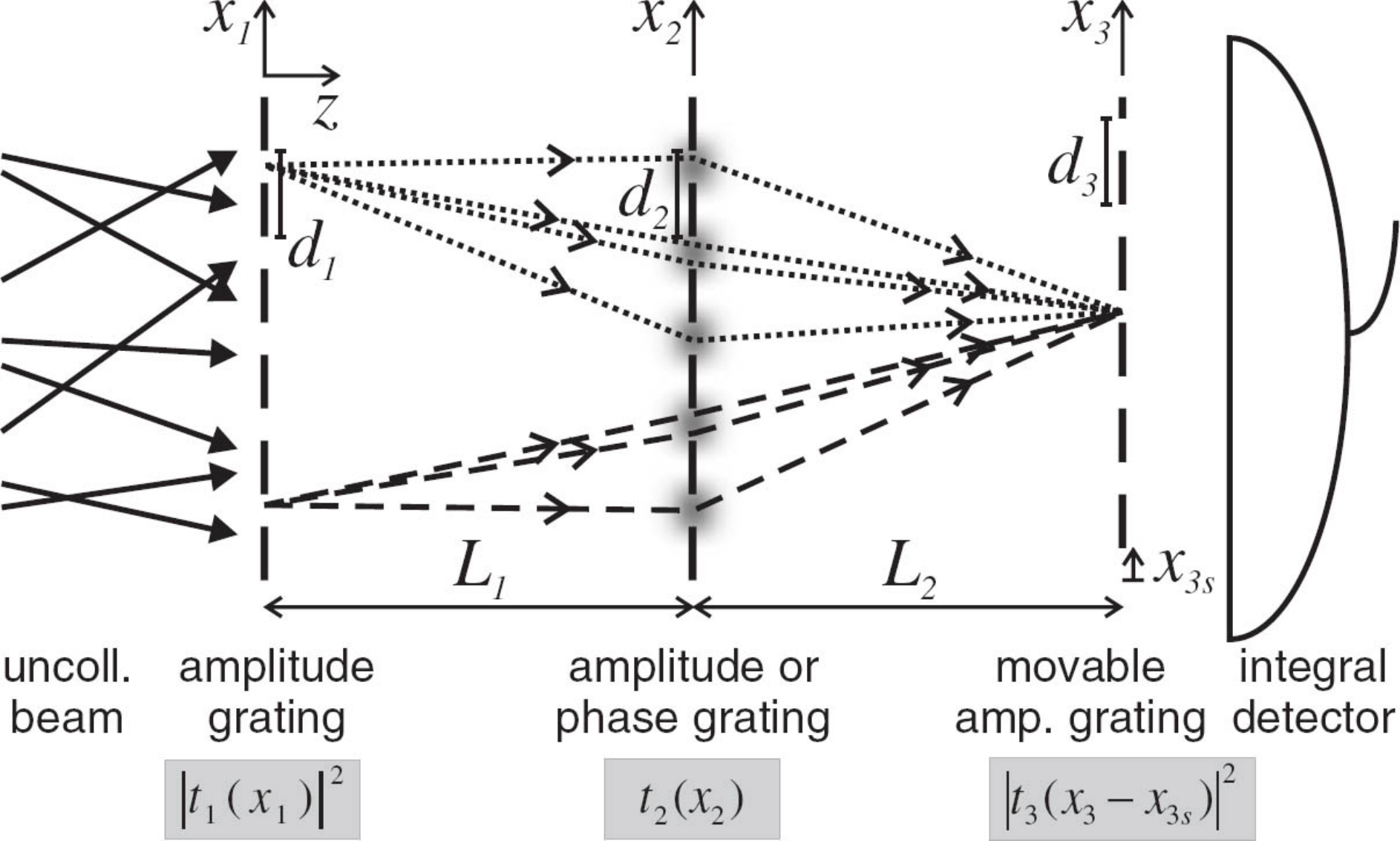}    
\caption{A sketch of the Talbot-Lau interferometer setup consisting of three gratings.  The first grating is illuminated by an uncollimated molecular beam.  Still, coherent interference occurs between all paths that originate from one point at the first grating and meet at the a point on the third grating.  By varying the grating position $x_{3}$, a periodic pattern in the molecular distribution can be detected. Figure from ref.~\cite{BAZ03}.
\label{fig:TLIsketch}}
\end{center}
\end{figure}

A high degree of spatial coherence is needed to create recurring self-images of a grating due to near-field diffraction (the Talbot effect).  But completely incoherent light can still produce fringes downstream of a grating pair (the Lau effect).  Two gratings with equal period $d$ separated by a distance $L_1$ create a the Laue fringe with period $d'$ at a distance $L_2$ beyond the second grating:   
\beq 
L_2 = \frac{L_1 L_T \frac{n}{2m}}{L_1 - L_T \frac{n}{2m}} \quad \quad  d' = d \frac{L_2 + L_1}{m L_1} 
\eeq 
Here $L_T=2d^2/\lambda_{dB}$ is the Talbot length and the integers $n$ and $m$ refer to the $n^{th}$ revival of the $m^{th}$ order Fourier image. If a $3^{rd}$ grating with period $d'$ is used as a mask to filter these fringes, then a single large-area integrating detector can be used to monitor the fringes. In such a 3-grating Talbot-Lau Interferometer (TLI) the contrast is unaffected by the beam width and a large transverse momentum spread in the beam can be tolerated, hence much larger count rates can be obtained. The TLI does not separate the orders - components of the wave function are only displaced by one grating period at the Talbot length, it is still sensitive to inertial forces, decoherence, and field gradients.

In a TLI the relationship $L_1=L_2=L_T/2$ means that the maximum grating period is $d < \sqrt{L_1 \lambda_{dB}} \sim M^{-1/4}$ where $M$ represents mass for a thermal beam.  In comparison, for a MZI with resolved paths the requirement is $d< \lambda_{dB} L / (\Delta x) \sim M^{-1/2}$ where $\Delta x$ is the width of the beam and $L$ is the spacing between gratings. Thus the TLI design is preferable for demonstration of interference for large mass particles (see lectures by M.Arndt).

A Talbot-Lau interferometer was first built for atoms by John Clauser~\cite{CLL94}. Using a slow beam of potassium atoms and gratings with a period of $d$=100 $\mu$m, and a count rate of $\langle I \rangle = 4 \times 10^7$ atoms/sec was achieved, even though the source brightness was 2500 times weaker than in the 3 grating Mach Zehnder interferometer at MIT, but the signal was about 3000 times stronger.  Because of its attractive transmission features, and the favorable scaling properties with $\lambda_{dB}$, the TLI has been used to observe interference fringes with complex molecules such as C$_{60}$, C$_{70}$, C$_{60}$F$_{48}$, and C$_{44}$H$_{30}$N$_4$~\cite{BHU02,HUH03}.

\subsubsection{Selected experiments with beam interferometers}  \label{sssec:Experiments}

\paragraph{Examples of phase shifts}

We will now briefly discuss typical phase shifts that can be observed using an interferometer.  In the JWKB approximation the phase shift induced by an applied potential is given by:
\beq \label{e:phase_shift0}
        \phi (k_0) = \int_{\Gamma_{cl}}(k(x)-k_0(x)) \, dx =
        \int_{\Gamma_{cl}}\Delta k(x) \, dx
         = k_0\int_{\Gamma_{cl}}(n(x)-1)\, dx
\eeq
$\Gamma_{cl}$ is the classical path and $k_0$ and $k$ are the unperturbed and perturbed $k$-vectors, respectively and $n$ is the refractive index as given in equation~\ref{e:refIndex_pot}.  If the potential $V$ is much smaller than the energy of the atom $E$ (as is the case for most of the work described here) the phase shift can be expanded to first order in $V/E$.   
If $V$ is time-independent, one can furthermore transform the integral over the path into one over time by using $t = x/v$.
\beq \label{e:phase_shift2}
        \Delta \varphi (k_0) = -\frac{1}{\hbar v}\int_{\Gamma_{cl}}V(x) \, dx   = -\frac{1}{\hbar}\int_{t_a}^{t_b}V(t) \, dt
\eeq
where $n$ is the refractive index as given in Eq.~\ref{e:refIndex_pot}.

A constant scalar potential $V$ applied over a length $L_{int}$ results in a phase shift $\phi (k_0) = -\frac{m}{\hbar_2 k_o} V L_{int}$ .  The power of atom interferometry is that we can measure these phase shifts very precisely. A simple calculation shows that 1000 m/s Na atoms acquire a phase shift of 1 rad for a potential of only $V=6.6 \times 10^{-12}$\,eV in a 10\,cm interaction region.  Such an applied potential corresponds to a refractive index of $|1-n| = 2.7 \times 10^{-11} $.  Note that positive $V$ corresponds to a repulsive interaction that reduces $k$ in the interaction region, giving rise to an index of refraction less that unity and a negative phase shift.

Equation \eqref{e:phase_shift2} further more shows that the phase shift associated with a constant potential depends inversely on velocity and is therefore dispersive (it depends linearly on the de~Broglie wavelength). If, on the other hand, the potential has a linear velocity dependence, as in $U = \ve{\mu} \cdot \frac{1}{c} (\ve{v} \times \ve{E})$ for a magnetic dipole $\ve{\mu}$ in an electric field $\ve{E}$, the phase shift becomes independent of the velocity $\ve{v}$~\cite{Cimmino92, Sangster94, Zeiske95, Gorlitz95}.  Similarly, a potential applied to all particles for the same length of time, rather than over a specific distance, will produce a velocity independent phase shift $\phi = -\frac{1}{\hbar}\int_{t_a}^{t_b}V(t) \, dt $.  The latter is related to the scalar Aharonov Bohm effect~\cite{Zeilinger_ScalAB, Allm93, Badu93}.

\paragraph{Electric polarizability }

\begin{figure}[t]
\center \includegraphics[width = 10cm]{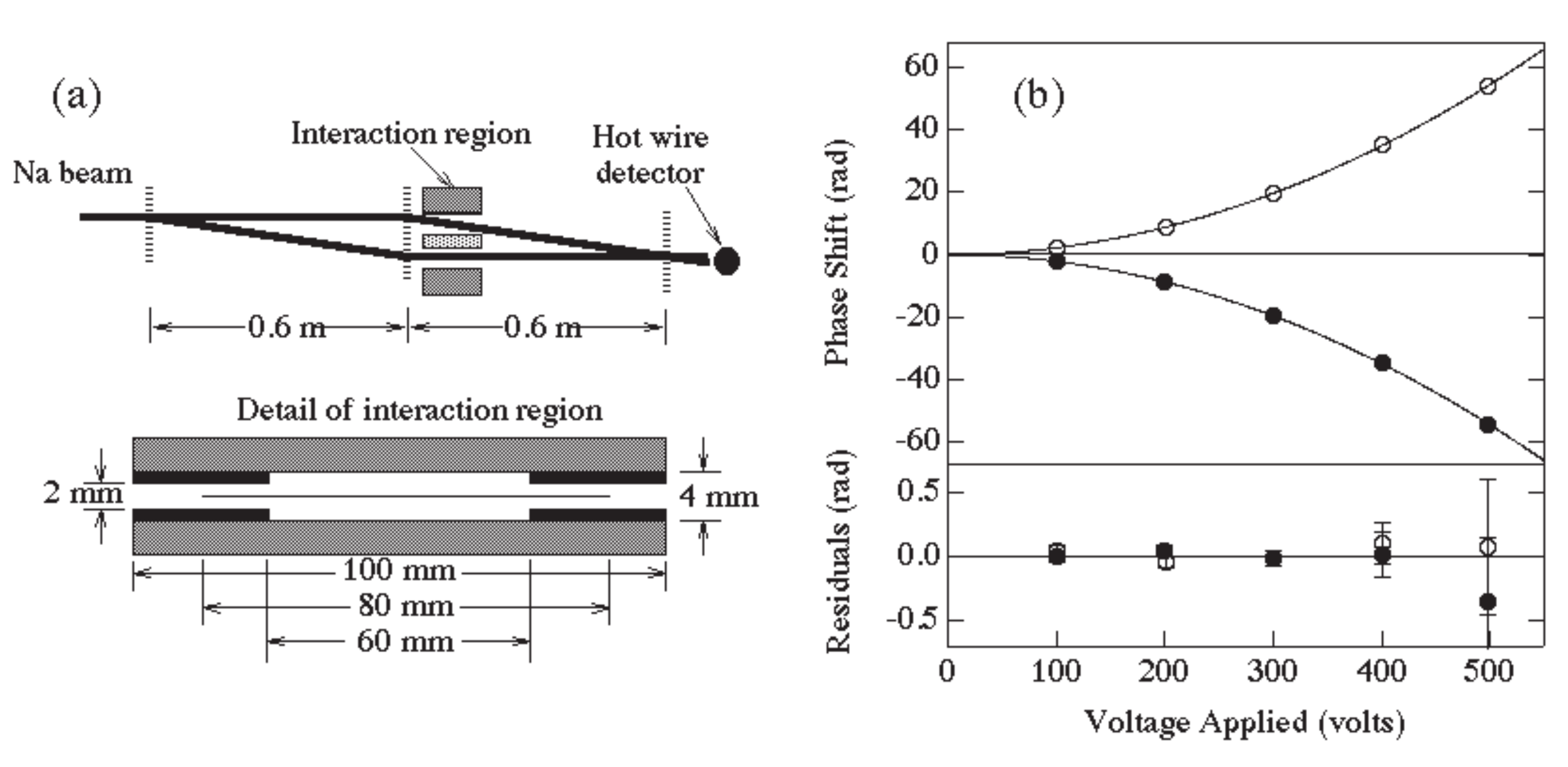}    
\caption{Measurement of atomic polarizability. (a) Schematic of the interaction region installed behind the second grating. 
(b) Measured phase shifts vs. applied voltage. The two different signs of the phase shift stem from the voltage being applied on either the left (open circles) or the right (filled circles) side of the interaction region (arm of the interferometer).  The fit is to a quadratic and the residuals are shown on the lower graph. Figure from ref.~\cite{BER97}.} 
\label{fig:polarizability}
\end{figure}

By inserting a metal foil between the two separated arms, as shown in figure~\ref{fig:polarizability}, an a uniform electric field $E$  can be applied to to one of the separated atomic beams, shifting its energy by the Stark potential $U=-\alpha E^2/2$. The static scalar ground-state polarizability $\alpha_{pol}$ can then be determined from the phase shift, $\phi$, of the interference pattern by
\beq
\alpha_{pol} = \left(\frac{\phi} {V^2} \right) \left( \frac{D^2}{L_\text{eff}} \right) (2\hbar v), \label{eq:polphase}
\eeq 
where $V$ is the voltage applied to one electrode in the interaction region, $D$ is the distance between the electrode and the septum,  $v$ is the mean velocity of the atomic beam, and $L_{eff}$ is the effective length of the interaction region defined as $\left( \frac{V}{D} \right)^2 L_\text{eff} \equiv \int E^2 dz$.

For an accurate determination of electric polarizability, the three factors in equation~\eqref{eq:polphase} must each be determined precisely.  They are (1) the phase shift as a function of applied voltage, (2) the geometry and fringing fields of the interaction region, and (3) the velocity of the atoms.  In ref.~\cite{ESC95} the uncertainty in each term was less than 0.2\%. This allowed to extract the static ground-state atomic polarizability of sodium to $\alpha_{pol}=24.11\times 10^{-24}$ cm$^3$, with a fractional uncertainty of 0.35\%~\cite{ESC95}. Similar precision has been demonstrated for $\alpha_{He}$ by the Toennies group~\cite{TOE01} and $\alpha_{Li}$ with a precision of 0.66\% by the Vigu\'{e} group~\cite{MJB06,MJB06b}. These experiments offer an excellent test of atomic theory.

\paragraph{Refractive index}

\begin{figure}[t]
\center\includegraphics[width = 10cm]{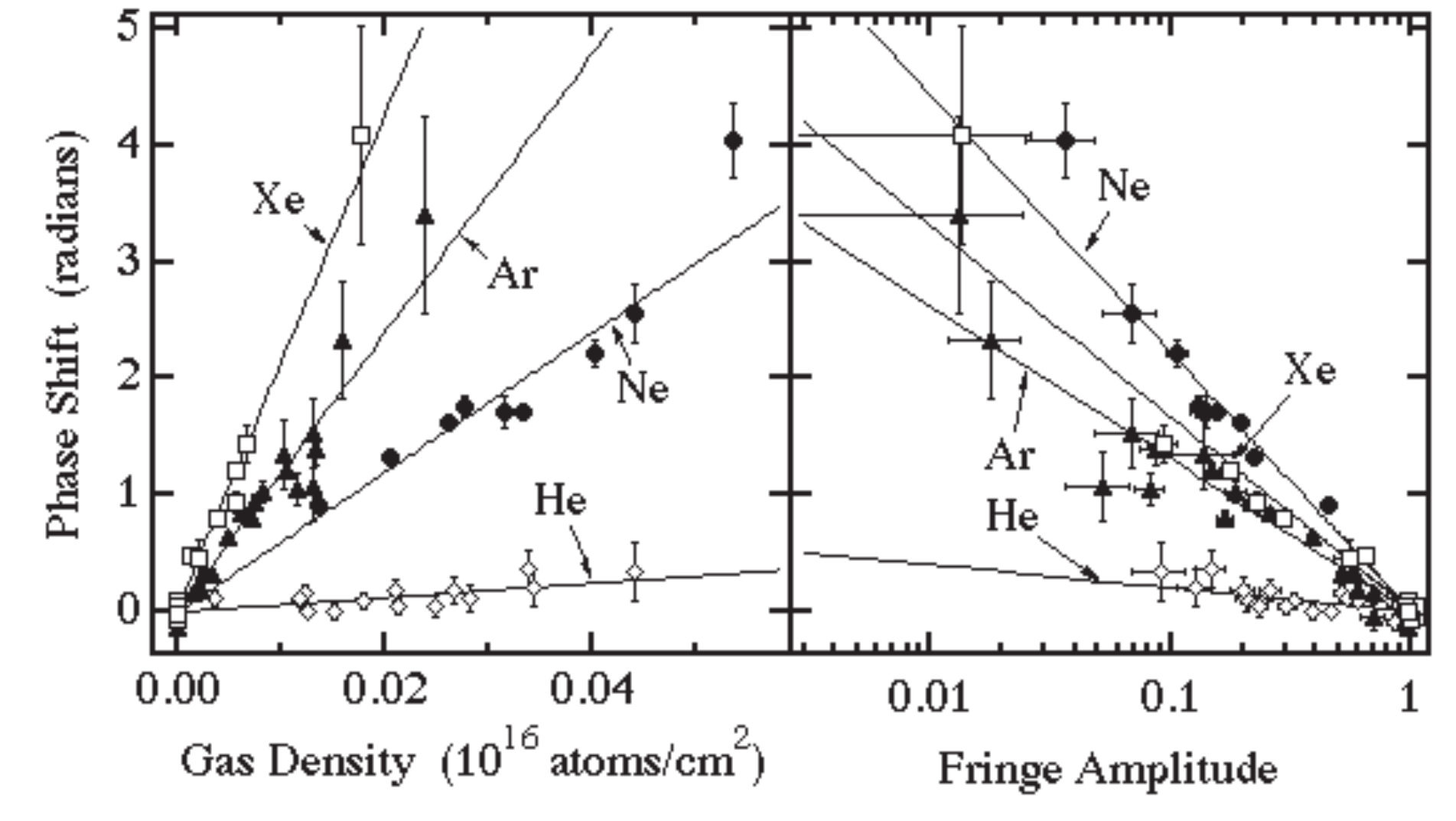}    
\caption{(Left) Phase shift $\phi$ as a function of gas density $N$ for different gas samples. (Right) Phase shift vs Fringe amplitude.  The fringe amplitude is proportional to $e^{-N\sigma_{tot}z/2}$. Figure from ref.~\cite{BER97}.
\label{fig:gas-index}}
\end{figure}

A physical membrane separating the two paths allows to insert a gas into one path of the interfering wave.
Atoms propagating through the gas are phase shifted and attenuated by the index 
\beq  \psi(z) = \psi(0)e^{inkz} =\psi(0)e^{ikz} e^{i \phi(N,z)} e^{-\frac{N}{2} \sigma_{tot} z}.  \eeq
The phase shift due to the gas, 
\beq \phi(N,z) =(2\pi N k z / k_{cm}) \mathrm{Re}[f(k_{cm})], \eeq 
is proportional to the \emph{real} part of the forward scattering amplitude, while the attenuation is related to the \emph{imaginary} part. Attenuation is described by the total scattering cross section, and this is related to $\mathrm{Im}[f]$ by the optical theorem
\beq \sigma_{tot} = \frac{4 \pi}{k_{cm}} \mathrm{Im}[f(k_{cm})]\label{eq:optical-theorem}.\eeq
Measurements of phase shift as a function of gas density are shown in figure~\ref{fig:gas-index}.

The ratio of the real and imaginary parts of the forward scattering amplitude is a natural quantity to measure and compare with theory.  This ratio, 
\beq \rho(k) = \frac{\phi(N)}{ln[A(N)/A(0)]} = \frac{\mathrm{Re}[f(k)]}{\mathrm{Im}[f(k)]}.\eeq
where $A$ is the fringe amplitude, gives orthogonal information to the previously studied total scattering cross section.  In addition it is independent of the absolute pressure in the scattering region and therefore much better to measure.

The motivation for studying the phase shift in collisions is to add information to long-standing problems such as inversion of the scattering problem to find the interatomic potential $V(r)$, interpretation of other data that are sensitive to long-range interatomic potentials, and description of collective effects in a weakly interacting gas \cite{CHS89,BMT93,CMH94,LHP93,WAF94,MOV94,MOS94,STO91,SHV94}. These measurements of $\rho$ are sensitive to the shape of the potential near the minimum, where the transition from the repulsive core to the Van der Waals potential is poorly understood. The measurements of $\rho(k)$ also give information about the rate of increase of the interatomic potential $V(r)$ for large $r$ independently of the strength of $V(r)$. The real part of $f$ was inaccessible to measurement before the advent of separated beam atom interferometers.  

\paragraph{Measurement of the Coherence length} 

\begin{figure}[t]
\begin{center}
\includegraphics[width = 8cm]{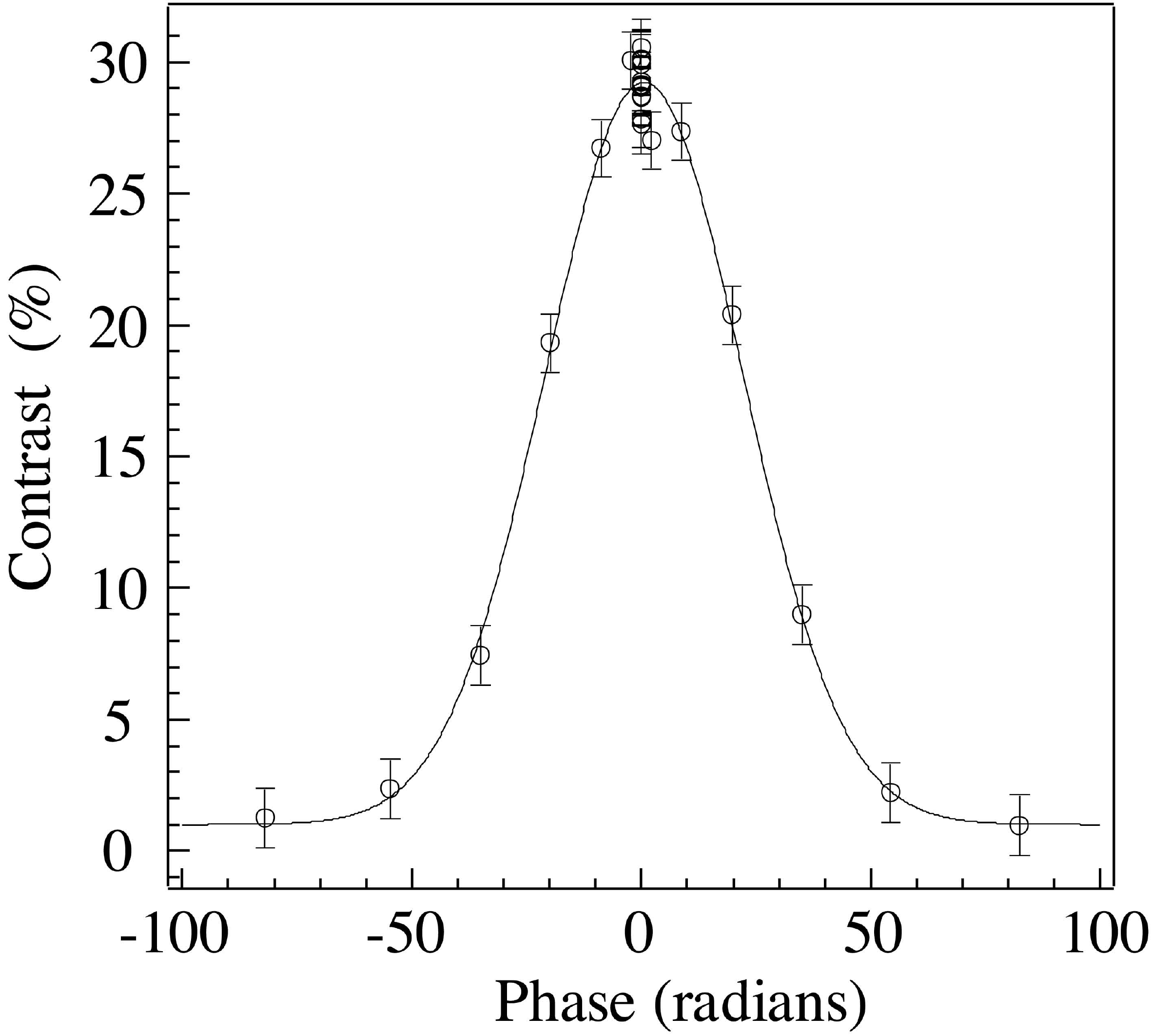} 
\caption{Measurement of the coherence length of a Na atomic beam with de~Broglie wavelength of $\lambda_\text{dB}=170$\,pm in an atom interferometer. The phase shift was realized using the electric polarizability of the atoms, by applying an electric field. \label{fig:CohLengthNa}}
\end{center}
\end{figure}

The coherence length $\ell_{c}$ of an atomic beam is related to its momentum distribution $\sigma_{p}$ by eq.\ref{e:lcoh_long} to $ \ell_c  = \frac{1}{\sigma_k} = \frac{\langle \lambda_\text{dB} \rangle}{2 \pi} \frac{\langle p \rangle}{\sigma_p}$. The longitudinal coherence length $\ell_{lcoh}$ limits the size of optical path difference between two arms of an interferometer before the contrast is reduced. 

The coherence length $\ell_{l}$ can be measured directly in an atom interferometer.  Applying a classical potential in one path results in a phase shift $\phi$ and simultaneously in a spatial shift of the wave function relative to the other path by $\delta x = \frac{\phi}{2 \pi} \lambda_{dB}$. This allows a direct measurement of the first order coherence function 
$g^{(1)}(\delta x)  =  \langle \psi(x) | \psi(x + \delta x) \rangle    =  \int{ \psi^{*}(x) \psi(x + \delta x) \, d x}$. An example of a measurement for a Na atomic beam is shown in figure \ref{fig:CohLengthNa}. 

\subsection{Einsteins recoiling slit: a single photon as a coherent beamsplitter}  \label{sssec:SPhBS}.  Up to now when using ligh fields to manipulate atomic motion we were using classical light which is described by a classical electro magnetic wave.  We now discuss the other extreme: An experiment where a single emitted photon is used as a beam splitter

In spontaneous emission an atom in an excited state undergoes a transition to the ground state and emits a single photon. Associated with the emission is a change of the atomic momentum due to photon recoil~\cite{SPBS:Milonni_Book} in the direction opposite to the photon emission. The observation of the emitted photon direction implies the knowledge of the atomic momentum resulting from the photon-atom entanglement. 

\begin{figure}[t]
\center\includegraphics[width = 12cm]{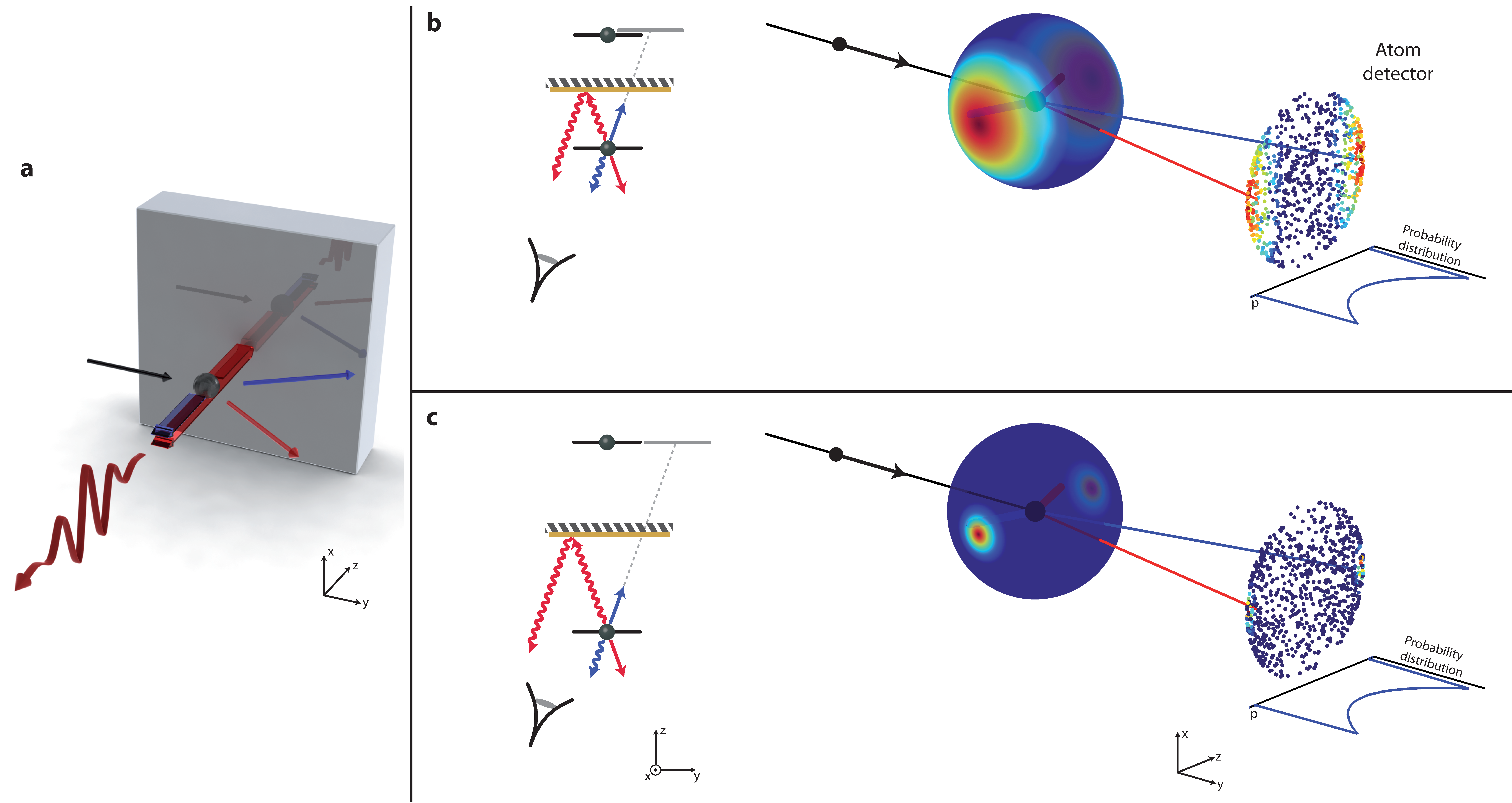}    
\caption{Motional coherence generated by a single spontaneous emission event. (a) An atom in front of a mirror spontaneously emits a single photon. For emission perpendicular to the mirror surface an observer can in principle not distinguish if the photon has been reflected or not. Momentum conservation in the atom-photon system implies that the atom after the emission is in a coherent superposition of two different momentum states separated by twice the photon recoil. 
(b) With the spatial extension of the atom corresponding to the optical absorption cross section, indistinguishability can be estimated by the projected overlap of atom and its mirror-image. This overlap is represented color coded on a sphere for all emission directions (red: full coherence, blue: no coherence) and on the pattern at the atom detector. 
(c) For large distance between the atom and the mirror the coherent fraction drastically reduces, approaching the limit of vanishing coherence in free space. Figure from ref.~\cite{SPB_natphys}.}
\label{fig:SPB_fig1}
\end{figure}

If the spontaneous emission happens very close to a mirror the detection of the photon does not necessarily reveal if it has reached the observer directly or via the mirror. For the special case of spontaneous emission perpendicular to the mirror surface the two emission paths are in principle in-distinguishable for atom-mirror distances $d \ll c/\Gamma$  with $c$ the speed of light and $\Gamma$ the natural line-width. In this case the photon detection projects the emission in a coherent superposition of two directions and the atom after this emission event is in a superposition of two motional states. Consequently the photon can be regarded as the ultimate lightweight beamsplitter for an atomic matter wave (figure~\ref{fig:SPB_fig1}a).  Consequently spontaneous emission is an ideal model system to implement the original recoiling slit Gedanken experiment by Einstein~\cite{SPBS:Bohr_1949}. 
 
\begin{figure}[t]
\center\includegraphics[width = 10cm]{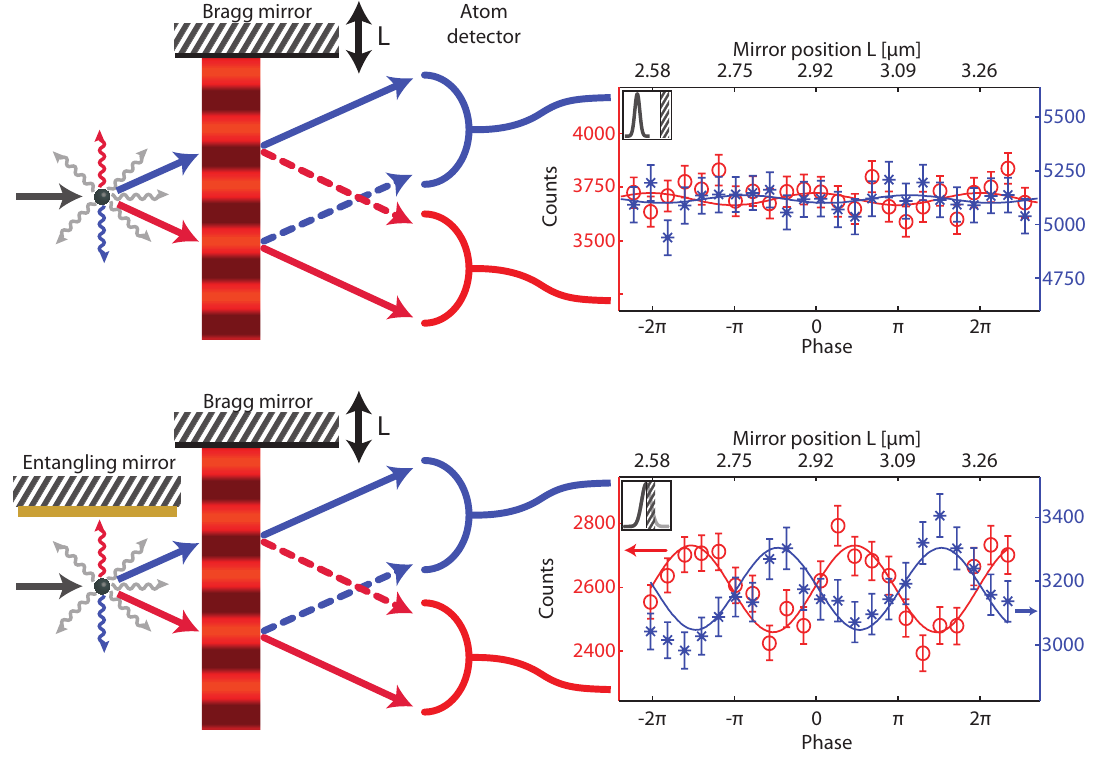}    
\caption{Experimental confirmation of coherence induced by spontaneous emission employed as the first beamsplitter of an atom interferometer. The recombination is accomplished by Bragg scattering from a standing light wave. The relative phase of the two paths can be changed by moving the ''Bragg'' mirror as indicated. 
\em{(top)} For a large  mean distance between atoms and the mirror of $54 \ \mu m$ no interference signal is observed confirming the free space limit. The inset depicts the position of the mirror relative to the atomic beam.
\em{(bottom )} For a mean distance of $2.8 \ \mu m$ the two complementary outputs of the interferometer reveal an interference pattern with a maximal visibility of $5.9\% \pm 1.1 \% $. Figure adapted from~\cite{SPB_natphys}.}
\label{fig:SPB_fig2}
\end{figure}

This reasoning can easily be generalized to the case of tilted emission close to the mirror surface. One finds residual coherence for emission angles where the optical absorption cross section of the atom and the mirror-atom observed by a fictitious observer in the emission direction still overlap (figure~\ref{fig:SPB_fig1}b). For larger distance to the mirror, the portion of coherent atomic momentum is strongly reduced (figure~\ref{fig:SPB_fig1}c).

The coherence can be probed by superposing the two outgoing momentum states using Bragg scattering at a far detuned standing light wave on a second mirror~\cite{OAB99, MOM88}. One observes an interference pattern as function of a phase shift  $\phi_{\mathrm{B}}$ applied by translating the Bragg standing light wave by moving the retro-reflecting mirror. The two outermost momentum states, which represent maximum momentum transfere due to photon emission in the back to back directions orthogonal to the mirror surface are expected to show the highest coherence.

The experiment of ref.~\cite{SPB_natphys} was performed with a well collimated and localized beam of  $^{40}$Ar atoms in the metastable $1s_5$ state (for the level scheme see figure~\ref{fig:ArLevel}) (figure \ref{fig:LightPotential}). In order to ensure the emission of only a single photon we induce a transition $1s_5 \rightarrow 2p_4$ ($\lambda_E = 715$ nm). From the excited state $2p_4$ the atom predominantly decays to the metastable $1s_3$ state via spontaneous emission of a single photon ($\lambda_{SE} = 795$ nm, branching ratio of $1s_5/1s_3 = 1/30$). The residual $1s_5$ are quenched to an undetectable ground state with an additional laser. Choosing the appropriate polarization of the excitation laser the atomic dipole moment is aligned within the mirror plane. The interferometer is realized with a far detuned standing light wave on a second mirror. Finally the momentum distribution is detected by a spatially resolved multi channel plate approximately $1m$ behind the spontaneous emission enabling to distinguish between different momenta~\cite{SPB_natphys}.

Figure \ref{fig:SPB_fig2} shows the read-out of the interference pattern for two distances between the atom and the mirror surface. The upper graph depicts the results obtained for large distances \mbox{($> 54$ $\mu$m)}  i.e.\ an atom in free space. In this case no interference is observed, and thus spontaneous emission induces a fully incoherent modification of the atomic motion. For a mean distance of $2.8$ $\mu$m clear interference fringes are observed demonstrating that a single spontaneous emission event close to a mirror leads to a coherent superposition of outgoing momentum states.

It is interesting to relate this experiment to the work by Bertet et al.~\cite{Bertet2001} where photons from transitions between internal states are emitted into a high fines cavity. there the transition happens from from indistinguishability when emission is into a large \emph{classical} field to distinguishability and destruction of coherence between the internal atomic states when emission is into the vacuum state of the cavity. Using the same photon for both beamsplitters in an internal state interferometer sequence, coherence can be obtained even in the empty cavity limit. In this experiment the photon leaves the apparatus and one observes coherence only when the photon cannot carry away which-path information. This implies that the generated coherence in motional states is robust and lasts. In this sense it is an extension of Einstein's famous recoiling slit Ge\-danken experiment~\cite{SPBS:Bohr_1949}. In free space the momentum of the emitted photon allows to measure the path of the atom. This corresponds to a well defined motional state of the beamsplitter i.e.\ no coherence. Close to the mirror the reflection renders some paths indistinguishable realizing a coherent superposition of the beamsplitter. The large mass of the mirror ensures that even in principle the photon recoil cannot be seen. Thus the atom is in a coherent superposition of the two paths.

\section{Interferometry with Bose-Einstein condensates in double-well potentials}  \label{sec:IFMTrappedBEC}

It was recognized as early as 1986 by J.\ Javanainen~\cite{Javanainen1986}, almost 10 years before the experimental production of the first Bose-Einstein condensates (BECs) in dilute gases~\cite{Anderson1995, Davis1995}, that BECs trapped in double-well potentials shared common features with solid states Josephson junctions~\cite{Josephson1962}. In the former case, the cooper pairs are replaced by neutral atoms, and the thin insulating layer by a tunnel potential barrier that can be realized either optically~\cite{Shin2004, Albiez2005, Levy2007, Folling2007}, magnetically~\cite{Hall2007}, or with hybrid traps such as radio-frequency dressed magnetic potentials~\cite{Schumm2005, Schumm2005a, Hofferberth2006a, Jo2007a, Berrada2013}. Nevertheless, the contact interaction between atoms, and the absence of leads connecting to an external circuit modifies the physics compared to standard Josephson junctions.

In section~\ref{sec:single_particle}, we start by neglecting interactions and introduce the basic concepts associated to this system: its reduction to a two-level problem (\ref{sec:2_mode}), its dynamics (\ref{sec:Rabi}), and the ways to measure it and use it for interferometry (sections \ref{sec:TOF_recombination}, \ref{sec:trapped_recombination} and \ref{sec:phase_shifts}). In section \ref{sec:interactions}, we analyze the effects arising from interactions: the emergence of the nonlinear Josephson dynamics (\ref{sec:MF}), the possibility to control the quantum fluctuations (\ref{sec:fluctuations}) by splitting a condensate (\ref{sec:splitting2}) and connections to interferometry.

\subsection{A Bose-Einstein condensate in a double-well potential: a simple model}
\label{sec:single_particle}

\begin{figure}[ht]
\begin{center}
\includegraphics[width=\linewidth]{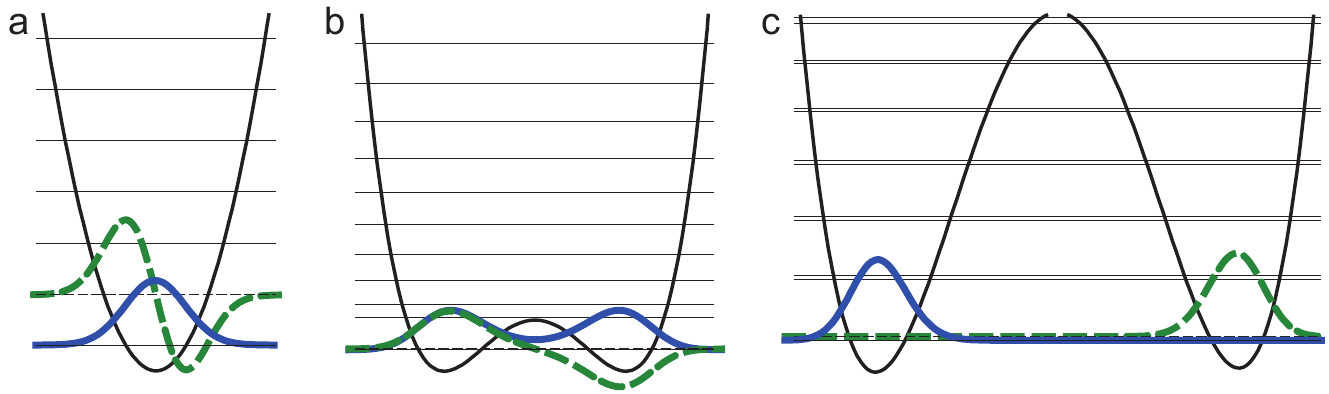}
\caption[A simple model describing a BEC in a double well potential]{\label{fig:simple_double_well} A simple model describing a BEC in a double well potential. The horizontal lines correspond to the single-particle eigenenergies. The two lower lying orbitals are also depicted. \textbf{a,} In the single-well limit, the states are approximately equally spaced in energy. \textbf{b,} In the presence of two minima, the lower-lying states gather by pairs, which become completely degenerate in the strongly split limit. \textbf{c,} A slight asymmetry of the trap is enough to lifts the degeneracies, and the two lower-lying states are localized in the wells.}
\end{center}
\end{figure}

Let us first assume that interactions between particles are negligible, and that a condensate containing $N$ atoms is trapped in a one-dimensional (1D) tunable potential, which can be turned into a double-well potential~\cite{ACF02}. We assume that the temperature is negligible, i.e.\ that the system is initially in the ground state. In practice, experiments generally involve interacting atoms at finite temperature, which are trapped in real three-dimensional potentials, but the above simplifications will help us tackle the most important features of the true system. The effect of interactions will be the subject of section~\ref{sec:interactions}. The geometry is depicted in figure~\ref{fig:simple_double_well}.

\subsubsection{Single-particle approach}

Upon neglecting interactions and temperature effects, the condensate wave function (the ground state of the $N$-particle system) can be written as a product of single particle ground states:
\begin{equation}
\Psi_\text{BEC}(x_1, x_2, \dots, x_N) = \prod_{j=1}^N \psi_\text{g}(x_j) .
\label{eq:GS_no_interactions}
\end{equation}
Here, $\psi_\text{g}$ is the single particle ground state, i.e.\ the solution of $\hat h \psi = E \psi$ having the smallest possible energy $E = E_\text{g}$. The single particle Hamiltonian $\hat h$ is given by
\begin{equation}
\hat h = -\frac{\hbar^2}{2 m} \frac{\partial^2}{\partial x^2} + V(x) .
\label{eq:SP_Hamiltonian}
\end{equation}
In a word, because the system is non-interacting and prepared at zero temperature, it is formally equivalent to $N$ independent single particles prepared in the same initial ground state of the double well potential. We can thus understand everything by studying the case of a single particle trapped by the potential of figure~\ref{fig:simple_double_well}.

The spectrum of $\hat h$ is displayed in figure~\ref{fig:simple_double_well}. When the two wells are not separated and the barrier height small, the spectrum resembles that of a harmonic oscillator, that is, the levels are roughly equidistant with a splitting $E \sim \hbar \omega$, where $\omega$ is typically the trap frequency. If one is able to prepare the system at a temperature $T \ll E/k_\text{B}$, the system is almost purely in the ground state. When the spacing between the traps (or the barrier height) is increased, the eigenstates group by pairs, until the pairs become completely degenerate which corresponds to a probability to tunnel from one side to the other that has become negligible (cf.\ figure~\ref{fig:simple_double_well}).

\subsubsection{Two-mode approximation} \label{sec:2_mode}

Guided by the fact that the system can be prepared in the ground state, we here simplify the full problem (Schr\"odinger equation with the Hamiltonian \eqref{eq:SP_Hamiltonian}) to an effective one, which will describe the physics at low energy. The most simple description taking into account deviations from the ground state is obtained by keeping only the two lower lying states $\psi_\text{g}$ and $\psi_\text{e}$. In the next sections, we will distinguish different cases depending on the precise shape of the trapping potential $V(x)$. We will first consider a strictly symmetric trap: $V(-x) = V(x)$, and then discuss the differences arising from an asymmetry. We note also that one can go beyond the two-mode approximation~\cite{Ananikian2006}, but this goes beyond the scope of these lectures.

\paragraph{Symmetric traps}

The symmetry with respect to the barrier ensure that each eigenstate of the system is itself symmetric or antisymmetric. Eigenstates thus correspond to an equal probability of finding the particle in one well or the other (the particles are not localized around one potential minimum as one would expect in classical physics). It seems more natural to describe the system in terms of modes that are localized on each side of the barrier, consequently we introduce the left/right basis\footnote{We have chosen the wave functions $\psi_\text{g}(x)$ and $\psi_\text{e}(x)$ real, and with the proper sign, such that $\psi_\text{l}(x)$ and $\psi_\text{r}(x)$ as defined by equations~\eqref{eq:left_mode} and \eqref{eq:right_mode} are indeed localized around the left and right wells respectively.}:
\begin{gather}
\psi_\text{l} = \frac{1}{\sqrt{2}}\left(\psi_\text{g} + \psi_\text{e} \right) , \label{eq:left_mode}\\
\psi_\text{r} = \frac{1}{\sqrt{2}}\left(\psi_\text{g} - \psi_\text{e} \right) . \label{eq:right_mode}
\end{gather}
This corresponds to a $45^\circ$ rotation in the 2-dimensional Hilbert space spanned by the initial basis $\{\psi_\text{g}, \psi_\text{e}\}$ describing the problem.

From the Hamiltonian \eqref{eq:SP_Hamiltonian}, we can derive a simple effective Hamiltonian describing the system in the left/right basis introduced above. We approximate the wave function $\psi(x,t)$ by the Anstaz
\begin{equation}
\psi(x, t) = c_\text{l}(t) \psi_\text{l}(x) + c_\text{r}(t) \psi_\text{r}(x)
\end{equation}
which corresponds to a two-mode approximation. Following the approach of refs.~\cite{Javanainen1986, Milburn1997, Smerzi1997, Raghavan1999} we obtain the two-mode Hamiltonian
\beq
\hat h_\text{2m} = \begin{pmatrix} 
 E & -J \\ 
-J &  E
\end{pmatrix} \label{eq:eff_Hamiltonian_sym}
\eeq
for the evolution of the state vector $\tilde \psi = (c_\text{l}, c_\text{r})$. The matrix elements of this Hamiltonian are
\begin{gather}
E = \bra{\psi_\text{l}} \hat h \ket{\psi_\text{l}} = \bra{\psi_\text{r}} \hat h \ket{\psi_\text{r}} = \frac{1}{2}(E_\text{g} + E_\text{e}) , \\
J = -\bra{\psi_\text{l}} \hat h \ket{\psi_\text{r}} = -\bra{\psi_\text{r}} \hat h \ket{\psi_\text{l}} = \frac{1}{2} (E_\text{e} - E_\text{g}) ,
\end{gather}
$E_\text{g,e}$ being the eigenenergies of the ground and first excited eigenstates respectively.

\paragraph{Sensitivity to trap asymmetry} \label{sec:asymmetry}

A problem arising when splitting a non-interacting gas is the sensitivity to the asymmetry of the trap. If we assume that the right well is higher by an energy $\delta >0$ compared to the left well, the Hamiltonian~\eqref{eq:eff_Hamiltonian_sym} becomes\footnote{Compared to equation \eqref{eq:eff_Hamiltonian}, we have removed the constant energy shift $E$ which does not play any role.}
\beq
\hat h_\text{2m} = \begin{pmatrix} 
-\delta/2 & -J \\ 
-J        & \delta/2 
\end{pmatrix} . \label{eq:eff_Hamiltonian}
\eeq
In the absence of asymmetry ($\delta = 0$), the eigenstates are the symmetric and antisymmetric superposition of left and right modes as explained above. Nevertheless, a small asymmetry is responsible for the collapse of the eigenstates in the two wells. This situation was studied in a box potential in reference \cite{Gea-Banacloche2002}, but the discussion below will be valid for any potential. The eigenvalues of \eqref{eq:eff_Hamiltonian} are
\beq
E_\text{e,g} = \pm \sqrt{J^2 + \delta^2/4} = \pm \frac{\delta}{2 \cos(\theta)} , \label{eq:eigenvalues}
\eeq
where the ``$+$'' corresponds to the excited state, and the ``$-$'' to the ground state. Here we have introduce an auxiliary angle $\theta$ defined by $\tan \theta = -2 J/\delta$. The eigenstates are given by (see ref.~\cite{Cohen-Tannoudji1973}, Volume 1, Complement B$_\text{IV}$)
%
\begin{align}
\psi_\text{g} &= \cos \frac{\theta}{2} \psi_\text{l} + \sin \frac{\theta}{2} \psi_\text{r}  , \\
\psi_\text{e} &= - \sin \frac{\theta}{2} \psi_\text{l} + \cos \frac{\theta}{2} \psi_\text{r}  .
\end{align}
We see that if there is no coupling, $J=0$, and in the absence of a bias, $\delta \neq 0$, then $\theta = 0$ and the eigenstates $\psi_\text{g}$ and $\psi_\text{e}$ ``collapse'' to the left and right wells respectively. From the expression of the eigenvectors, one could also compute the population of the two modes as a function of the ratio $J/\delta$. This sensitivity of the double well to an asymmetry is illustrated in figure~\ref{fig:simple_double_well}c, where a small gradient has been added to a symmetric double-well potential, leading the two lower lying states to collapse into the left and right wells.

\subsubsection{Rabi dynamics} \label{sec:Rabi}

We now focus on the dynamics that can occur when the system is not prepared in an eigenstate, or when the potential is changed such that the eigenstates are modified, which brings the system out of equilibrium and triggers dynamics. In the left/right basis, we recognise in \eqref{eq:eff_Hamiltonian_sym} a simple two-level Hamiltonian with off-diagonal coupling, which will exhibit oscillations of the populations $|c_i|^2$ at the (angular) Rabi frequency
\beq
\Omega_\text{R} = 2J/\hbar .
\eeq
If the system is initially prepared in a superposition of the two eigenstates, the particles will tunnel back and forth between the two wells at this frequency. The corresponding time $\tau_\text{R} = h/2J$ is the typical time scale associated to tunnelling of atoms. This oscillation of the populations is accompanied by an oscillation of the relative phase $\phi = \arg(c_\text{l}^*c_\text{r})$ at the same frequency. While the oscillation of the populations is strictly sinusoidal: $|c_\text{l}(t)|^2 = [1 + A\cos(\Omega_\text{R} t + \varphi)]/2$, with $-1<A<1$, the oscillation of the phase becomes anharmonic as soon as the amplitude $|A|$ is not much less than 1. Note that when the initial state is localized in one of the wells, the amplitude is maximal $|A| = 1$, which means that all the atoms tunnel back and forth between the two wells.

\subsubsection{Splitting}

\paragraph{Principle of coherent splitting} \label{sec:splitting1}

From the results of section~\ref{sec:single_particle}, we see that a condensates can be coherently split by adiabatically changing the potential from a single well to a double well. This was for example studied numerically in ref.~\cite{Andersson1999} and realized experimentally~\cite{Andrews1997, Schumm2005, Jo2007a, Hofferberth2008, Berrada2013}. Adiabaticity ensures that the system, initially prepared in the ground state, will stay in the ground state during the transformation, which, analysed in the right left basis, is a coherent superposition state of the type
\beq
\psi = \frac{1}{\sqrt{2}} \left( \psi_\text{l} + \psi_\text{r} \right) .
\eeq
In order for the final state to be a coherent superposition after splitting, the additional constrain that the ground state must be delocalized in the two wells must be fulfilled. As explained in section~\ref{sec:asymmetry}, this condition is quite constraining, as even a very slight asymmetry would lead the ground state to collapse in the well having the lowest energy. This is modified in the case of an interacting Bose-Einstein condensate, in which a new energy scale related to interactions, the chemical potential, will play a significant role.

\paragraph{Condition for adiabatic splitting}

We model the splitting process as a slow and monotonic change of the parameters $J(t)$ and $\delta(t)$ which are now time-dependent. The condition for adiabatic evolution is given by \cite{Comparat2009}
\beq
\left| \frac{\hbar \bra{\psi_\text{e}} \partial_t \ket{\psi_\text{g}}}{E_\text{e} - E_\text{g}} \right| \ll 1 , \label{eq:adiabatic}
\eeq
Here the two eigenstates of \eqref{eq:eff_Hamiltonian} have the energies given by equation \eqref{eq:eigenvalues}. When evaluated, condition \eqref{eq:adiabatic} yields
\beq
\hbar \frac{|\dot J + \dot \delta/2|}{4 J^2 + \delta^2} \ll 1 . \label{eq:adiabaticity1}
\eeq
In the symmetric case discussed above ($\delta = 0$) we see that, during splitting, because $J \rightarrow 0$, the splitting must become slower and slower in order to stay adiabatic. It will not be strictly speaking possible to stay adiabatic all the way to $J=0$.

\paragraph{Non-adiabatic splitting}

This opens the possibility of splitting non-adiabatically: if the ground state wave function is still delocalized when adiabaticity starts to break down, then the wave function will stay essentially the same as the splitting process goes on despite the fact that the eigenstates may change very abruptly. In this case, a coherent superposition of left and right modes is also created.

\subsubsection{Time-of-flight recombination} \label{sec:TOF_recombination}

\begin{figure}[t]
\begin{center}
\includegraphics[width=0.5\linewidth]{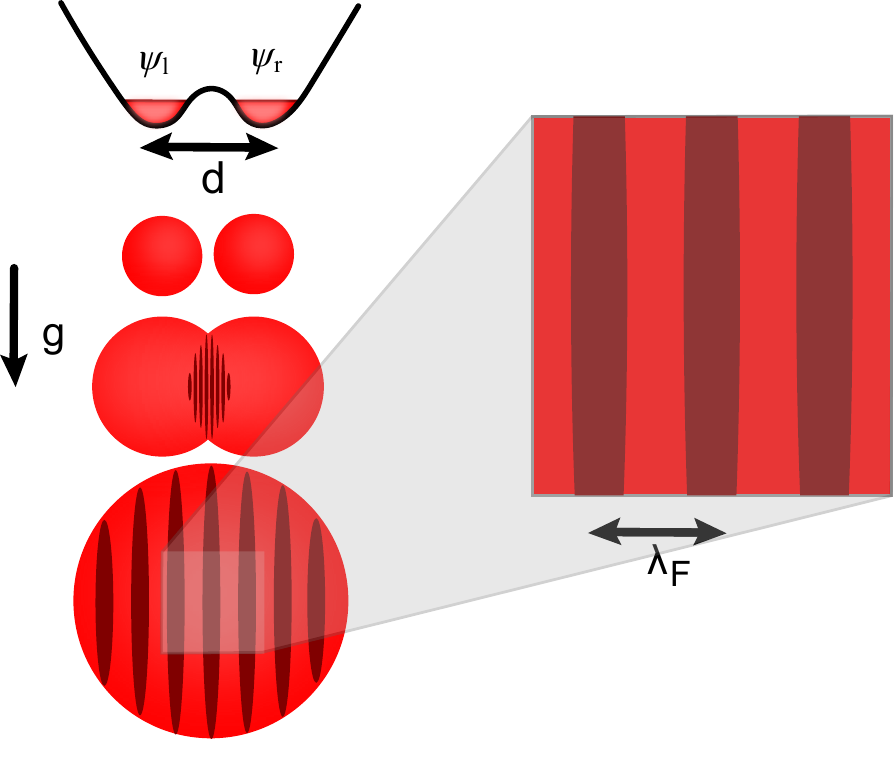}
\caption{\label{fig:interference} Principle of time-of-flight recombination. After a long time of flight, the two wave packets overlap and interfere. Figure adapted from Tim Langen's PhD thesis~\cite{LangenThesis}.}
\end{center}
\end{figure}

\begin{figure}[t]
\begin{center}
\includegraphics[width=\linewidth]{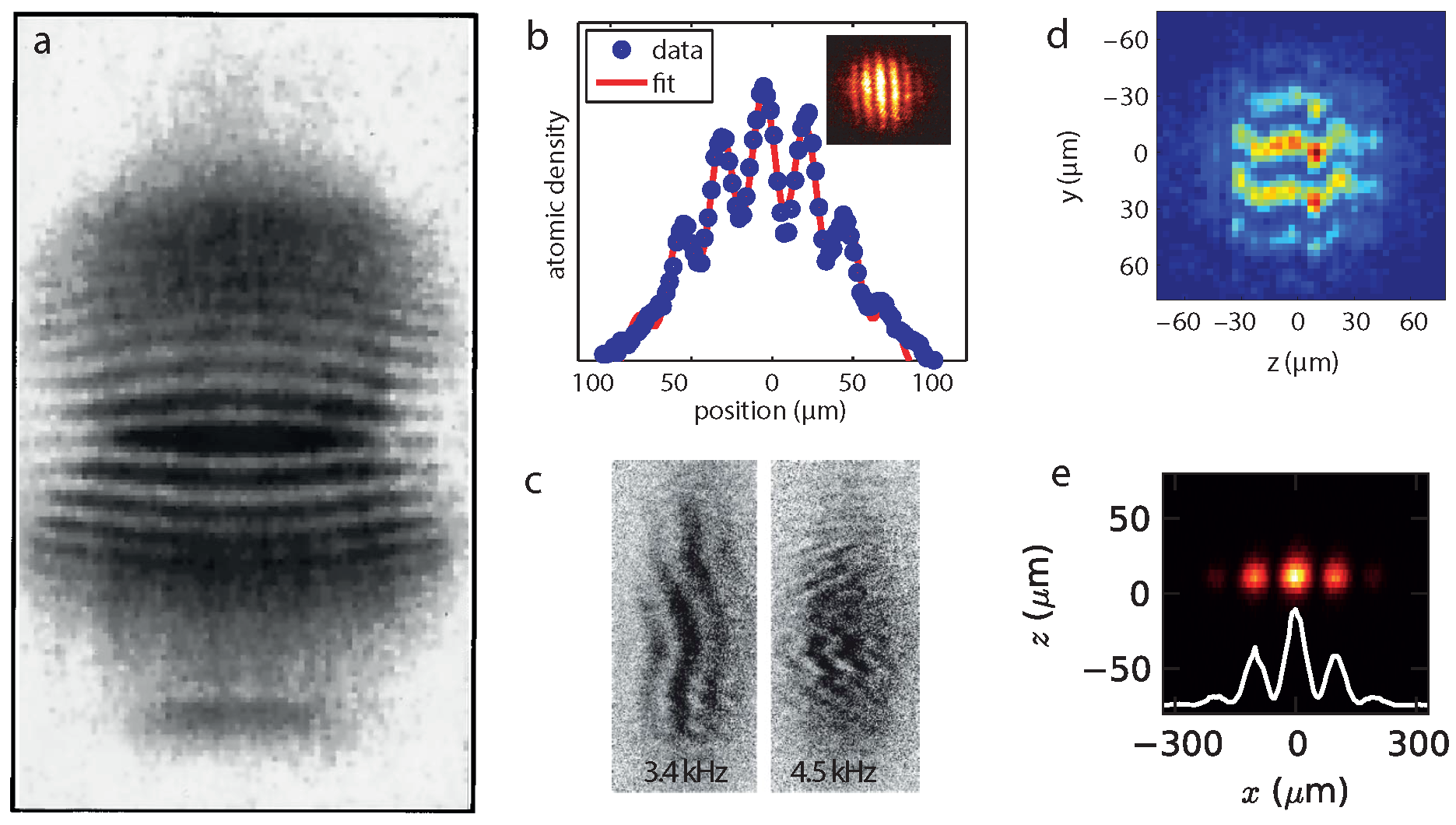}
\caption{\label{fig:interference_patterns} Collection of Bose-Einstein condensates interference patterns. \textbf{a,} First BEC interference pattern observed by Andrews et al.~\cite{Andrews1997}: the BEC was trapped magnetically and split with a repulsive optical barrier, \textbf{b, c, d, e} interference pattern of Schumm et al.~\cite{Schumm2005}, Jo et al.~\cite{Jo2007a}, Hofferberth et al.~\cite{Hofferberth2008} and Berrada et al.~\cite{Berrada2013}: the BECs were trapped and split in radio-frequency dressed magnetic traps. The patterns of c and d display a relative phase which varies along the long axis of the BEC (see section~\ref{sec:III} for a discussion of such 1D effects).}
\end{center}
\end{figure}

A coherent superposition state can be measured by recombining the two ``arms'' of the interferometer (in our case the two BECs trapped in the two wells) and measuring the interference. Time of flight recombination is similar to a double split experiment. The matter-waves in each potential minima are released and freely expand until they overlap, forming a typical double-slit interference pattern (cf.\ examples of such patterns on figure \ref{fig:interference_patterns}). Nevertheless, as surprising as it may seem, the observation of interference fringes between two overlapping Bose-Einstein condensates is not a proof of coherence. Indeed, it was shown that two condensates having each a fixed, well defined, atom number (Fock states), and therefore no fixed relative phase (vanishing off-diagonal terms in the one-body density matrix) would display interference when overlapped and measured~\cite{Castin1997}. The fringes that appear can be seen as the result of the measurement-induced collapse of the system's state to coherent state (cf.\ section~\ref{sec:interactions}) having a random relative phase. Therefore, first order coherence is proven when the phase can be measured repeatedly and always yields the same value, that is when the fringes observed are identical from shot to shot.

In the current and next sections (sections \ref{sec:TOF_recombination} and \ref{sec:trapped_recombination}) we assume that a condensate has been split symmetrically, and that a relative phase $\phi$ has been accumulated. The initial wave function can therefore be written
\beq
\psi(x, t=0) = \frac{1}{\sqrt{2}}\left(\psi_\text{l}(x) + e^{i\phi} \psi_\text{r}(x)\right) \label{eq:initial_WF}
\eeq
with $\psi_\text{r}(-x) = \psi_\text{l}(x) \equiv f(x, t = 0)$\footnote{Note that in quantum mechanics, the global phase of the state \eqref{eq:initial_WF} can be chosen arbitrarily.}. The generalization to an imbalanced splitting yielding different populations in the left and right modes is straightforward.

When the two trap minima are well separated, one can generally approximate them by harmonic potentials, and the left and right modes are therefore well modeled by minimal uncertainly Gaussian wave packets. Let us assume a symmetric trap with trapping frequencies $\omega$. When the trapping potential is abruptly turned off, the dynamics can be decomposed into
\begin{enumerate}
\item the center of mass movement, which will follow a classical trajectory: $z_\text{com} = - g t^2/2$,
\item the relative movement in the free-falling frame, witch is identical to that in the absence of any gravity.
\end{enumerate}
We can thus neglect gravity. For simplicity, we assume that the 3D potential is separable, and just solve the problem in the splitting direction $x$.

\paragraph{Expansion of a single wave packet}

The function $f$ is initially
\beq
f(x,t=0) = \left(\frac{1}{\pi \sigma_0^2}\right)^{1/4} \exp\left(-\frac{(x-x_\text{l})^2}{2 \sigma_0^2}\right) , 
\eeq
with $x_\text{l} = -d/2$, $d$ being the distance between the trap minima. Equivalently, in momentum space representation\footnote{We have chosen the definition of the Fourier transform and its inverse as $\tilde g(p,t) = (2 \pi \hbar)^{-1/2} \int e^{-i px/\hbar} g(x,t) \, \mathrm{d} x$, and $g(x,t) = (2 \pi \hbar)^{-1/2} \int e^{i px/\hbar} \tilde g(p,t) \, \mathrm{d} p$.}
\beq
\tilde f(p,t=0) = \left(\frac{\sigma_0^2}{\pi \hbar^2}\right)^{1/4} \exp\left(-\frac{\sigma_0^2 p^2}{2 \hbar^2}\right) \exp\left(-i \frac{p x_\text{l}}{\hbar}\right) . \label{eq:GausGS_p}
\eeq
Here $\sigma_0 = \sqrt{\hbar/m\omega}$ is the spatial width of the ground state in each well. The width in momentum space is $\hbar/\sigma_0 = \sqrt{m \hbar \omega}$. The free expansion after trap switch off is described by the evolution operator $\hat U(t) = \exp(-i \hat H t/\hbar) = \exp(-i \hat p^2 t/(2 \hbar m))$, which is diagonal in momentum representation thanks to the absence of a potential in $\hat H$. The wave function is therefore given at time $t$ by multiplying the initial one \eqref{eq:GausGS_p} by $\hat U$ and applying an inverse Fourier transform to recover the spatial wave function. We recover the well known results that a Gaussian minimal uncertainty state stays Gaussian with a width increasing as
\beq
\sigma(t) =  \sigma_0 \left( 1 + \omega^2 t^2 \right) . \label{eq:Gaus_width}
\eeq
The time-dependent wave function is given by
\beq
f(x,t) = \left(\frac{1}{\pi \sigma(t)^2}\right)^{1/4} \exp\left(-\frac{(x-x_\text{l})^2}{2 \sigma(t)^2}\right) \exp( i \varphi(x,t)) ,
\eeq
with the spatially-dependent phase accounting for the velocity distribution of the atoms
\beq
\varphi(x,t) = \frac{\hbar t}{2 m \sigma_0^2}\left(\frac{(x-x_\text{l})^2}{\sigma(t)^2} - 1\right) .
\eeq

\paragraph{Interference of two expanding wave packets}

In the case of a double-well potential, the two wave packets follow the evolution described above, and, using~\eqref{eq:initial_WF} as the initial state, we obtain an interference pattern when they start to overlap. The interference term is
\beq
2 |f(x,t)||f(-x,t)| \cos \left( 2 k(t) x + \phi \right),
\eeq
where the time-dependent wave vector $k(t)$ is
\beq
k(t) = \frac{\hbar t d}{2 m \sigma_0^2 \sigma(t)^2} .
\eeq
At long times ($\omega t \gg \sqrt{d^2/(4\sigma_0^2 - 1}$), the width of the wave functions is much larger than their initial distance $d$, and the two envelopes coincide $f(-x,t) \simeq f(x,t)$. The density is approximately
\beq
N |\psi(x,t)|^2 \simeq N |f(x,t)|^2 \left(1 + \cos \left( 2 k(t) x + \phi \right)\right) .
\eeq
We see that the fringe spacing $\lambda_\text{F} = \pi/k(t)$ coincides with $\lambda_\text{F} = \lambda_\text{dB}/2$, half the de Broglie wavelength of an atom initially position at a trap minimum, and having a velocity $v = \pm d/2t$ such that it could reach the central point $x = 0$ after a time $t$.




\begin{figure}[t]
\begin{center}
\includegraphics[width=0.7\linewidth]{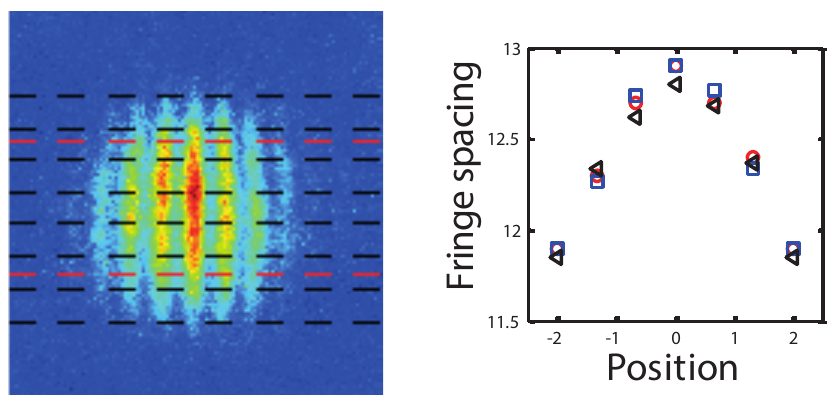}
\caption{\label{fig:bent_fringes} Effect of mean-field repulsive interaction on the fringe spacing: The atom wave moving expanding from one well moves through the expanding atom wave of the other well and accumulates a phase shift. Consequently the fringe spacing becomes position dependent, leading to bent fringes.}
\end{center}
\end{figure}

\subsubsection{In-trap recombination} \label{sec:trapped_recombination}

Reading out the relative phase of two waves by overlapping them and observing the interference pattern, as described above, is not the standard way used in optics. The beams are rather recombined on a beam splitter. The two input modes are ``mixed'' and the resulting two output modes have different intensities, depending on the relative phase of the input modes. A beam splitter is therefore an object which converts phase differences into intensity differences and vice versa. This strategy can by extended to Bose-Einstein condensates trapped in double-well potentials~\cite{Andersson1999, Berrada2013}.

\paragraph{Recombining with Rabi oscillations}

A first possibility is to adiabatically recouple the two traps by reducing the barrier height or well spacing~\cite{Andersson1999}. In this case the coupling $J$, which is initially negligible, increases, and Rabi oscillations are triggered (cf.\ section~\ref{sec:Rabi}). A quarter of a Rabi oscillation will convert the phase into a population imbalance between the two wells, which can then be measured. This requires to couple the wells for a time $\tau \simeq h/8J$ (assuming that Rabi oscillations have not started during the recombination stage).

Nevertheless, we will see that interactions will reduce the efficiency of this approach, as they tend to reduce the amplitude of the oscillations of the populations (see section \ref{sec:MF}, equation \eqref{eq:zmax}).

\paragraph{Non-adiabatic recombination}

\begin{figure}[t]
\centering
\includegraphics[width=0.6\linewidth]{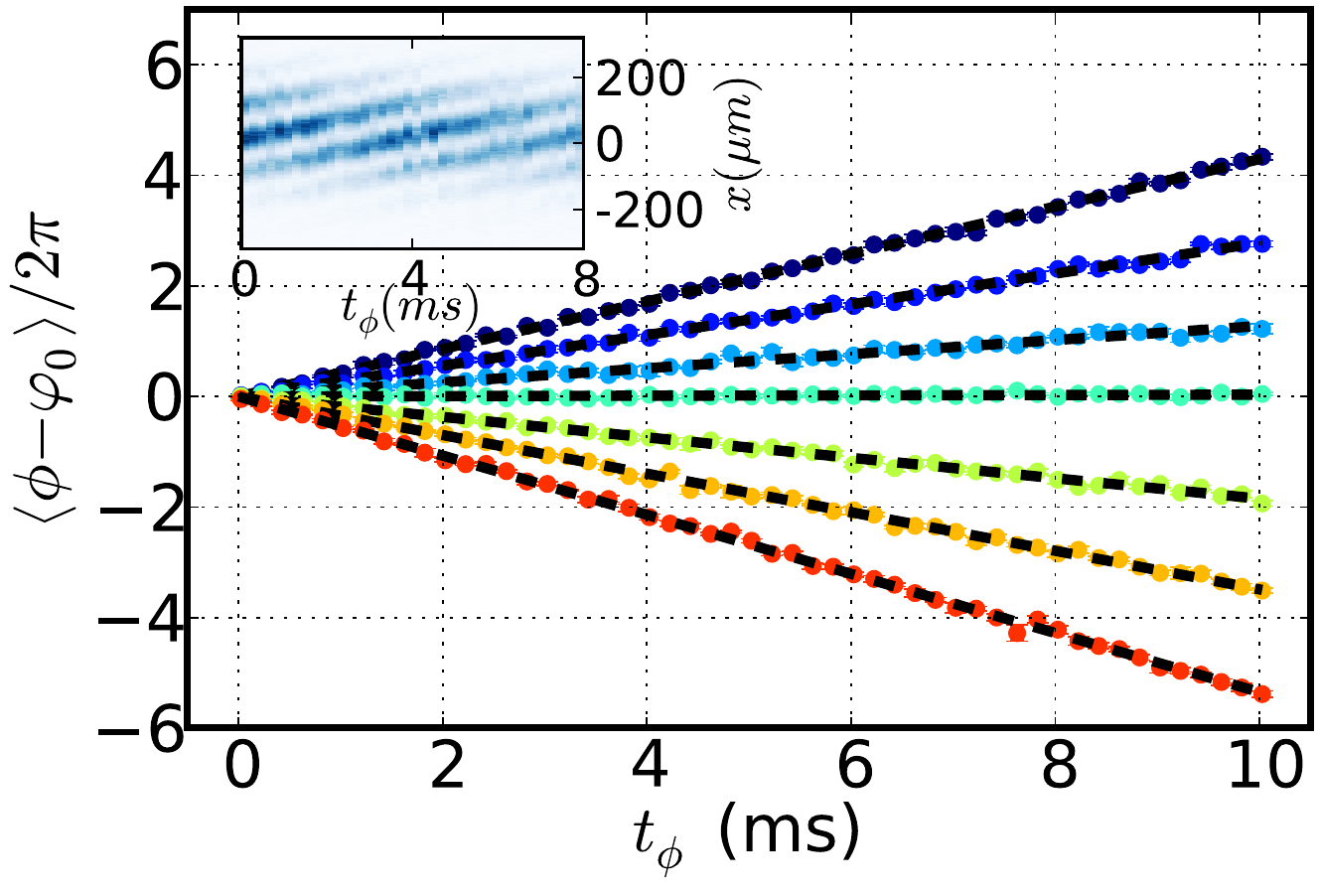}
\caption{\label{fig:phase_accumulation} Accumulation of a relative phase due to an energy difference $\delta \neq 0$ (cf.\ equation~\eqref{eq:phase_shift1}). The phase is measured by time of flight recombination. Inset: evolution of the interference pattern (see figure~\ref{fig:interference_patterns}e) from which the phase is measured. Figure adapted from ref.~\cite{Berrada2013}.}
\end{figure}

\begin{figure}[t]
\centering
\includegraphics[width=\linewidth]{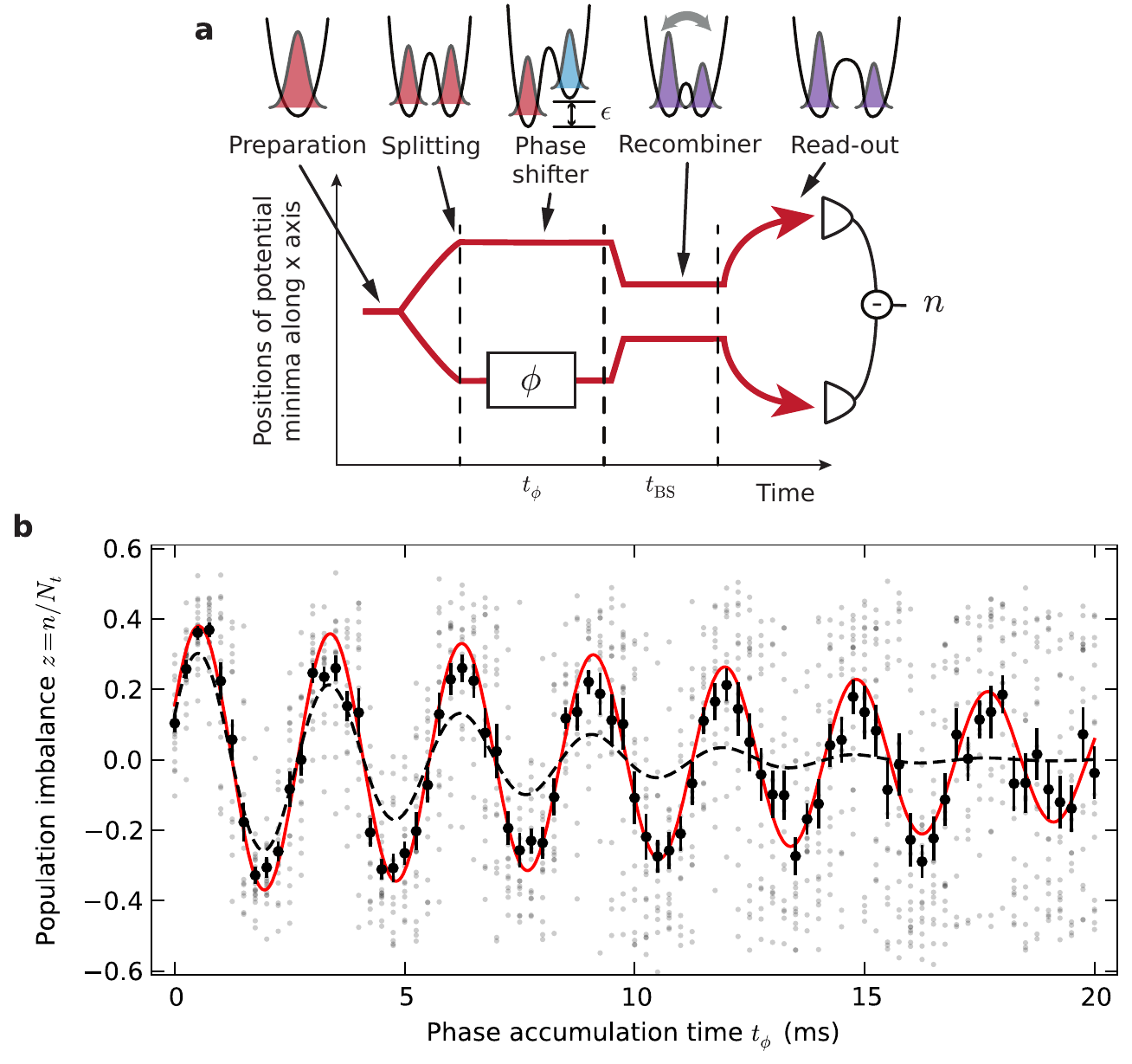}
\caption{\label{fig:MZI_fringes} Mach-Zehnder interferometer with a Bose-Einstein condensate in a double-well potential. \textbf{a,} Nearly adiabatic splitting creates a coherent superposition of left and right modes. The recombination is performed non-adiabatically. \textbf{b,} At the end of the sequence, the atom number difference between the two traps show interference fringes. Grey dots: single realizations, black dots: ensemble average. Figure adapted from ref.~\cite{Berrada2013}.}
\end{figure}

An alternative approach is to modify the trap in a non-adiabatic way. Here we rely on the symmetry of the trapping potential $V(-x,t) = V(x,t)$. Before recombination, the wave function describing the two condensates has the same symmetry, except for a relative phase (cf.\ equation \eqref{eq:initial_WF}). From the symmetry of the potential and propagating with the Schr\"{o}dinger equation, one finds that the BEC wave function can be written at all time as
\begin{equation}
\psi(x,t) = \frac{1}{\sqrt{2}}\left[f(x,t) + e^{i\phi} f(-x,t)\right] .
\label{Eq:WFRecomb}
\end{equation}
The imbalance $z \equiv \int_{x<0}|\psi|^2 \, \mathrm{d}x - \int_{x>0}|\psi|^2 \, \mathrm{d}x$ can readily be calculated from Eq.~\eqref{Eq:WFRecomb}, which gives 
\begin{equation}
z(t) = C(t) \sin \phi .
\label{Eq:contrast}
\end{equation}
Note here that the argument is very general, and has not assumed a particular time-dependence or shape of the potential $V(x,t)$. It only relies on the symmetry of the potential and wave function, and on the superposition principle. The contrast
\beq
C(t) = 2\int_0^\infty \Im\left[ f(-x, t) f^*(x, t)\right] \mathrm{d}x 
\eeq
depends on the details and can be maximized by optimizing both the shape and the way the potential is dynamically varied. As a whole, this sequence turns an initial state which has equal population on both sides of the barrier and a relative phase $\phi$, into a state whose populations depend on the sine of the phase $\phi$. This is very similar to an optical beam splitter.

Note that other strategies have been attempted, such as measuring the phase-dependent heating produced when two condensates are merged~\cite{Jo2007b}. Nevertheless, this technique strongly relies on interactions, as the heating was interpreted in terms of soliton creation, which could not occur in the absence of interactions.

\subsubsection{Phase shifts} \label{sec:phase_shifts}

Here we assume that the two wells are not coupled ($J=0$), and that the condensate has been split symmetrically. If a force $F$ is applied along the splitting direction during a time $t$ (or equivalently if the system is subject to an acceleration $a = -F/m$), the energy shift $\delta = F d$ between the two wells gives rise to a phase shift, given at first order by
\beq
\phi_\text{force} = \frac{F d t}{\hbar} , \label{eq:phase_shift1}
\eeq
where $d$ is the distance between the two trap minima. This illustrates how such an atom interferometer could in principle be used to measure inertial effects. An example of such phase accumulation is displayed in figure~\ref{fig:phase_accumulation} for different values of $\delta$. This assumes that both traps stay harmonic with the same frequencies. If they have a different trapping frequency $\omega_\text{l} = \omega_\text{r} + \Delta \omega$, then there is an additional phase shift
\beq
\phi_\text{asym.} = \frac{1}{2} \Delta \omega \, t .
\eeq
This illustrates the difficulty of performing precision measurements with such interferometers, because this would require a very good knowledge of the trap shape to know its response to an external force.

\subsection{Effect of interactions: Josephson dynamics, squeezing, and dephasing}  \label{sec:interactions}

In this section, we extend the results of section \ref{sec:single_particle} to the case of interacting particles. A simple way to take into account interactions is to still assume that the particles are distributed among two modes, say the left and right modes introduced in section \ref{sec:2_mode}, and to work with the second quantization formalism. To each mode is associated a pair of creation and annihilation operators. It is convenient to work in the basis of Fock states, in which the number of atoms in each mode is well defined (i.e.\ has no quantum fluctuations). Let us write $\ket{n_\text{l}, n_\text{r}}$ the Fock state having $n_\text{l}$ atoms in the left-localised mode and $n_\text{r}$ atoms in the right-localized mode. For example, the vacuum state (no atoms) is written $\ket{0, 0}$. If we assume that the total atom number $N$ is fixed, then the Hilbert space has $N+1$ dimensions. This can be seen because the set of Fock states $\{\ket{N-k,k}, k\in\{0, N\}\}$ contains $N+1$ vectors, and is an obvious basis. If we write $\hat l$ and $\hat l^\dagger$ the annihilation and creation operators associated to the left mode, and $\hat r$, $\hat r^\dagger$ those associated to the right mode, we have the usual relations
\begin{align}
[\hat l, \hat l^\dagger] = 1, \ &[\hat r, \hat r^\dagger] = 1, \ [\hat l^{(\dagger)}, \hat r^{(\dagger)}] = 0 \\
\hat l \ket{n_\text{l}, n_\text{r}} &= \sqrt{n_\text{l}} \ket{n_\text{l} - 1, n_\text{r}} \\
\hat l^\dagger \ket{n_\text{l}, n_\text{r}} &= \sqrt{n_\text{l} + 1} \ket{n_\text{l} + 1, n_\text{r}} \\
\hat r \ket{n_\text{l}, n_\text{r}} &= \sqrt{n_\text{r}} \ket{n_\text{l}, n_\text{r} - 1} \\
\hat r^\dagger \ket{n_\text{l}, n_\text{r}} &= \sqrt{n_\text{r} + 1} \ket{n_\text{l}, n_\text{r} + 1}
\end{align}
and $\hat l \ket{0,n_\text{r}} = 0$, $\hat r \ket{n_\text{l}, 0} = 0$. The operators $\hat l^\dagger \hat l = \hat n_\text{l}$ and $\hat r^\dagger \hat r = \hat n_\text{r}$ give the number of atoms in the left and right modes respectively.

\subsubsection{Bose-Hubbard model with two sites}

Once again we focus on the most simple model which will display the basic features arising from interactions: the Bose-Hubbard model.

It has become increasingly popular in the past decade, thanks to the development of optical lattices in cold atom experiments. A double well potential can be seen as the most simple lattice: one containing only two sites. The usual lattice approach will therefore breakdown, due to the very small size of the system, but we will see that the basic features known in the case of large lattices, such as the existence of a superfluid phase and a Mott insulator phase, have an analogy in the case of only two sites. To properly derive the Bose-Hubbard model as an effective theory describing the real system (3D trapping potential), one can use a two-mode approach as in section \ref{sec:single_particle}, but this time on the field operator describing the system of interacting bosons. The field operator is thus approximated by
\beq
\hat \psi(x) = \hat l \psi_\text{l}(x) + \hat r \psi_\text{r}(x) .
\eeq
We refer the interested reader to refs.~\cite{Milburn1997, Smerzi1997, Raghavan1999, Garcia-March2012} for an example of such a derivation.

\paragraph{Bose-Hubbard Hamiltonian}

The Bose-Hubbard Hamiltonian reads
\begin{equation}
\hat H_\text{BH} = -J \left( \hat l^\dagger \hat r + \hat l \hat r^\dagger \right) + \frac{U}{2} \Big[ \hat n_\text{l} (\hat n_\text{l} - 1) + \hat n_\text{r} (\hat n_\text{r} - 1) \Big] + \frac{\delta}{2}(\hat n_\text{r} - \hat n_\text{l})
\label{eq:BH_Hamiltonian}
\end{equation}
where the new energy scale $U$, which did not appear in section~\ref{sec:single_particle}, is related to interactions. The first term is a kinetic energy, and allows tunneling between the two wells. The third term corresponds to an energy difference $\delta$ between the two modes (in the absence of tunneling), and the second term corresponds to the interaction energy associated with contact interaction between the atoms ($s$-wave scattering). We note that this interaction is local, i.e.\ the atoms in the left-localized well do not interact with those in the right-localized well. In the following, we will always assume repulsive interactions $U > 0$.

In order to relate $U$ to a physical value that can be measured in the experiment, we can analyze the Hamiltonian \eqref{eq:BH_Hamiltonian} in the absence of tunneling and bias ($J = \delta = 0$). The energy of the condensate in the left well is then given by
$
E_\text{l} = U n_\text{l} (n_\text{l} - 1)/2
$. 
The chemical potential of this BEC is thus 
\beq
\mu_\text{l} = \frac{\partial E_\text{l}}{\partial n_\text{l}} \simeq U n_\text{l} , \label{eq:mul}
\eeq
therefore we see that the Bose-Hubbard model assumes that the chemical potential scales linearly with the atom number, and we can also make the identification
\beq
U = \left.\frac{\partial \mu_\text{l}(\mathcal{N})}{\partial \mathcal{N}}\right|_{\mathcal{N} = n_\text{l}} , \label{eq:def_U}
\eeq
$\mu_\text{l}(\mathcal{N})$ being the chemical potential of the left-localized condensate, seen as a function of its atom number\footnote{Note that the expression \eqref{eq:mul} cannot exclude the expression $U = \mu_\text{l}/n_\text{l}$ compared to equation~\eqref{eq:def_U}. An answer to this ambiguity is obtained by using the full many-body Hamiltonian of a BEC in a 3D potential. In this case, the chemical potential does not scale linearly with $n_\text{l}$, which allows to check that equation \eqref{eq:def_U} is the right form.}. Therefore an underlying assumption in equation~\eqref{eq:BH_Hamiltonian} is that the trap is sufficiently symmetric and the populations not too different, such that the interaction energy $U$ is the same in both wells.

\subsubsection{Mean-field treatment} \label{sec:MF}

To get a first intuition on this Hamiltonian, we can apply a mean field approach to obtain the approximate dynamics of mean values of the operators~\cite{Milburn1997, Smerzi1997, Raghavan1999, Garcia-March2012}. For that, one replaces the annihilation operators by complex numbers:
\beq
\hat l \rightarrow c_\text{l} \in \C, \ \hat r \rightarrow c_\text{r} \in \C \label{eq:mean_field}
\eeq
and the creation operators by their conjugate: $\hat l^\dagger \rightarrow c_\text{l}^*, \hat r^\dagger \rightarrow c_\text{r}^*$. The two c-numbers $c_\text{l}$ and $c_\text{r}$ play the role of a (single particle) wave function, which is just defined on the two sites. This wave function is normalized to $N$: $|c_\text{l}|^2 + |c_\text{r}|^2 = N$, such that $|c_\text{l}|^2 \simeq \langle \hat n_\text{l} \rangle = n_\text{l}$ and $|c_\text{r}|^2 \simeq \langle \hat n_\text{r} \rangle$. We can write these numbers in polar representation:
\begin{align}
c_\text{l} &= \sqrt{n_\text{l}} e^{i\phi_\text{l}} , \label{eq:cl_polar} \\
c_\text{r} &= \sqrt{n_\text{r}} e^{i\phi_\text{r}} . \label{eq:cr_polar}
\end{align}
It is then convenient to define the population imbalance $z$ and relative phase $\phi$ of the Josephson junction as
\begin{align}
z &= \frac{1}{N}(n_\text{l} - n_\text{r}) = \frac{1}{N}(c_\text{l}^* c_\text{l} - c_\text{r}^* c_\text{r}) , \\
\phi &= \phi_\text{r} - \phi_\text{l} = \arg(c_\text{l}^* c_\text{r}) .
\end{align}
Substituting equations~\eqref{eq:mean_field}, \eqref{eq:cl_polar} and \eqref{eq:cr_polar} in the Hamiltonian~\eqref{eq:BH_Hamiltonian}, and rewriting it in terms of the variables introduced above, yields the classical Hamiltonian\footnote{We have use the relations: $\langle \hat n_\text{l} \rangle = (1+z)N/2$, $\langle \hat n_\text{r} \rangle = (1-z)N/2$, $\sqrt{\langle \hat n_\text{l} \rangle \langle \hat n_\text{r} \rangle} = \sqrt{1-z^2}N/2$, and $N = \hat n_\text{l} + \hat n_\text{r}$.}
\beq
H_\text{MF} = -\frac{\delta}{\hbar} z + \frac{UN}{2\hbar} z^2 - \Omega_\text{R} \sqrt{1-z^2} \cos \phi .
\eeq
In the absence of bias $\delta = 0$, it resembles the Hamiltonian of a pendulum with variable length in which $z$ plays the role of momentum, and $\phi$ is the angle with respect to the acceleration of gravity $g$: $H = p^2/(2m) - mg\ell \cos \phi$, $\ell$ being the length of the pendulum, and $m$ its mass. A difference is that the length $\ell$ gets shorter with higher momenta: $H_\text{MF}$ is the so-called ``momentum-shortened'' pendulum.

\begin{figure}[t]
\centering
\includegraphics[width=\linewidth]{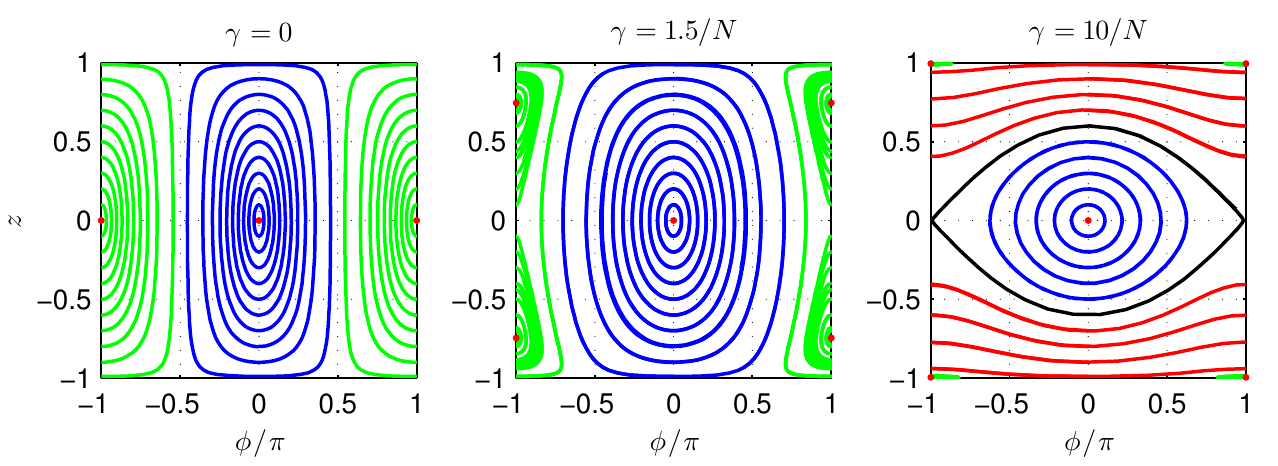}
\caption{\label{fig:phase_portrait} Phase portraits of the bosonic Josephson junction illustrating Rabi oscillations ($\gamma = 0$), Josephson oscillations ($\gamma > 1/N$, blue lines) and macroscopic self-trapping ($\gamma > 2/N$, red lines). The red dots indicate the fixed points (see text). Figure adapted from T.~Berrada's PhD thesis~\cite{Berrada2014}.}
\end{figure}

The evolution of $z$ and $\phi$ is obtained through Hamilton's equations $\dot \phi = \partial H_\text{MF}/\partial z$, and $\dot z = - \partial H_\text{MF}/\partial \phi$, that is 
\beq
\begin{cases}
\dot{z} &= - \Omega_\text{R} \sqrt{1-z^2} \sin \phi ,\\
\dot{\phi} &= - \frac{\delta}{\hbar} + \frac{UN}{\hbar} z + \Omega_\text{R} \frac{z}{\sqrt{1-z^2}} \cos \phi .
\end{cases} \label{eq:Josephson_set}
\eeq
The dynamics described by these coupled nonlinear equations has been extensively studied theoretically~\cite{Milburn1997, Smerzi1997, Raghavan1999, Leggett2001}, observed in experiments with BECs trapped in double-well potentials~\cite{Albiez2005, Levy2007, Berrada2013}, and beautifully verified with BECs in two different hyperfine states~\cite{Zibold2010}\footnote{In this experiment, two BECs share a common spatial wave function, but the two modes correspond to two different hyperfine states that are ``tunnel coupled'' by a two-photon transition. If the left and right modes are replaced by the two hyperfine states, the system is formally equivalent, as it is described by the same Hamiltonian. The tunnel coupling energy $J$ corresponds to the Rabi frequency of the two-photon transition, and interactions can be tuned thanks to a Feshbach resonance.}.

We shall just underline the existence of three distinct regimes, depending on the value of $\gamma = U/2J$. These regimes are apparent of figure~\ref{fig:phase_portrait}, which shows the phase space trajectories described by the system of equations~\eqref{eq:Josephson_set}\footnote{For simplicity here we set the bias to $\delta = 0$.}:
\begin{enumerate}
\item for $\gamma \ll 1/N$, interactions are essentially negligible, and we recover the Rabi dynamics described in section~\ref{sec:Rabi}. In particular the angular frequency of these oscillations is the Rabi frequency $\Omega_\text{R} = 2 J/\hbar$. Close to $\gamma = 1/N$, the oscillation frequency around the fixed point $(\phi = \pi, z = 0)$ is significantly modified: $\omega_\pi^{-} = \Omega_\text{R}\sqrt{1 - \gamma N}$.
\item for $1/N < \gamma < 2/N$, the fixed point $(\phi = \pi, z = 0)$ becomes unstable, and two new fixed points appear at $(\phi = \pi, z = \pm \sqrt{1 - (\gamma N)^{-2}})$. Depending on the initial conditions $(\phi(0), z(0))$, the population imbalance and phase will either oscillate around 0 (so-called ``Josephson oscillations'', blue trajectories in figure~\ref{fig:phase_portrait}) at the ``Josephson'' or ``plasma'' frequency
\beq
\omega_\text{p} = \Omega_\text{R}\sqrt{1 + \gamma N} , \label{eq:plasma_freq}
\eeq
which is higher than the Rabi frequency, or around the two other fixed points. In this later case the oscillation frequency is $\omega_\pi^{+} = \Omega_\text{R}\sqrt{1 - (\gamma N)^{-2}}$.
\item for $\gamma > 2/N$ the modes discuss above still exist, but a new behavior appears, in which the phase is not a periodic function of time (it increases almost linearly), and the population imbalance oscillates around an non-zero value (red lines in figure~\ref{fig:phase_portrait}). This is the so-called ``macroscopic quantum self-trapping'' regime. Interactions inhibit full population transfer. There is a smooth transition to the trivial case of two decoupled coherent condensates ($\gamma \rightarrow \infty$), for which the phase accumulation is strictly linear, and the population imbalance constant, as atoms cannot tunnel anymore.
\end{enumerate}

\paragraph{Maximal amplitude of the Josephson oscillations}

The fact that a bifurcation separates Josephson oscillations from macroscopic quantum self-trapping also imposes a maximal amplitude to the Josephson oscillations, given by
\beq
z_\text{max} = \max_t |z(t)| = \frac{2\sqrt{\gamma N - 1}}{\gamma N} . \label{eq:zmax}
\eeq
We see that if Rabi/Josephson oscillations are used to perform a beam-splitter operation, as discussed in section~\ref{sec:trapped_recombination}, then interactions will limit the contrast of the beam splitter to $z_\text{max}$.

\subsubsection{Fluctuations and interferometry} \label{sec:fluctuations}

\paragraph{Shot noise limit}

Fluctuations play an important role in interferometry as they are related to the fundamental sensitivity attainable. To illustrate that, let us assume that $N$ photons are sent into an interferometer one after another. Let us assume that the mean number of photons that escape through one output, say the ``left output port'' L, is given by $\langle \hat n_\text{l}\rangle = (N/2)(1 + \sin \phi)$, $\phi$ being an adjustable phase shift. If we set $\phi$ to 0, then on average, $50\%$ of the photons escape in each output port. This means that the photons are each in an equal superposition of being in both outputs of the interferometer. If one tries to detect them, they will be found with a probability $p = 1/2$ in each output. For one photon, the variance of the number of photons which has escaped through L is
$
\text{Var}(\hat n_\text{l}^{(1)}) = p \times 0^2 + p \times 1^2 - p^2 = 1/4 ,
$
the last term being the mean photon number squared. For $N$ photons, the experiment is repeated independently $N$ times, therefore the variances just add up, such that $\text{Var}(\hat n_\text{l}) = N \times \text{Var}(\hat n_\text{l}^{(1)}) = N/4 = \Delta \hat n_\text{l}^2$. The sensitivity to a phase shift, giving the ability to distinguish between nearby values of the phase, is
\beq
\delta \phi = \frac{\Delta \hat n_\text{l}}{\partial \langle \hat n_\text{l}\rangle/\partial \phi} = \frac{1}{\sqrt{N}} .
\eeq
This is the so-called \emph{shot noise limit} of interferometry, also called ``standard quantum limit'', or just ``standard limit''. Here we have derived it assuming a perfect interferometer (of maximal contrast), and independent particles\footnote{Note that other values of the phase would give worse sensitivities, as the fluctuations would be reduced, but the slope of $\langle \hat n_\text{l}(\phi) \rangle$ as well.}. We see that if one is now able to introduce correlations between the particles, one may be able to improve the sensitivity. For instance, if the probability of finding a particle in the ``right'' output gets larger when a particle has escaped through the ``left'' one, the sensitivity may be brought below the shot noise limit: $\delta \phi < 1/\sqrt{N}$.

\paragraph{Mapping to a spin $N/2$ system}

In order to analyze the fluctuations properties of the Bose-Hubbard model, we first introduce the mapping of the system to an ensemble of $N$ spin $1/2$. For that we define new operators form the creation/annihilation operators introduced above:
\begin{align}
\hat S_x &\equiv \frac{1}{2}\left(\hat l^\dagger \hat r + \hat l \hat r^\dagger \right) \label{eq:Sx} \\
\hat S_y &\equiv \frac{1}{2i}\left(\hat l^\dagger \hat r - \hat l \hat r^\dagger \right) \label{eq:Sy} \\
\hat S_z &\equiv \frac{1}{2}\left(\hat l^\dagger \hat l - \hat r^\dagger \hat r \right) = \frac{1}{2}(\hat n_\text{l} - \hat n_\text{r}) . \label{eq:Sz}
\end{align}
One can readily check that they satisfy spin commutation relations: $[\hat S_l, \hat S_m] = i \hat S_p$, where $(l,m,p)$ is any circular permutation of $(x,y,z)$.

\paragraph{``Coherent'' atomic states}

Among the many quantum states that such a complex system can have, a class of states is particularly interesting: the coherent states. They can be parametrized by two angles, setting the mean direction in which the spin is pointing. They are defined as the eigenstates of a spin component in the $(\theta, \phi)$ direction, i.e.\ the eigenstates of $\hat S_{\theta, \phi} = \hat S_x \sin \theta \cos \phi + \hat S_y \sin \theta \sin \phi + \hat S_z \cos \theta$, with eigenvalue $N/2$, where $\theta$ and $\phi$ denote the polar and azimuth angles \cite{Kitagawa1993}. They can be written
\beq
\ket{\theta, \phi} = \frac{1}{\sqrt{2^N N!}} \left(\cos\frac{\theta}{2} \hat l^\dagger + e^{i \phi} \sin\frac{\theta}{2} \hat r^\dagger\right)^N \ket{0,0} , \label{eq:coherent_state}
\eeq
form which underlines the fact that these states are generated by ``stacking'' particles one after another into the same (single particle) coherent superposition. 

To better understand these states, we can calculate the number distribution corresponding to a coherent state pointing on the equator of the Bloch sphere. By developing the expression \eqref{eq:coherent_state} we obtain the probability of finding $k$ particles in one wells and $N-k$ in the other
\beq
|\langle k,N-k|\pi/2,\phi\rangle|^2 = \frac{1}{2^N} \begin{pmatrix} N \\ k \end{pmatrix} .
\eeq
We recognize a binomial distribution of atoms in the two wells.

\paragraph{Bloch sphere representation}

\begin{figure}[t]
\centering
\includegraphics[width=0.3\linewidth]{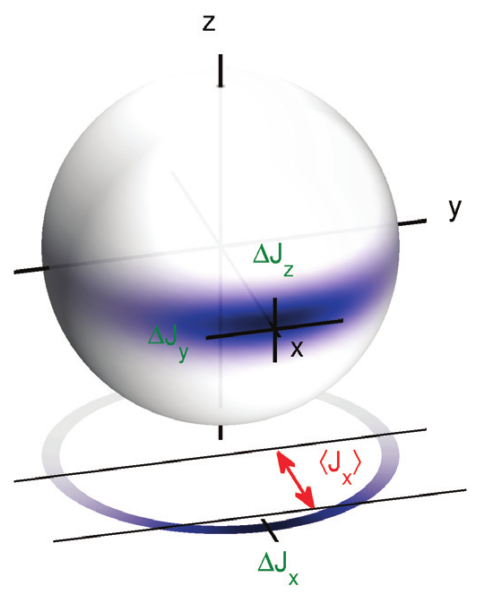}
\caption{\label{fig:bloch_sphere}Quasiprobability of a number-squeezed state with squeezing factor $\xi_N \simeq 0.2$ represented on the Bloch sphere. The z-axis corresponds to the number difference. Figure from ref.~\cite{Grond2010}.
}
\end{figure}

The possible states of a single spin 1/2 are parametrized by the polar and azimuthal angles and can therefore be represented as points on the surface of the Bloch sphere. This representation can be extended to a spin $N/2$. For this, one can represent the many-body state $\ket{\Psi}$ using the quasiprobability distribution~\cite{Kitagawa1993}
\beq
Q(\theta, \phi) = |\langle \theta, \phi | \Psi \rangle|^2 ,
\eeq
which can be plotted on the surface of a ``generalized'' Bloch sphere.

\paragraph{Bose-Hubbard Hamiltonian}

Rewriting the Bose-Hubbard Hamiltonian~\eqref{eq:BH_Hamiltonian} with the operators \eqref{eq:Sx}--\eqref{eq:Sz} we obtain\footnote{We have assumed a closed system with constant atom number $N = \hat n_\text{l} + \hat n_\text{r}$, and removed the constant terms from the Hamiltonian (constant energy shifts).}
\beq
H_\text{BH} = -2J \hat S_x + U \hat S_z^2 - \delta \hat S_z .
\eeq
This from is particularly appealing, because it has a nice interpretation in terms of trajectories on the Bloch sphere. Indeed, if one considers the first term only (setting $U = \delta = 0$), it generates an evolution operator $\hat U(t) = \exp(2 i J \hat S_x t/\hbar)$, that is a rotation around the $x$ axis of the Bloch sphere at the angular frequency given by the Rabi frequency $\Omega_\text{R} = 2J/\hbar$. Similarly, the third terms drives rotations around the $z$ axis at an angular frequency $\delta/\hbar$ (phase accumulation). The interaction term (second term) is different, as it is $\propto \hat S_z^2$. Classically, it would be a rotation around the $z$ axis, whose frequency depends on the position on the $z$ axis (making the approximation $\hat S_z^2 \simeq \langle \hat S_z \rangle \hat S_z$). Contrary to the other, this non-linear term, which stems from interactions, is responsible for non-trivial modification of the state and leads to dephasing, squeezing, and the generation of strongly entangled states \cite{Kitagawa1993, Piazza2008}.

\paragraph{Fluctuations}

For simplicity, let us consider the coherent state $\ket{\pi/2, 0}$, i.e.\ the eigenstate of $\hat S_x$. All the other coherent states can be transformed into this one by applying an appropriate rotation. By construction this state has a length $N/2$ along the $x$-axis and no fluctuations: $\Delta \hat S_x^2 = 0$. Nevertheless, it has fluctuations along the other axes $y$, and $z$, that can be seen as a consequence of adding $N$ $1/2$ spins having the same direction (and strong fluctuations). The uncertainty relations associated with the commutation relations satisfied by the components of the spin are $\Delta \hat S_k  \, \Delta \hat S_m \geq |\langle \hat S_n \rangle|/2$. Applied to the coherent state above with $(k,m,n) = (y,z,x)$ gives $\Delta \hat S_y \,   \Delta \hat S_z \geq N/4$. Coherent states are characterized by a saturation of the uncertainty relation above and equal fluctuations along these two orthogonal axis: $\Delta \hat S_y = \sqrt{N}/2$ and $\Delta \hat S_z = \sqrt{N}/2$. Since $\hat S_z$ corresponds to half the population difference (cf.\ equation \eqref{eq:Sz}), we see that coherent states have binomial number fluctuations $\Delta \hat n = \sqrt{N}$\footnote{Here we used the notation $\hat n = (\hat n_\text{l} - \hat n_\text{r}) = 2 \hat S_z$}. If we express the fluctuations along $y$ as fluctuations of the angle $\phi$, we obtain $\Delta \phi \simeq (\sqrt{N}/2)/(N/2) = 1/\sqrt{N}$. This is assuming $\Delta \phi \ll \pi$, i.e.\ that the atom number be not too small.

\paragraph{Squeezing as a resource for sub-shot-noise interferometry}

Squeezing is defined as a redistribution of the fluctuations along two axes orthogonal to the mean spin direction \cite{Kitagawa1993}. One can thus talk about \emph{number squeezing} when the number fluctuations are reduced compared to the binomial case. For a state close to the coherent state $\ket{\pi/2, 0}$\footnote{We choose this state because it is the ground state of the Bose-Hubbard Hamiltonian in the absence of interactions.} this is quantified by the \emph{number squeezing factor} \cite{Kitagawa1993}
\beq
\xi_N^2 = \frac{\Delta \hat S_z^2}{N/4} ,
\eeq
It compares the number fluctuations ($\Delta \hat S_z^2$) to that of a coherent state ($N/4$). Therefore when $\xi_N < 1$, the fluctuations are reduced compared to shot noise, and the state is said to be squeezed.

One could as well define squeezing along any axes of the Bloch sphere \cite{Kitagawa1993}. For instance, phase squeezing is quantified by the \emph{phase squeezing factor} $\xi_\phi^2 = 4 \Delta \hat S_y^2/N$.

Squeezing has been shown to be a resource for interferometry below the shot noise limit\footnote{Cf.\ ref.~\cite{Wineland1994} on the applications to spectroscopy, and references therein concerning the use of squeezed states in optical interferometry. For a more general review on the use of non-classical states for quantum-enhanced measurements, cf.\ ref.~\cite{Giovannetti2004} and references therein.}. As we saw a the beginning of section \ref{sec:fluctuations}, the shot noise stems from the measurement of coherent states at the output of the interferometers. Since this corresponds to having no correlations between particles, or in other words, to sending the particles in the interferometer one after another, this limit is often considered ``classical''. Interactions can introduce correlations between the particles, an example being squeezing, which can be used to obtain squeezed state at the output of the interferometer, thus decreasing the ``shot noise'' along an axis most appropriate for the measurement.

The improvement of signal to noise ratio of an interferometer using a squeezed state is given, not directly by the squeezing factors introduced above, but by the so-called \emph{spin squeezing factor} \cite{Wineland1994, Sorensen2001}
\beq
\xi_S^2 = \frac{N\Delta \hat S_\mathbf{n_1}^2}{\langle \hat S_\mathbf{n_2} \rangle^2 + \langle \hat S_\mathbf{n_3} \rangle^2} ,
\eeq
where $\hat S_\mathbf{n} = \mathbf{n} \cdot \mathbf{\hat S}$ and the $\mathbf{n}$s are mutually orthogonal unit vectors \cite{Sorensen2001}. For states close to the ground state of the Bose-Hubbard Hamiltonian, this expression reduces to\footnote{This corresponds to the choice $(\mathbf{n_1}, \mathbf{n_2}, \mathbf{n_3}) = (\mathbf{z}, \mathbf{x}, \mathbf{y})$, and noticing that that, for this particular state, $\langle \hat S_y \rangle = 0$ (or equivalently $\langle \phi \rangle = 0$). We could have chosen another direction than $z$ to define the squeezing factor, but this choice is governed by the fact that the ground state is maximally squeezed along this axis.}
\beq
\xi_S = \frac{\sqrt{N} \Delta \hat S_z}{|\langle \hat S_x \rangle|} .
\eeq
It can be interpreted as the ratio of the number squeezing factor $\xi_N$ introduced earlier, and the \emph{coherence factor}, $2 \langle \hat S_x \rangle/N$, characterizing the coherence of the state. Indeed, in the classical limit, we have $2 S_x/N = \cos \phi$, and since the mean phase is $\langle \phi \rangle = 0$, $2 \langle S_x\rangle/N$ will be close to 1 when the phase distribution is peaked around 0 (very coherent state), and close to 0 when the phase is uniformly distributed in $]-\pi, \pi]$ (no coherence).

Finally, we note that the spin squeezing factor has been related to the degree of entanglement present in the system \cite{Sorensen2001, Sorensen2001a, Hyllus2012}.

\subsubsection{Squeezing during splitting} \label{sec:splitting2}

We return to the splitting problem in a double-well potential, but now analyze how the fluctuations are modified, and show how number-squeezed states are generated. 

\paragraph{Ground-state fluctuations of the Bose-Hubbard Model}

\begin{figure}
\centering
\includegraphics[width=\linewidth]{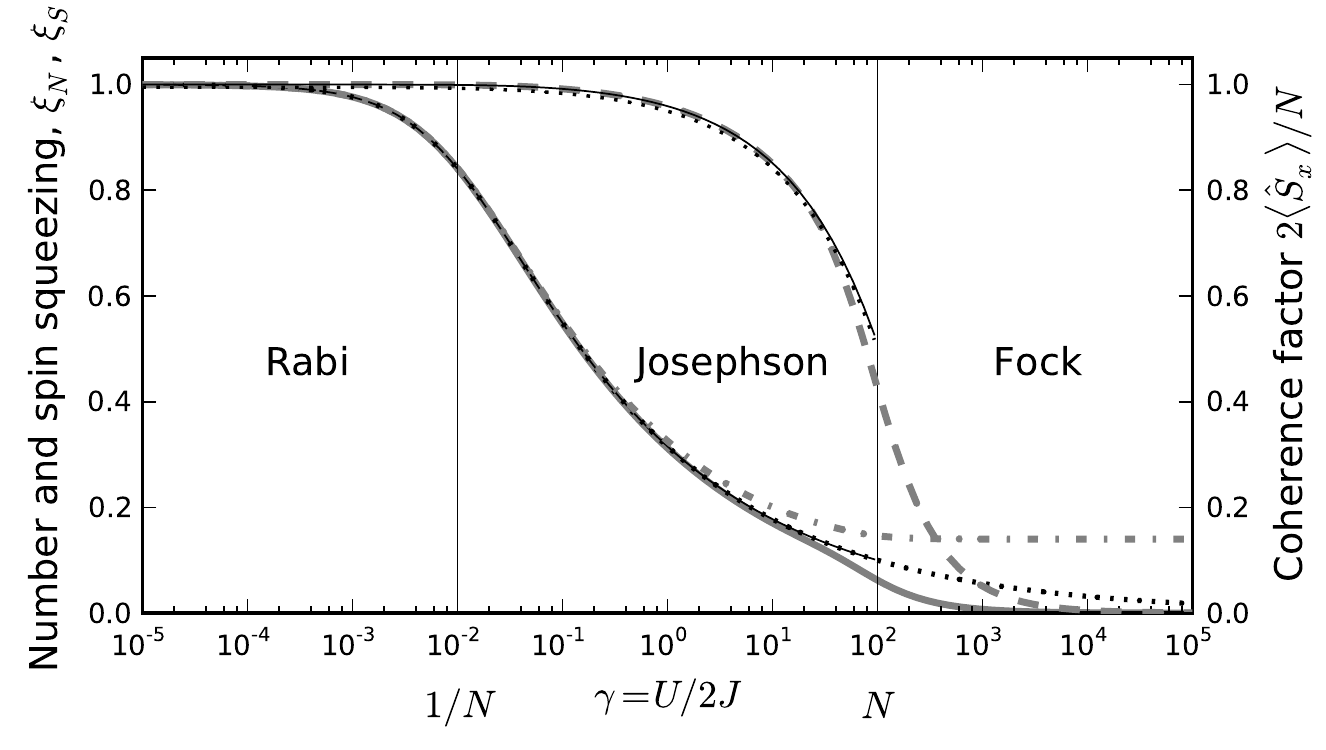}
\caption{\label{fig:GS_BH_H}Number fluctuations and coherence of the ground state of the Bose-Hubbard Hamiltonian vs strength of interactions. The thick grey lines are obtained from an exact computation of the ground state of the Hamiltonian~\eqref{eq:BH_Hamiltonian} with $N = 100$ atoms and no bias ($\delta = 0$). Solid line: $\xi_N$, dashed: coherence factor, dot-dashed: $\xi_S$. The thin solid lines are the results of the Bogoliubov treatment detailed in ref.~\cite{Guery-Odelin2011}, valid only in the Rabi and Josephson regimes. The dotted lines are obtained from the harmonic approximation of section~\ref{sec:harmonic} (see equations~\eqref{eq:harmonic_number} and \eqref{eq:harmonic_coherence}). The vertical lines at $\gamma = 1/N$ and $\gamma = N$ separate the three regimes.}
\end{figure}

As introduced in section~\ref{sec:splitting1}, double-well potentials essentially allow adiabatic splitting. The system stays in the ground state as long as adiabaticity is not broken by the fact that the typical time scale of the problem diverges when the tunnel coupling approaches $J \rightarrow 0$. It is therefore interesting to analyse the fluctuations of the ground state, which can be found by exact diagonalization of the Hamiltonian \eqref{eq:BH_Hamiltonian} as long as the particle number is not too large. In figure~\ref{fig:GS_BH_H}, the fluctuations are shown as a function of the ratio of interactions to tunneling energy $\gamma = U/2J$. The exact calculation is compared to the result of a Bogoliubov treatment of the Josephson junction, not detailed here~\cite{Guery-Odelin2011}, and to the approximate quadratic model detailed in section~\ref{sec:harmonic}.

We see three distinct regimes:
\begin{enumerate}
\item $\gamma \ll 1/N$: the interactions are negligible. This is the \emph{Rabi regim}, in which the dynamics is essentially the non-interacting dynamics presented in section~\ref{sec:Rabi} and \ref{sec:MF}, and the ground state is close to a coherent state. The squeezing factor is close to 1 (no squeezing) and the phase distribution has a typical width of $1/\sqrt{N}$.
\item $1/N < \gamma < N$: this is the \emph{Josephson regime}. The dynamics was already discussed in section~\ref{sec:MF}, in particular the small amplitude Josephson oscillations occur at the plasma frequency~\eqref{eq:plasma_freq}. In this regime the number fluctuations are reduced by repulsive interactions compared to the binomial case, and the coherence is also decreased but still high.
\item $\gamma > N$: the number fluctuations are low, approaching 0 for Fock states ($\gamma \rightarrow \infty$). This is the \emph{Fock regime}. The coherence is almost completely lost. In this regime, we expect the mean-field treatment to be invalid because phase fluctuations are large.
\end{enumerate}

\paragraph{Squeezing during adiabatic splitting}

Once again, splitting can be understood as the adiabatic following of the ground sate, which breaks when the adiabaticity condition cannot be fulfilled any more because of the divergence of $1/J(t)$ \cite{Leggett1991, Leggett1998}. In the many-body case, the Rabi frequency is replaced by the plasma frequency~\eqref{eq:plasma_freq}. One can show that the lower-lying many-body states are equally spaced by an energy $\hbar \omega_\text{p}$ in the Rabbi and Josephson regimes. The adiabatic condition~\eqref{eq:adiabaticity1} becomes:
\beq
| \dot \omega_\text{p} | \ll \omega_\text{p}^2 .
\eeq
If the Josephson frequency is initially high, and splitting performed slowly enough, then one can reach the Josephson regime adiabatically, before adiabaticity breaks down (when $|\dot \omega_\text{p}| \sim \omega_\text{p}^2$). In this regime, the squeezing factors introduced above are on the order of $\xi_N \sim (\gamma N)^{-1/4}$, and $\xi_\phi \sim (\gamma N)^{1/4}$, which corresponds to a number-squeezed state, because in the Josephson regime $\gamma > 1/N$ (see ref.~\cite{Leggett1998} and reference thereinn, and section\ref{sec:harmonic} for a derivation).

Once adiabaticity is broken and $J \rightarrow 0$, no tunneling is possible and the number distribution is frozen. Therefore the number fluctuations cannot change anymore. The best spin squeezing factor attainable with this adiabatic strategy (thick grey dash-dot line in figure~\ref{fig:GS_BH_H}) has been calculated in ref.~\cite{Pezze2005} to be
\beq
\xi_S^\text{min} = \sqrt{\frac{2}{N}} ,
\eeq
which corresponds to a phase sensitivity following the Heisenberg scaling $\delta \phi \propto 1/N$.

Concerning the relative phase distribution, we will see in the next section that it keeps evolving, yielding interesting effect.

\subsubsection{Phase diffusion}

\begin{figure}[t]
\centering
\includegraphics[width=\textwidth]{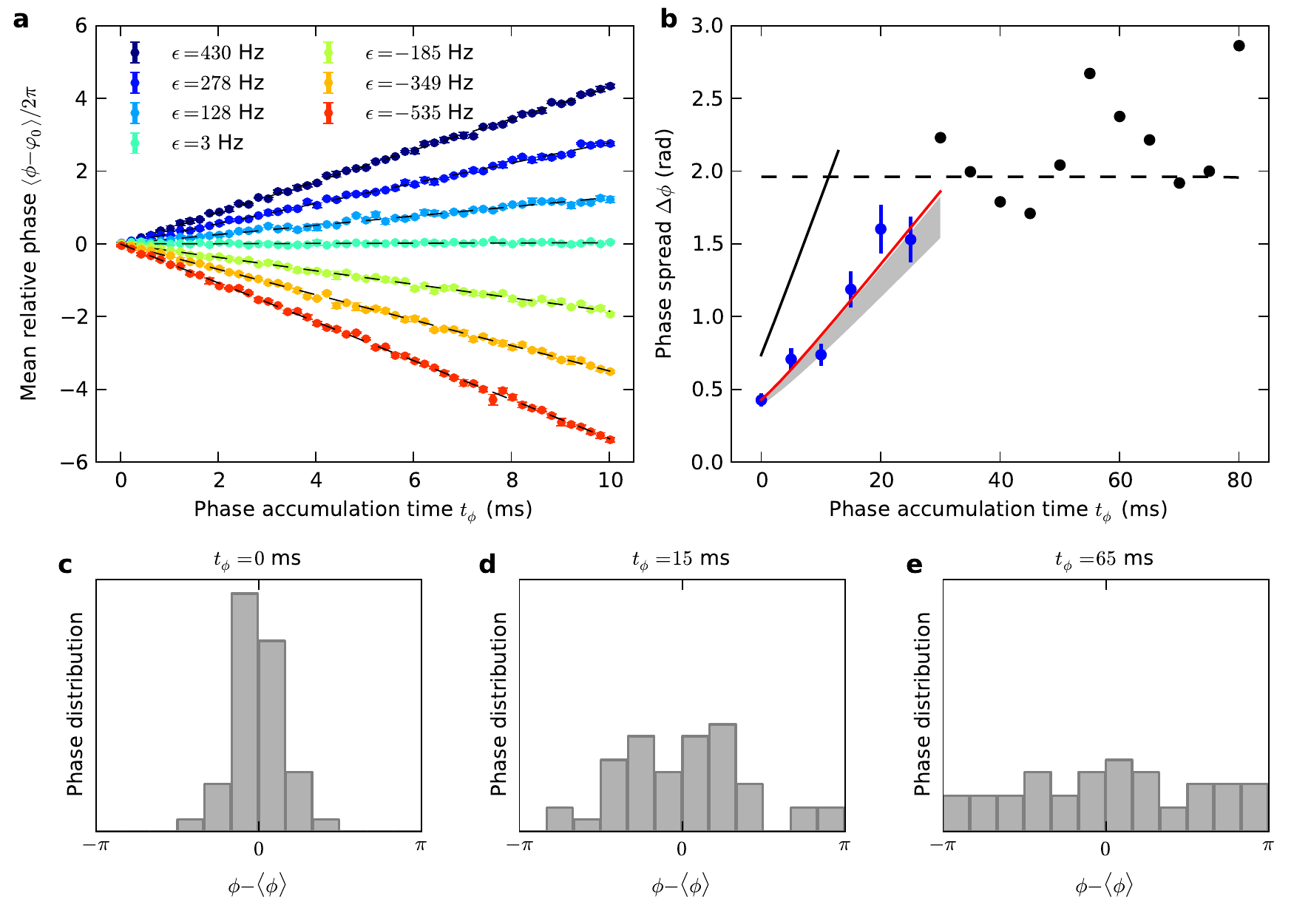}
\caption{Evolution of the relative phase and its fluctuations in a decoupled double well. \textbf{a,} Linear evolution of the phase for various energy differences induced by tuning the angle between the two wells. \textbf{b,} Evolution of the circular standard deviation of the phase $\Delta \phi$ corresponding to the orange curve of panel a. It exhibits phase diffusion. The red line is a fit to the data with the model of equation~\eqref{eq:phase_diffusion}. Shaded area: theoretical prediction (see details in ref.~\cite{Berrada2013}) \textbf{c,d,e,} Measured phase distributions vs time after decoupling the two wells. Figure adapted from ref.~\cite{Berrada2013}.
\label{fig:MZI_fig3}}
\end{figure}

We now discuss the evolution of the relative phase after the condensate has been fully split ($J=0$) in a symmetric way. In this case, the eigenstates of the system are the Fock states. Phase diffusion is essentially a dephasing effect between the Fock states involved in the superposition state describing the system~\cite{Leggett1991, Kitagawa1993, Castin1997, Javanainen1997}. We assume that the state after splitting is close to a coherent state (possibly squeezed) and therefore has small number fluctuations $2\Delta\hat S_z  = \xi_N \sqrt{N}$ with $\xi_N \lesssim 1$. The eigenenergy $E(n)$ of the Fock state $\ket{(N+n)/2, (N-n)/2}$ 
%
%
%
is calculated from equation~\eqref{eq:BH_Hamiltonian} to be
\beq
E(n) = E(0) -\frac{\delta}{2} n + \frac{U}{4} n^2 .
\eeq
One can show that the linear terms leads to a global phase accumulation at the rate $\mathrm{d}{\langle \phi \rangle}/\mathrm{d} t = \delta t/\hbar$, and that the quadratic term, which is finite in the presence of interactions ($U \neq 0$), leads to a randomization of the phase. At short times, the variance of the phase evolves as~\cite{Leggett1991, Castin1997, Javanainen1997, Leggett1998}
\beq
\Delta \phi^2 = (\Delta \phi^2)_{t=0} +  R^2 t^2 , \label{eq:phase_diffusion}
\eeq
where the phase diffusion rate is
\beq
R = \frac{1}{\hbar} \xi_N \sqrt{N} U .
\eeq
It scales as the product of the derivative of the chemical potential with respect to the number of atoms $U$, and the number fluctuations $\Delta \hat S_z \propto \xi_N \sqrt{N}$.

Phase diffusion is not limited to Bose-Einstein condensates trapped in double-well potential, but rather occurs in a variety of interacting systems. It is one of the main factors limiting the use of dense samples for matter-wave interferometry, as the presence of interactions sets a limit to the coherence time of the system. Nevertheless, we stress that the effect can be strongly reduced by using more dilute systems, or by tuning interactions: either by canceling them \cite{Gustavsson2008, Fattori2008}, or by reversing their sign in order to realize an analog of a spin echo in inhomogeneously broadened systems. 

\begin{figure}[t]
\centering
\includegraphics[width=0.6\textwidth]{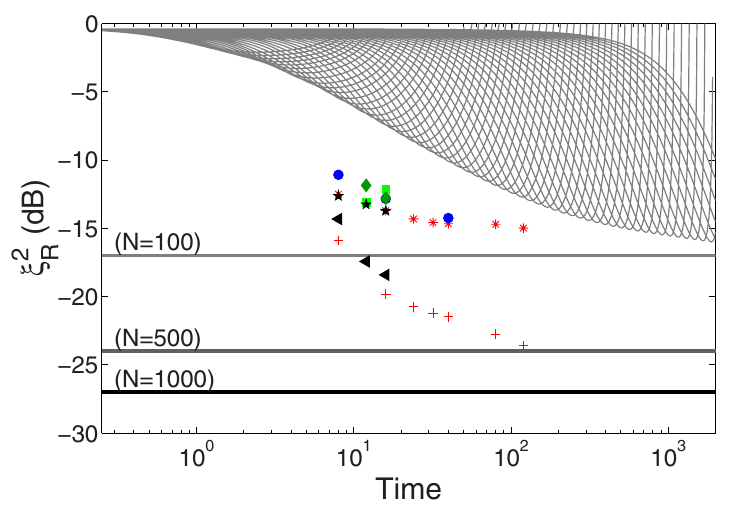}
\caption{Spin squeezing $\xi_R$ for optimized (symbols) and exponential splitting (gray lines). The quasiadiabatic exponential splitting
is shown for several time scales for N=100. After $\xi_R$ reaches a minimum, it rises again due to loss of phase coherence caused by interaction induced de-phasing. With optimized splitting, better spin squeezing (lower $\xi_R$) is obtained at much shorter time scales. OCT calcualtions are for N=100 except: black triangle: N=500  and red + : N=1000. The horizontal lines show the Heisenberg limit $\xi_R=\sqrt{\frac{2}{n}}$. Figure adapted from ref.~\cite{Grond2009a}.
\label{fig:OCT_SpinSqueezing}}
\end{figure}

A diferent approach is to optimize the dynamics of the splitting process using optimal control theory in order to counteract phase diffusion~\cite{Grond2009, Grond2009a, Grond2010, Grond2011a}.  The optime controll inspired splitting sequences of J. Grond et al. ~\cite{Grond2009, Grond2009a} will allow to speed up the splitting process by mre then an order of amgnitude compared to an adiabatic splitting and reaches the same or better number squeezing.  The faster splitting leads to much reduced phase diffusion and consequently to much better spin squeezing $\xi_S$ (figure~\ref{fig:OCT_SpinSqueezing}). The simplest procrdure is: first splitting fast to finite tunnel coupling, then keeping the double well for an extended time so that number squeezing can be establishd, and then splitting fast into two separated wells. 

We also underline that phase diffusion is similar to the generation of squeezing in the one-axis twisting scheme proposed in ref.~\cite{Kitagawa1993} and demonstrated in the group of M.~K.~Oberthaler~\cite{Gross2010}. The variance of the phase increases, but simultaneously the fluctuations are reduced along other directions on the surface of the Bloch sphere, as squeezing is generated~\cite{Kitagawa1993}.

Finally, we note that at longer times, the phase will in theory display revivals~\cite{Castin1997} similar to those observed in optical lattices~\cite{Greiner2002a}, and the system is expected to reach strongly non-classical states, for example corresponding to a superposition of two coherent states having a relative phase which differ by $\pi$~\cite{Piazza2008}.

\subsubsection{Effective single-particle Hamiltonian} \label{sec:harmonic}

We now show that the many-body Hamiltonian \eqref{eq:BH_Hamiltonian} can, to some extend, be approximated by a single particle Hamiltonian. Similar approaches have been described in refs.~\cite{Paraoanu2001, Pezze2005, Julia-Diaz2012a}. For this we follow the same approach as in section \ref{sec:MF}, but keeping the population imbalance and phase as quantum mechanical operators\footnote{Note that a phase operator $\hat \phi$ cannot be defined rigorously, see for instance refs.\ \cite{Lerner1968, Leggett1991, Leggett2001}.}.
From the classical limit of section \ref{sec:MF}, we have seen that $z$ and $\phi$ are canonically conjugate, therefore we are tempted to introduce the following commutation relation for $\hat S_z = (\hat n_\text{r} - \hat n_\text{l})/2$ and $\hat \phi$:
\beq
[\hat \phi, \hat S_z] = i . \label{eq:commutator}
\eeq
The constant $1 \times i$ is chosen to be dimensionless and such that the Heisenberg uncertainly relation yields minimal uncertainly states which resemble the coherent states introduced in section \ref{sec:fluctuations}. That is $\Delta \hat S_z \, \Delta \hat \phi \geq 1/2$, and thus $\Delta \hat S_z = \sqrt{N}/2 \Rightarrow \Delta \hat \phi \geq 1/\sqrt{N}$.

We substitute $\hat l = \sqrt{\hat n_\text{l}}$ and $\hat r = \sqrt{\hat n_\text{r}} e^{i \hat \phi}$ in equation \eqref{eq:BH_Hamiltonian}. The Hamiltonian \eqref{eq:BH_Hamiltonian} becomes
\begin{equation}
H = -JN \sqrt{1 - 4(\hat S_z/N)^2} \cos \hat \phi + U \hat S_z^2 + \delta \hat S_z , \label{eq:number_phase}
\end{equation}
where constant terms have again been removed. 

\paragraph{Harmonic approximation}

Assuming no bias ($\delta = 0$) zero imbalance ($\langle \hat S_z \rangle = 0$) and small fluctuations ($\Delta \hat S_z \ll N$, $\Delta \hat \phi \ll 1$), one can develop \eqref{eq:number_phase} to second order in $\hat S_z$ and $\hat \phi$ to obtain the quadratic Hamiltonian
\begin{equation}
H = \left(U + \frac{2 J}{N}\right) \hat S_z^2 + \frac{JN}{2} \hat \phi^2 .
\end{equation}
If we assume that the operators $\hat S_z$ and $\hat \phi$ have continuous spectra (which is strictly speaking not the case), then they are equivalent to position and momentum operators because they have a similar commutation relation \eqref{eq:commutator}, and one can therefore describe the many-body system by a single-particle wave function $\psi(\phi, t)$. Here the phase plays the role of a position and the number difference, that of a momentum. In this ``$\ket{\phi}$'' representation, the number difference operator takes the form
\begin{equation}
\hat S_z = - i \frac{\partial}{\partial \phi} ,
\end{equation}
and the wave function satisfies the Schr\"{o}dinger equation
\beq
i \hbar \, \frac{\partial\psi}{\partial t}  = \left[- \left(U + \frac{2 J}{N}\right) \frac{\partial^2}{\partial\phi^2} + \frac{JN}{2} \phi^2 \right] \psi , \label{eq:harmonic}
\eeq
This is the Schr\"{o}dinger equation of a single particle having a mass $m \propto (U + 2J/N)^{-1}$, in a potential having an angular frequency $\omega_p = 2 J \sqrt{1 + U N/2 J}/\hbar$. We recover the expression of the plasma frequency \eqref{eq:plasma_freq}.

Rabi and Josephson oscillations will therefore correspond to a dipole oscillation of the ``phase wave function'' $\psi(\phi)$ in an effective harmonic potential.

\paragraph{Ground-state fluctuations}

The number and phase fluctuations can also be computed. For instance the ground sate is the well known Gaussian minimal uncertainty state, which has number and phase fluctuations given by
\begin{align}
\Delta \hat S_z &= \frac{\sqrt{N}}{2}(1 + \gamma N)^{-1/4} , \label{eq:harmonic_number} \\
\Delta \hat \phi &= \frac{(1 + \gamma N)^{1/4}}{\sqrt{N}} \label{eq:harmonic_phase} .
\end{align}
From the second equation, the coherence, introduced in section \ref{sec:fluctuations}, can be calculated as
\beq
\langle \cos \hat \phi \rangle \simeq \cos \Delta \hat \phi \simeq 1 - \frac{\Delta \hat \phi^2}{2} , \label{eq:harmonic_coherence}
\eeq
provided that $\gamma \ll N$. The expressions \eqref{eq:harmonic_number} and \eqref{eq:harmonic_coherence} are represented as dotted lines in figure~\ref{fig:GS_BH_H}. We see that they agree well with both the exact calculation and the Bogoliubov expressions in the Rabi and Josephson regimes.

\paragraph{Phase diffusion}

Phase diffusion can also be recovered with this model. A Gaussian wave packet having an initial ``width in momentum space'' $\Delta \hat S_z$, and left in a flat potential ($J = 0$) will spread as
\beq
\Delta \phi^2 = (\Delta \phi^2)_{t=0} +  4 N  \Delta \hat S_z^2 \frac{U^2 t^2}{\hbar^2} , \label{eq:phase_diffusion2}
\eeq
which is the same expression as equation~\eqref{eq:phase_diffusion}.

%
%
%
%
%
%
%
%
%
%

\section{Probing Many-Body Physics by Interference}  \label{sec:ProbingManyBody}

A general understanding of quantum many-body systems is an important unsolved problem, touching systems from diverse fields such as cosmology, high-energy physics or condensed matter. In particular, the question of why and how isolated quantum systems relax toward equilibrium states has so far only been studied for a very limited number of special cases (for a review, see ref.~\cite{Polkovnikov11}). For example, when dealing with integrable systems exhibiting many constants of motion, thermalization can be completely absent or strongly inhibited.

A key challenge in such studies is the lack of techniques for characterizing complex quantum many-body states. Moreover, investigations of non-equilibrium dynamics are challenged by the scarcity of quantum many-body systems that are both well isolated from the environment and accessible to experimental study. In the following we show that the toolbox of atomic physics and matter-wave interferometry that has been presented in the previous lectures is ideally suited to tackle such problems. 

\subsection{Interference of 1D Bose gases}  \label{ssec:1DIFM}

\subsubsection{The 1D Bose gas}  \label{sssec:1D}

The experimental studies that we present are performed using trapped 1D Bose gases. Such systems offer two unique advantages for quantum many-body physics: firstly, on the experimental side, realizing them with ultracold atoms facilitates a precise preparation and probing of the system. Secondly, on the theoretical side, 1D Bose gases offer a model system which contains complex many-body physics but can still be captured with reasonable theoretical effort, in particular due to the existence of effective models which allow to describe the essential physics in a relatively simple way~\cite{Giamarchi04}. Furthermore, the homogeneous 1D Bose gas with repulsive contact interactions is an example of a fully integrable quantum system~\cite{Lieb63,Yang69}. The approximate realization of such a system in experiments thus allows the study of thermalization in the vicinity of multiple conserved quantities and hence the study of the interplay between integrability, many-body dynamics and thermalization. In this lecture we present a short overview of the topic. Further details can be found in the PhD thesis of Tim Langen~\cite{LangenThesis}. 

\paragraph{Theoretical description}

In cold atom experiments, a 1D Bose gas can be realized using anisotropic magnetic or optical trapping potentials, where the confinement in two directions is strong enough such that the temperature and the chemical potential of the system are smaller than the excited energy levels of the trapping potential. This can be expressed by the condition 
\begin{equation}
k_B T,\mu\lesssim\hbar \omega_\perp,\label{eq:1Dcondition}
\end{equation}
where $\omega_\perp$ denotes the harmonic trap frequency in the two strongly confining directions. In the following, we will use the convention that the strongly confining trap directions are the $x$-direction and the $y$-direction, such that $\omega_\perp=\omega_x=\omega_y$. In contrast to the strongly confining directions many momentum modes can be occupied in the weakly confining direction. This leads to markedly different behavior than in 3D BECs, where only the lowest momentum mode is macroscopically occupied. These many momentum modes in 1D Bose gases are the origin of strong density and phase fluctuations. It has been shown rigorously by Mermin, Wagner and Hohenberg~\cite{Mermin66,Hohenberg67} that because of this enhanced role of fluctuations no off-diagonal long-range order and thus no BEC can exist in ideal 1D Bose gases, even at zero temperature. 

The fact that the 1D nature of the gas is achieved by strongly confining it in two directions of the 3D space is reflected in the scattering properties of the atoms. These are still 3D for the parameters reached in our experiments with 1D Bose gases. For temperatures below the degeneracy temperature $T_D=\hbar^2\nOneD^2/(2mk_B)$ and sufficiently high density, a mean-field description is applicable to model these properties. In this case, the 1D dynamics can be described by integrating out the two strongly confining directions, leading to an effective scattering potential $U(z-z^\prime)=g\delta(z-z^\prime)$. Here, 
\begin{equation}
g = 2\hbar a_s \omega_\perp,\label{eq:g1D}
\end{equation}
is the 1D interaction parameter, with $a_s$ denoting the s-wave scattering length, and we have assumed $a_s\ll \sqrt{\hbar/m\omega_\perp}$~\cite{Olshanii98,Haller10b}.

The homogeneous 1D Bose gas with such delta function interactions is one of the prime models of mathematical and statistical physics and can be described by the Hamiltonian
\begin{align}
\hat H = \int dz&\,\hat\Psi^\dagger(z,t)\left(-\frac{\hbar^2}{2m}\frac{\partial^2}{\partial z^2}\right)\hat\Psi(z)+\nonumber\\
&+\frac{1}{2}\int dz\, dz^\prime\, \hat\Psi^\dagger(z,t)\hat\Psi^{\dagger}(z^\prime,t)g\delta(z-z^\prime)\hat\Psi(z^\prime,t)\hat\Psi(z,t)\label{eq:LiebLiniger}.
\end{align}

An exact solution based on the Bethe Ansatz~\cite{Bethe31} was found by Girardeau, Lieb and Liniger~\cite{Lieb63,Girardeau60,Lieb63b}. This exact solution was used by Yang and Yang to describe the system at finite temperature~\cite{Yang69,KorepinBook}. 
Note that the existence of this exact solution implies that the system is integrable, which is expected to have important consequences on its non-equilibrium dynamics~\cite{KorepinBook,Rigol09}. In the exact solution, the strength of the interactions is parametrized by the Lieb-Liniger parameter $\gamma$ which is defined as
\begin{equation}
\gamma = \frac{m g}{\hbar^2n_{1D}}.
\end{equation}
The 1D Bose gas thus becomes more strongly interacting for lower density. For $\gamma\gg 1$ the system is in the strongly-correlated Tonks-Girardeau regime, for $\gamma\ll 1$ it is in the weakly-interacting regime. The complete diagram of states at finite temperature was studied in ref.~\cite{Kheruntsyan03}. For the typical parameters of our experiments, $\gamma$ is on the order of $10^{-2}-10^{-3}$ and the temperature is approximately $10^{-2} T_D$. The gas is thus a weakly-interacting quasi-condensate characterized by suppressed density fluctuations. The phase, on the other hand, strongly fluctuates. The corresponding correlation function decays exponentially, with the thermal coherence length $\lambda_T=2\hbar\nOneD/(mk_BT)$ being directly related to the temperature $T$ and the 1D line density $\nOneD$ of the system. 

\begin{figure}
\centering
\includegraphics[width=.70\textwidth]{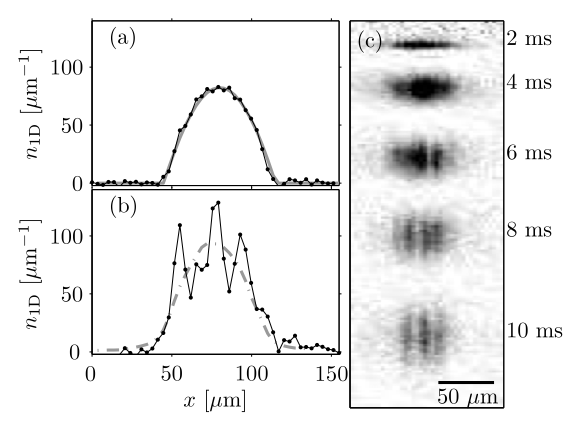} 
\caption{(a) Linear density profile of a 1D Bose gas after a short expansion time ($2\,$ms), where density fluctuations have not yet developed. (b) Linear density profile after $10\,$ms of free expansion, showing high contrast density fluctuations, where the averaged density profile (gray dashed line) is smooth. Studying the statistical properties of these density fluctuations enables the measurement of the temperature of the gas. (c) Example images of density fluctuations emerging during free expansion. Taken from~\cite{Manz10}.}
\label{fig:ripples}
\end{figure}

In experiment, these phase fluctuations of a quasi-condensate can be observed in two different ways. First, the spatially varying phase $\theta(z)$ corresponds to a velocity field $\bm v  = \nabla \theta(z)$ for the atoms. In time-of-flight expansion the phase fluctuations will thus turn into density fluctuations, similar to an optical speckle pattern~\cite{Imambekov09,Manz11,Langen13D}. An example is shown in figure~\ref{fig:ripples}. As the gas rapidly expands in the radial direction, interactions can be neglected in the expansion and it is thus possible to directly relate the correlation properties of these \textit{density ripples} density ripples to the \textit{in situ} phase fluctuations. Measuring these correlation properties can therefore be used to extract the temperature of the gas. This is particulary useful for highly-degenerate 1D Bose gases, where no thermal background is visible anymore. Second, two quasi-condensates can be made to interfere, revealing a fluctuating relative phase in their interference pattern. This situation will be discussed in detail in the next section. 

\subsubsection{Multi-Mode interference}  \label{sssec:MM}

The interference of 1D Bose gases in a double well can be described in a way that is very similar to the simple interference of the two Gaussian wave packets that was introduced in section~\ref{sec:IFMTrappedBEC}. We also point the reader to the lectures of Eugene Demler at the Varenna school 2006~\cite{Imambekov06}. To describe the many-body nature of the system the fields have to replaced with operators, leading to
\begin{align}
\left(\hat\Psi_l+\hat\Psi_r\right)^\dagger&\left(\hat\Psi_l+\hat\Psi_r\right)\nonumber\\
&=\langle|\hat\Psi_l|^2\rangle+\langle|\hat\Psi_r|^2\rangle + \hat\Psi_l{}^\dagger\hat\Psi_r{}+\hat\Psi_r{}^\dagger\hat\Psi_l{}.
\end{align}
Here, the field operators should be taken after expansion. Explicitly, $\hat\Psi(\bm r,z)= \hat\Psi_l(z)e^{i Q_{L}(\bm r+\bm d/2)-iQ_l^2t/2\hbar m}+\hat\Psi_r(z)e^{i Q_r(\bm r-\bm d/2)-iQ_r^2t/2\hbar m}$ denotes the total field operator after expansion, with $\bm Q_{l,r} = m(\bm r \pm \bm d/2)/\hbar t$ the momenta of atoms that are released from one of the condensates and detected at a point $\bm r$.~\cite{Polkovnikov06}. Again, the cross terms are responsible for the interference. To formally describe the interference pattern we introduce the operator~\cite{Polkovnikov06,Imambekov06}
\begin{equation}
\hat A(L)= \int_{L/2}^{L/2} dz\, \hat\Psi_l{}^\dagger(z,t)\hat\Psi_r{}(z,t),
\end{equation}
where $L$ denotes a length scale over which the interference pattern is integrated in the $z$-direction. The complex phase of this operator can be identified with the integrated phase $\phi(L)$ of the interference pattern, its magnitude $|\hat A|$ is connected to the interference contrast. For independent gases the expectation value of this operator vanishes, as the total phase is different in each individual realization. However, one can still observe high-contrast interference in these individual realizations. Consequently, the variance 
\begin{equation}
\langle|\hat A(L)|^2\rangle = \int_{L/2}^{L/2}\int_{L/2}^{L/2} dz dz^\prime \left\langle\hat\Psi_r{}^\dagger(z)\hat\Psi_l{}(z)\hat\Psi_l{}^\dagger(z^\prime)\hat\Psi_r{}(z^\prime)\right\rangle\label{eq:contrastvariance}
\end{equation}
is finite. The operator $\langle|\hat A(L)|^2\rangle$ is directly related to the mean squared contrast of the interference pattern $\langle{C^2(L)}\rangle=\langle|\hat A(L)|^2\rangle/\nOneD^2 L^2$, where $\nOneD = \langle|\Psi_1(z)|^2\rangle$ is the mean density in the two gases. Note that the argument of the integral in equation~(\ref{eq:contrastvariance}) can be identified with the two-point correlation function of the relative phase
\begin{align}
{\cal C}(z,z^\prime )= \frac{\langle\hat\Psi_r{}^\dagger(z)\hat\Psi_l{}(z)\hat\Psi_l{}^\dagger(z^\prime)\hat\Psi_r{}(z^\prime)\rangle}{\left\langle{|\Psi_l(z)|^2\rangle\langle|\Psi_r(z^\prime)|^2}\right\rangle}.
\end{align}
The mean squared contrast of the interference pattern is thus the double integral over the two-point correlation function. Neglecting the typically weak density fluctuations by using $\hat \Psi_{l,r} = \sqrt{\nOneD}\exp(\hat \theta_{l,r})$, the phase correlation function can be rewritten as
\begin{equation}
{\cal C}(z,z^\prime) = \langle e^{i\hat\phi(z)-i\hat\phi(z^\prime)}\rangle,\label{eq:corrEiPHI}
\end{equation}
with $\hat\phi(z)=\hat\theta_l(z)-\hat\theta_r(z)$ denoting the local relative phase. 

\paragraph{Full distribution functions}

Further information about the system beyond the two-point correlation function can be obtained from the shot-to-shot fluctuations of the contrast. Such noise measurements have a long and rich history in the characterization of various quantum systems~\cite{GringThesis}. Performing such measurements already deepened our understanding of quantum mechanics as it led to the discovery of the Hanbury-Brown-Twiss effect\,\cite{HBT,BROWN56} which triggered the development of modern quantum optics\,\cite{Glauber63}. Furthermore,  the study of current fluctuations led 
to important observations in quantum-Hall systems\,\cite{Reznikov97,Saminadayar97}. Recently in atomic physics the analysis of noise correlations revealed the coherence properties of atom lasers\,\cite{Ottl05} and enabled observations of the Hanbury-Brown-Twiss effect for massive fermions and bosons~\cite{Aspect08}. It was further suggested\,\cite{Altman04} and experimentally demonstrated\,\cite{Folling05,Spielman07,Rom06,Guarrera08,Perrin12} that noise correlations in time-of-flight can be used to probe strongly-correlated equilibrium states of quantum many-body systems.

The shot-to-shot fluctuations of the contrast can be characterized by the moments of the full distribution function (FDF) $P(\alpha)$ of fringe contrast~\cite{Gritsev06,Hofferberth2008}
\begin{equation}
\frac{\langle|\hat {A}|^{2m}\rangle}{\langle|\hat {A}|^{2}\rangle^m}\equiv\langle\alpha^m\rangle=\int_0^\infty P(\alpha)\alpha^m d\alpha,
\end{equation}
where $P(\alpha)d\alpha$ measures the probability to observe a contrast in the interval between $\alpha$ and $\alpha+d\alpha$. The normalized moments on the left hand side are each connected to a correlation function $\langle|\hat {A}|^{2m}\rangle$ of order $2m$, which is the reason why the FDF, in general, contains more information about the many-body state than the mean contrast. To calculate the full distribution function one has to calculate all moments, or equivalently, all even correlation functions. For the equilibrium situation powerful insights were obtained by mapping this problem onto a quantum impurity problem and to a generalized Coulomb gas model~\cite{Imambekov08}. Experimentally, the FDFs have successfully been used before to study 1D gases in thermal equilibrium~\cite{Hofferberth2008}, as well as the dynamics of an unstable quantum pendulum~\cite{Gerving12}. 

It is important to point out that the method of using FDFs to characterize a system requires the detection of \textit{single} realizations of the quantum system in question. If only ensemble averages can be measured in the experiment, the statistics of those values will always be Gaussian due to the central limit theorem. Consequently, the characteristic higher moments of the observable will not be accessible.

\subsection{Pairs of 1D Bose gases in equilibrium} 
\subsubsection{Theoretical description} 

A prerequisite for experiments is the ability to precisely prepare and characterize both non-equilibrium states, as well as the thermal equilibrium state of the system. One of the key advantages of coherently split 1D Bose gases is that both these states can be controlled and described with high precision. 

For the theoretical modeling the complex many-body dynamics of the Lieb-Liniger Hamiltonian (equation~(\ref{eq:LiebLiniger}) are captured using its phononic low-energy excitations. These excitations can be described using the Luttinger liquid formalism, which provides a universal framework to describe 1D quantum systems. For a single 1D Bose gas one finds the Hamiltonian
\begin{equation}
\hat H_{LL}= \frac{\hbar c}{2}\int dz\left[\frac{\pi}{K}\hat n^2(z) + \frac{K}{\pi}\left(\frac{\partial}{\partial z}\hat\theta(z)\right)^2\right]. \label{eq:luttingerhamiltonian2}
\end{equation}
The operators $\hat\theta(z)$ and $\hat n(z)$ describing density and phase fluctuations are coarse-grained in the sense that they represent the physics in the long-wavelength limit beyond a cutoff~\cite{Kitagawa11}. For 1D bosons the typical cutoff is defined by the inverse of the healing length $\xi_h = \hbar/mc$. 
 For a weakly-interacting 1D Bose gas, one uses the exact solutions of the Lieb-Liniger Hamiltonian to obtain the analytic expressions
\begin{align}
c &= \sqrt{\frac{g \nOneD}{m}} \quad,\quad  K = \hbar\pi\sqrt{\frac{\nOneD}{m g}} = \pi\gamma^{-1/2},
\end{align}
for the speed of sound $c$ and the Luttinger parameter $K$. Introducing bosonic creation and annihilation operators $\hat b^\dagger_k$ and $\hat b_k$, the physics of the Luttinger liquid Hamiltonian can be rewritten as a set of uncoupled harmonic oscillators
\begin{equation}
\hat H_{LL}= \sum_{k\neq0} \omega_k \hat b_k^\dagger \hat b_k,
\end{equation}
where $\omega_k=ck$ is the energy of the mode with momentum $k$. Here, we neglect the $k = 0$ mode, as the spatial correlations in 1D are determined by the modes with $k\neq0$. The effect of the $k=0$ modes, on the other hand, can be identified with the phase diffusion that was discussed in section~\ref{sec:IFMTrappedBEC} in the context of 3D Bose-Einstein condensates. The excitations of the momentum modes contain both a density and a phase quadrature
\begin{equation}
\hat b_k\sim \hat n_k(z) + \hat\theta_k(z).
\end{equation}

In the experiment we are interested in pairs of spatially separated 1D Bose gases. In this case, the excitations contain anti-symmetric and symmetric superpositions of phase and density in the individual gases. These superpositions are also refereed to as the relative and common degrees of freedom of the system. They are given by  
\begin{equation}
\hat\phi(z) = \hat\theta_l(z)-\hat\theta_r(z) \quad\textrm{,}\quad \hat\phi_\mathrm{com}(z)=\hat\theta_r(z)+\hat\theta_l(z)\label{eq:relativephase}
\end{equation}
for the phase, and 
\begin{equation}
\hat\nu(z)=\frac{\hat n_r(z)-\hat n_l(z)}{2} \quad\textrm{,}\quad \hat\nu_\mathrm{com}(z)=\frac{\hat n_r(z)+\hat n_l(z)}{2}
\end{equation}
for the density. Here $\hat \theta_{l,r}(z)$ describes the longitudinal phase profiles and $\hat n_{l,r}(z)$ the density of the left and the right gas, respectively. 

The relative phase $\hat\phi(z)$ between the two gases determines the phase of the interference pattern and can therefore be measured directly in experiment. One can show that within the Luttinger liquid formalism the dynamics of the relative degrees of freedom and the dynamics of the common degrees of freedom decouple~\cite{Kitagawa11}. Both the dynamics of the relative and the common degrees of freedom is thus described by an individual Luttinger liquid Hamiltonian. 

Writing this Hamiltonian in momentum space and solving the Heisenberg equations of motion for the relative phase operator, we obtain the phase variance
\begin{align}
\langle|\hat \phi_k(t)|^2\rangle=\frac{4m^2c^2}{\hbar^2k^2\nOneD^2}\langle|\hat \nu_k(0)|^2\rangle\sin^2(\omega_k t)+\langle|\hat \phi_k(0)|^2\rangle\cos^2(\omega_k t).\label{eq:phasevarianceDW}
\end{align}
This calculation describes both equilibrium and non-equilibrium situations by choosing appropriate initial values for $\langle|\hat \nu_k(0)|^2\rangle$ and $\langle|\hat \phi_k(0)|^2\rangle$. In thermal equilibrium, the excitations of the system are occupied according to Boltzmann statistics $\langle\hat b_k^\dagger \hat b_k\rangle=k_BT/\hbar\omega_k$. In terms of density and phase quadratures this corresponds to
\begin{align}
\langle|\hat \nu_k(0)|^2\rangle &= \frac{k_BT}{2g}\\
\langle|\hat \phi_k(0)|^2\rangle&=\frac{2mk_BT}{\hbar^2\nOneD k^2}.\label{eq:occupationnumbers}
\end{align}
Assuming Gaussian fluctuations the phase correlation function in equation~(\ref{eq:corrEiPHI}) is given by
\begin{equation}
{\cal C}(z,z^\prime) = e^{-\frac{1}{2}\left\langle{[\hat\phi(z)-\hat\phi(z^\prime)]^2}\right\rangle} = e^{-\int_0^\infty\frac{dk}{\pi}\langle|\hat \phi_k(t)|^2\rangle(1-\cos k|z-z'|)} = e^{-|z-z'|\kappa_T},\label{eq:correlationfunctionformulaDWequilibrium}
\end{equation}
where $\kappa_T=2/\lambda_T$. As in the case of a single gas, the thermal correlation function thus decays exponentially, the additional factor of two being a result of the definition of the relative phase $\phi(z)$ in equation~(\ref{eq:relativephase}).

\subsubsection{Measurement of the matter-wave interference contrast} 

\begin{figure}[tbp]
\centering
\includegraphics[width=.72\textwidth]{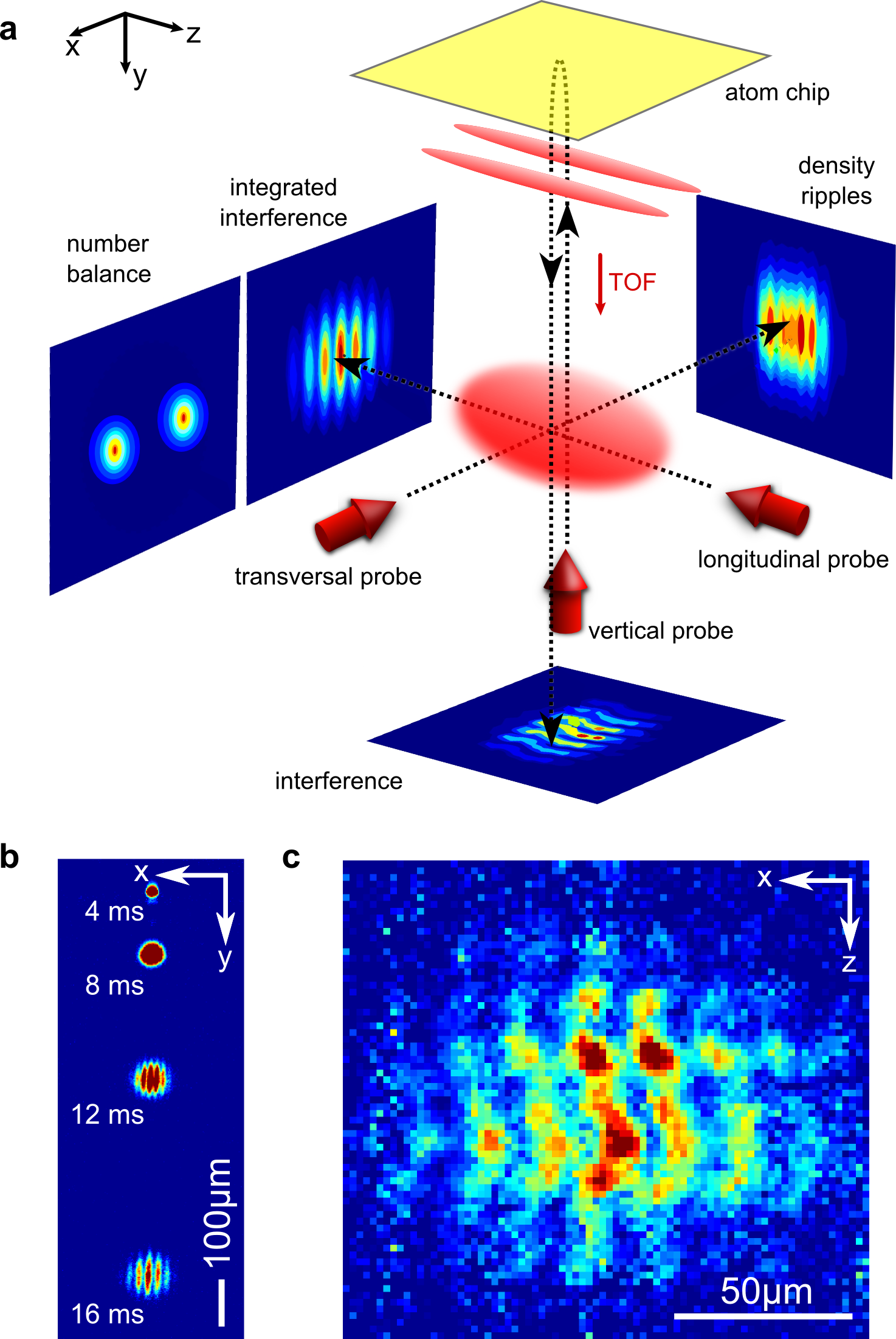} 
\caption{Imaging two interfering 1D Bose gases. (a) After turning off all trapping potentials, the clouds expand and form a matter-wave interference pattern. This pattern can be imaged in three spatial directions. Longitudinally, the interference pattern is integrated along its length. In the vertical direction the local fluctuations of the interference pattern are revealed. As the line of sight of the vertical imaging is blocked by the atom chip, the imaging beam is reflected before passing the atoms. Transversally, an image of the sum of the density ripples in both gases can be obtained. Alternatively, the double-well trap can be turned off before the static trap, such that the two matter-wave packets start to roll down the potential, picking up opposite momenta. After time-of-flight expansion (TOF), this leads to two well-separated clouds which can be individually resolved using the longitudinal imaging. This procedure can thus be used to count the number balance of atoms in the left or right gas. (b) Emergence of the interference pattern in time-of-flight in the longitudinal direction. (c) Example interference pattern (vertical imaging) that is used to extract phase and contrast by fitting equation (\ref{eq:FringeFitFunction}). Figure adapted from~\cite{LangenThesis}.}
\label{fig:doublewell}
\end{figure}

In a typical experimental sequence we prepare a thermal ultracold Bose gas using the microscopic trap of an atomchip~\cite{Reichel11}. This thermal gas is then split by deforming the harmonic trap into a double well potential. This deformation is realized via RF dressed-state potentials~\cite{Schumm2005}. Further evaporative cooling in the individual wells creates two completely independent degenerate 1D Bose gases. This situation corresponds to the thermal equilibrium situation. Subsequently, the gases are released from the trap and allowed to expand freely under gravity. After a certain time of flight they form an interference pattern, where the local position of the interference fringes fluctuates along the z-direction, as determined by $\phi(z)$. In a simple picture, every point along the length of the gas acts like an independent interference experiment. In every individual one of these experiments, the position of the fringes is determined by the local relative phase of the matter-waves emitted from the two wells. Stacking all these experiments together results in the observed interference pattern, as shown in figure~\ref{fig:doublewell}c. This simple picture neglects the effects of the expansion, but remains accurate for typical experimental parameters. 

The main experimental observables derived from this interference pattern are the contrast $C(L)$ integrated over a length $L$ and the local relative phase $\phi(z)$. To extract them, the interference pattern is integrated along a length $L$. This yields a line profile (see figure~\ref{fig:doublewell}) which is fitted with the cosine-modulated Gaussian function~\cite{Kuhnert13}
\begin{equation}
f_{L}(x)=A\cdot\exp\left(-\frac{(x-x_{0})^{2}}{2\sigma^{2}}\right)\cdot\left[1+C\left(L\right)\cos\left(\frac{2\pi(x-x_{0})}{\lambda_F}+\phi\left(L\right)\right)\right],\label{eq:FringeFitFunction}
\end{equation}
where $\sigma$ is the rms radius of the Gaussian profile, $x_0$ is its center of mass, and $\lambda_F$ is the fringe spacing. To extract the relative phase profile $\phi(z)$, the integration length $L$ is set to the size of one pixel. The results are shown in figure~\ref{fig:doublewell}. Repeated realizations of the experimental cycle allow us to build the time-dependent FDFs of the interference contrast $C(L)$ or the phase correlation function $\mathcal C(z,z')$. 

As in the case of a single 1D Bose gas, we can further image the atoms transversally to extract information using the resulting density fluctuations in time-of-flight. As the two gases are completely independent, the resulting density ripple pattern is an incoherent superposition of two single density ripple patterns and contains predominantly contributions from the symmetric mode. We simulate this situation using the Ornstein-Uhlenbeck process technique and use it to extract a temperature~\cite{Stimming10}. As expected for thermal equilibrium, we find exactly the same temperature as for the anti-symmetric mode.

\begin{figure}[htb]
\centering
\includegraphics[width=.64\textwidth]{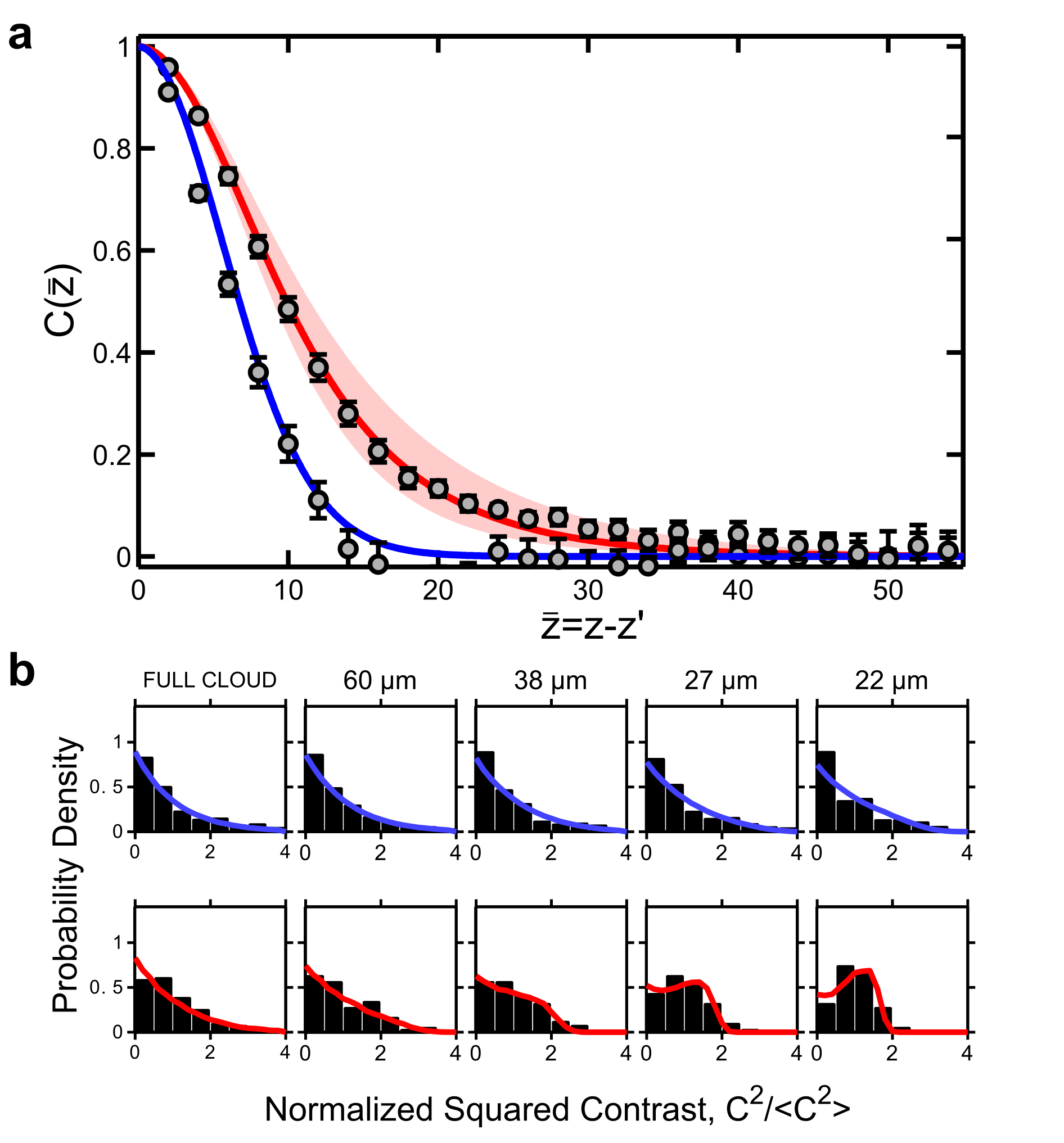} 
\caption{A pair of 1D Bose gases in thermal equilibrium. (a) Two-point phase correlation functions $\mathcal C(\bar z=z-z^\prime)$ for $\nOneD = 35/\mu$m, $T= (27\pm7)\,$nK (red) and for $\nOneD = 45/\mu$m, $T= (117\pm7)\,$nK (blue). Solid lines denote the theory predictions from equation~(\ref{eq:correlationfunctionformulaDWequilibrium}), including the optical resolution. Points are the experimental results, averaged over approximately $100$ realizations. The temperatures used for the theory lines have been independently determined using density ripples, demonstrating that the gases are in equilibrium. For the hotter dataset, the correlation function is completely determined by the optical resolution. Such high-temperature datasets thus enable an independent determination of the optical resolution. (b) The corresponding FDFs are exponentially decaying on all length scales for hot temperatures and show a crossover from exponentially decaying to Gumbel-like for lower temperatures~\cite{Hofferberth2008}, both in very good agreement with theory~\cite{Gritsev06,Stimming10}. Figure adapted from~\cite{LangenThesis}.}
\label{fig:equilibrium}
\end{figure}

\subsection{Probing relaxation in non-equilibrium 1D Bose gases} 

\subsubsection{Coherent splitting of a 1D Bose gas} 

\paragraph{Quenching a quantum many-body system}

\begin{figure}[htb]
\centering
\includegraphics[width=.9\textwidth]{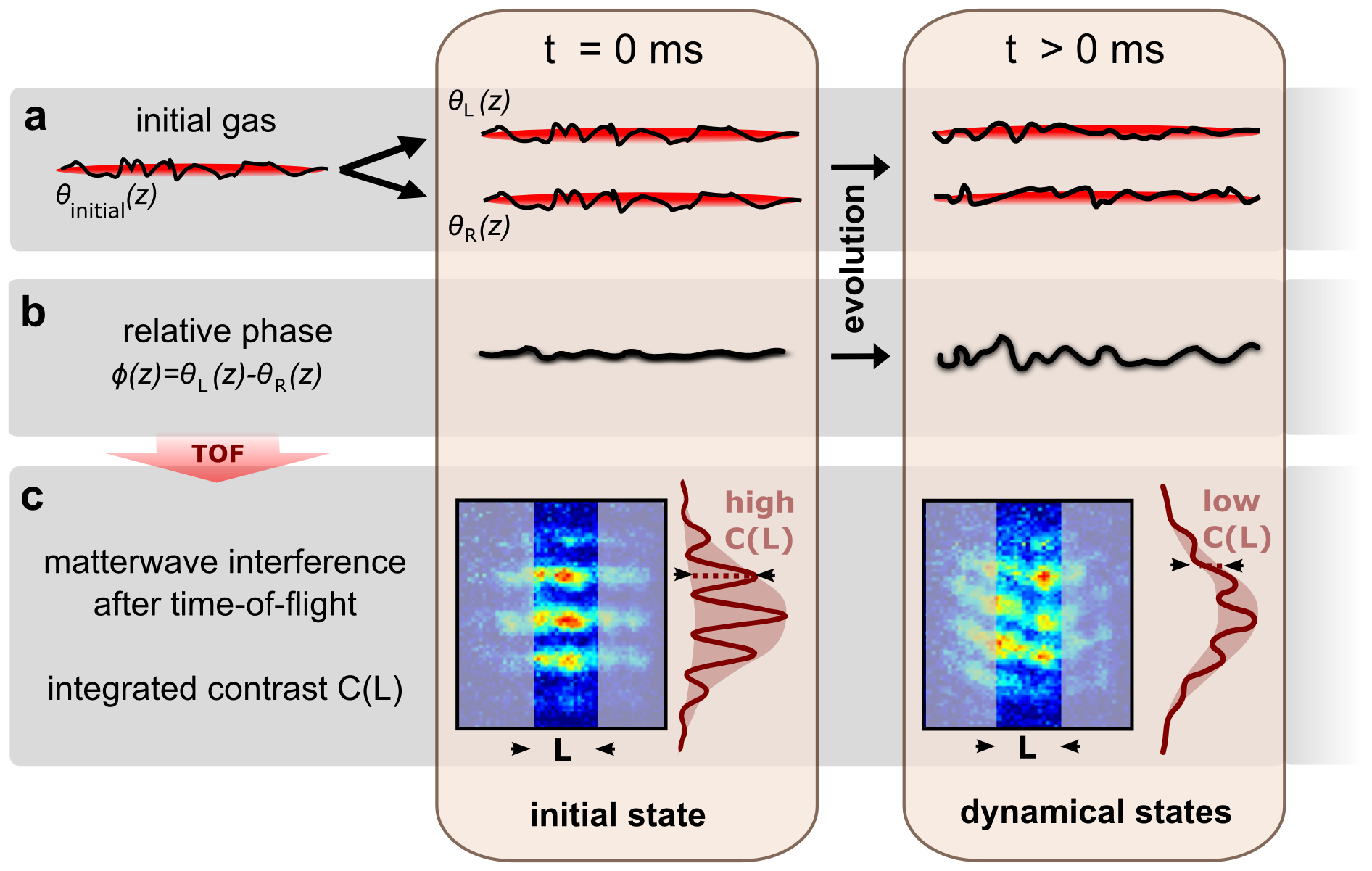} 
	\caption{Experimental scheme for the non-equilibrium experiments. (a) A single phase-fluctuating 1D Bose gas is coherently split into two uncoupled parts with almost identical phase distributions $\theta_l(z)$ and $\theta_r(z)$ (phases represented by the black solid lines). These evolve for a variable time $t$. (b) At $t = 0\,$ms, fluctuations in the local phase difference $\phi(z)$ between the two gases are very small, but start to randomize during the evolution. The question we study is if, and if yes, how this randomization leads to the thermal equilibrium situation of completely uncorrelated gases. (c) shows typical experimental matter-wave interference patterns obtained by overlapping the two gases in time-of-flight (TOF). Differences in the local relative phase lead to a locally displaced interference pattern. Integrated over a length $L$, the contrast $C(L)$ in these interference patterns is a direct measure of the strength of the relative phase fluctuations and thus enables the investigation of the dynamics. Figure taken from ref.~\cite{Langen13c}.}
\label{fig:noneqscheme}
\end{figure}

To study non-equilibrium dynamics we prepare a single degenerate 1D Bose gases in a harmonic trapping potential. A quench is realized by rapidly deforming this harmonic trapping potential into a double well potential, as shown in figure~\ref{fig:noneqscheme}. In general, such a quench is a rapid change of the Hamiltonian of a system, such that the system's new state is no longer an eigenstate of the Hamiltonian. As a consequence, the system will start to evolve under the influence of the new Hamiltonian. We are interested whether this leads to relaxation, the emergence of possible steady states or the emergence of thermal properties. Currently, such questions are under intense theoretical and experimental study, but in contrast to thermal equilibrium, no general framework exists yet to describe non-equilibrium quantum many-body systems. The general problem arises from the fact that the evolution of quantum systems is unitary, consequently it is impossible to reach a unique thermal state from different non-equilibrium states. Moreover, starting from a pure non-equilibrium state, quantum mechanics provides no straightforward connection to mixed thermal states~\cite{Polkovnikov11}. A possible mechanism to overcome these limitations in quantum many-body systems is the eigenstate thermalization hypothesis~\cite{Srednicki94,Deutsch91,Rigol08}. 

In the case of the 1D Bose gas the quench (i.e. the deformation of the potential) leads to a splitting of the gas into two uncoupled 1D gases with identical phase profiles. In terms of excitations, this means that all the thermal excitations are initially contained in the common degrees of freedom. The relative degrees of freedom, on the other hand, are initially populated only by quantum noise created in the splitting process. The quantum noise adds additional energy to the system, preparing it in a non-equilibrium state. The aim is now to probe how the initially almost perfect correlations of the relative phase become obscured over time and if the thermal equilibrium state discussed above, where common and relative degrees of freedom are described by the same temperature, is finally reached~\cite{Hofferberth2008,Betz11}. 

\subsubsection{Dynamics of the matter-wave interference contrast}

To this end, the two gases are allowed to evolve in the double-well potential for a varying evolution time $t$ before the relative phase correlations are probed via time-of-flight matter-wave interference (Fig~\ref{fig:noneqscheme}c). As described above, increasing fluctuations in the relative phase lead to a locally displaced interference pattern, making matter-wave interferometry a direct probe for the dynamics of the system~\cite{Polkovnikov06,Bistritzer07,Imambekov08,Schumm2005a,Hofferberth2007a}. 

\begin{figure}[tb]
\centering
\includegraphics[width=1\textwidth]{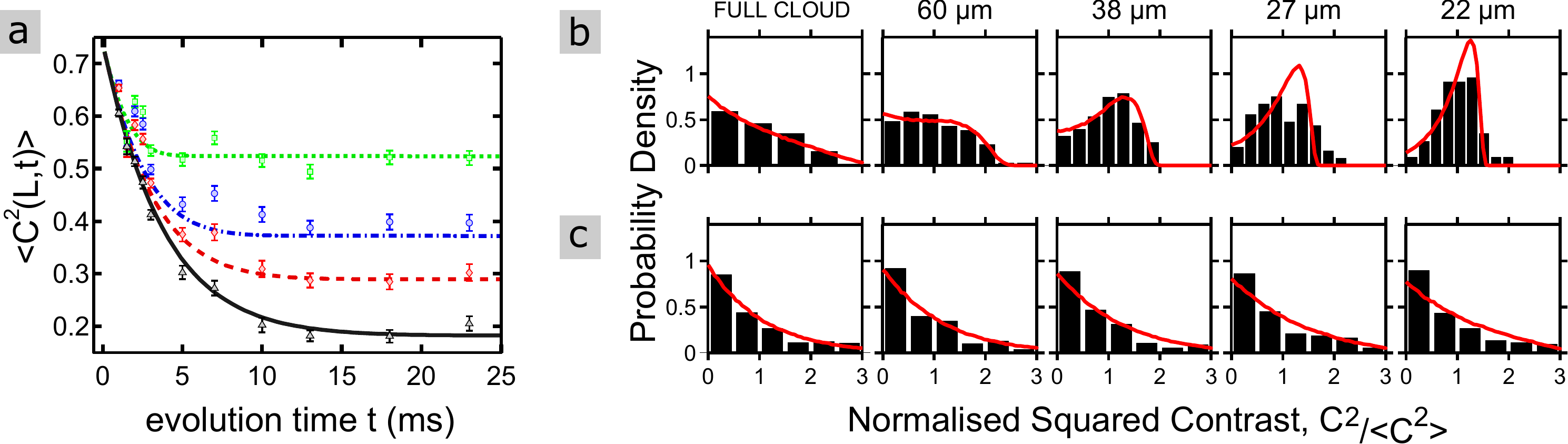} 
 \caption{Dynamics of a coherently split 1D Bose gas. (a) Measured values of the mean squared contrast for various integration lengths (points). From top to bottom: $L = 18, 40, 60, 100\,\mu$m. The lines show the results of a Luttinger liquid calculation for these integration lengths~\cite{Kuhnert13,Kitagawa11}. We observe a relaxation process in which a steady-state is established on a time-scale depending on L. (b) Experimental \textit{non-equilibrium distributions} (histograms) of the matter-wave interference contrast for this steady state. The solid red lines show theoretical \textit{equilibrium} distributions with an effective temperature of $T_\mathrm{eff}=14$\,nK, which is significantly lower than the true initial temperature of the gas ($T=120$\,nK). The prethermalized nature of the state is clearly revealed by comparing it to the vastly different thermal equilibrium situation shown in (c), which can be prepared by creating two completely independent 1D Bose gases. Figure adapted from~\cite{Langen13c}.}
\label{fig:dynamics}
\end{figure}

\begin{figure}[tb]
	\centering
		\includegraphics[width=0.95\textwidth]{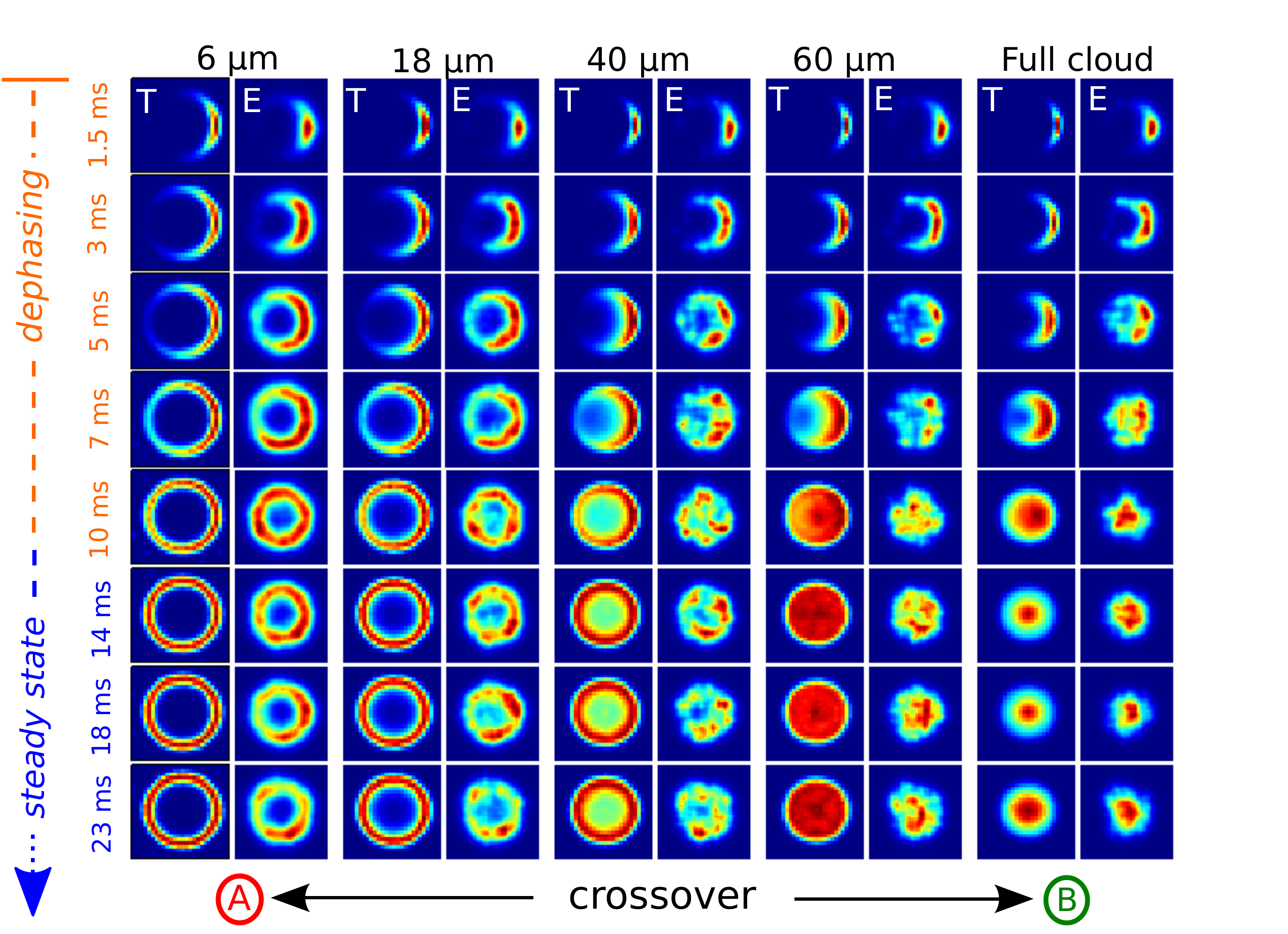}
	\caption{Multimode dynamics revealed by FDFs of matter-wave interference.
The probability density of contrast $C(L)$ and phase $\Phi(L)$ of interference patterns is measured for each integration length $L$ (horizontal axis) and evolution time $t$ (vertical axis). For each realization, they are plotted in a polar plot, where the radius denotes the contrast and the angle corresponds to the phase. Averaging over many realizations results in a probability distribution, where red and blue denote high and low probability respectively. For each value of $L$, the right (left) columns correspond to the experimental data (theoretical calculations), and full cloud to $L=100\,\mu$m. At $t=1.5\,$ms the high contrasts and small phase spreads demonstrate the coherence of the splitting process.
As time evolves, a steady state emerges and two distinct length-scale dependent regimes appear: a regime where only the phase diffuses (A) and a regime where the contrast decays strongly (B). For short (long) $L$, the phase is random and the probability of observing a high contrast is high (low), resulting in a ring (disk) shape in the density plot. Figure from ref.~\cite{Kuhnert13}.
}
		\label{fig:FDFs}
\end{figure}

In figure~\ref{fig:dynamics}a we show the result of this procedure for different integration lengths and different evolution times after the splitting~\cite{Kuhnert13}. We observe a rapid, length-dependent decay of the contrast on time-scales of $5\,$ms ($L=18\,\mu$m integration length) to $20\,$ms ($L=
100\,\mu$m integration length). These time-scales are much faster than the ones expected for thermalization, which is thought to be very slow in 1D Bose gases~\cite{Kinoshita06}. The observations demonstrate that the system relaxes but they do not reveal the physical mechanism behind the relaxation. 
To reveal the nature of the quasi-steady state, we go beyond mean values and measure the full distribution function (FDF) of the contrast. FDFs describing the evolution are shown in figure~\ref{fig:FDFs}. They reveal distinctively different behavior on different length scales, again demonstrating the multimode nature of 1D Bose gases.
The results for the steady state are shown in figure~\ref{fig:dynamics}b. For long integration lengths the distributions decay exponentially, reflecting the randomization of the phase and the decay of coherence. On short integration lengths, however, the histograms show a Gumbel-like distribution with a large probability for high contrasts. This reveals that some coherence from the initial state still remains in the system. The quasi-steady state can directly be compared with the thermal equilibrium state,
which can be prepared experimentally by splitting a thermal gas and cooling the resulting two gases independently to quantum degeneracy. The resulting FDFs are shown in figure~\ref{fig:dynamics}c. They decay exponentially for all length scales, as expected for a thermal equilibrium state in our temperature range, which typically exhibits very little coherence. From this strong difference, we conclude that the dynamically emerging quasi-steady state is not the thermal equilibrium state of the system. 

\subsubsection{Prethermalization}

Although it is not the thermal equilibrium state, the quasi-steady state still shows thermal properties. This is revealed by fitting the experimental data with the theoretically expected shape for a thermal equilibrium state of a certain temperature. For the experimentally prepared thermal equilibrium state this procedure yields, as expected, a temperature which corresponds to the temperature determined from the density ripples. In contrast to this, the quasi-steady state reveals thermal FDFs corresponding to an effective temperature that can be significantly lower than the temperature determined from the density ripples. For the data presented in figure~\ref{fig:dynamics}b we find $T_\mathrm{eff} =14\,$nK, almost an order of magnitude lower than the temperature $T = 120\,$nK of the initial gas.

\paragraph{Emergence of an effective temperature}
The nature of this very low effective temperature can be understood in the following way. The FDFs measure the temperature of the anti-symmetric (or relative) degrees of
freedom. As outlined above, these degrees of freedom are created during the splitting process and only contain quantum noise. This quantum noise that is introduced into the relative degrees of freedom has its origin in the binomial distribution of atoms into the left and right condensates during the splitting. 
We assume this splitting process to be instantaneous, i.e. $t_\mathrm{split}\ll \hbar/\mu = \sigma_h/c$, so that no information can propagate along the length of the system. Therefore the splitting of the individual atoms is completely uncorrelated, resulting in binomial statistics of the splitting process~\cite{Kitagawa11,Langen13}. For a segment with a length $\sigma_h$ containing $N=2\nOneD\sigma_h$ atoms, this leads to relative atom number fluctuations that are given by $N/4$. In terms of the density quadrature this can be expressed as
\begin{align}
\langle|\hat \nu_k(0)|^2\rangle &= \frac{\nOneD}{2}.
\end{align}
The phase quadrature follows from the uncertainty relation \mbox{$[\hat\phi_k^\dagger,\hat\nu_{k'}]=-i\delta_{k,k'}$} as
\begin{align}
\langle|\hat \phi_k(0)|^2\rangle&=\frac{1}{2\nOneD}.
\end{align}
For the excitation numbers this corresponds to
\begin{align}
\langle\hat b_k^\dagger \hat b_k\rangle=\frac{k_BT_\mathrm{eff}}{\hbar\omega_k},
\end{align}
where $k_BT_\mathrm{eff}=g\nOneD/2$ is the energy that is added to the system in the splitting process. These occupation numbers are vastly different than their counterparts in thermal equilibrium (see equation~(\ref{eq:occupationnumbers})). In detail density fluctuations are strongly enhanced, whereas phase fluctuations are strongly suppressed. However, as in thermal equilibrium, all momentum modes are initialized with the same amount of energy, independent of their momentum. Dephasing of the many momentum modes due to their different frequencies $\omega_k$ then leads to the thermal-like statistics of the FDFs. Probing the system on different length scales acts as a filter for the effect of the different modes. For a given integration length L only modes with a wavelength shorter than L can contribute to the randomization of the phase. Thus the level of contrast in the quasi-steady state is high for short L, but significantly lower for long L, where many modes can contribute. 

\paragraph{Evidence for a generalized Gibbs state}
The symmetric degrees of freedom of the system are still hot, because they are thermally populated from the beginning. This observation has two important consequences. Firstly, anti-symmetric and symmetric degrees of freedom have not equilibrated and the system has thus not thermalized. Secondly, the quasi-steady state can be identified with a prethermalized state, as predicted for heavy-ion collisions~\cite{Berges04}, condensed matter systems~\cite{Kollath07,Kollar08} or split 1D Bose gases~\cite{Kitagawa11}. In contrast to usual statistical mechanics, more than one temperature is needed to describe the system. This suggests that the relaxed state is an example of a generalized Gibbs state~\cite{Rigol07}, which is generally believed to describe relaxed states in systems with dynamical constraints, such as our 1D Bose gas.

\subsubsection{Light-cone-like emergence of thermal correlations} 
\begin{figure}[tbp]
\centering
\includegraphics[width=.65\textwidth]{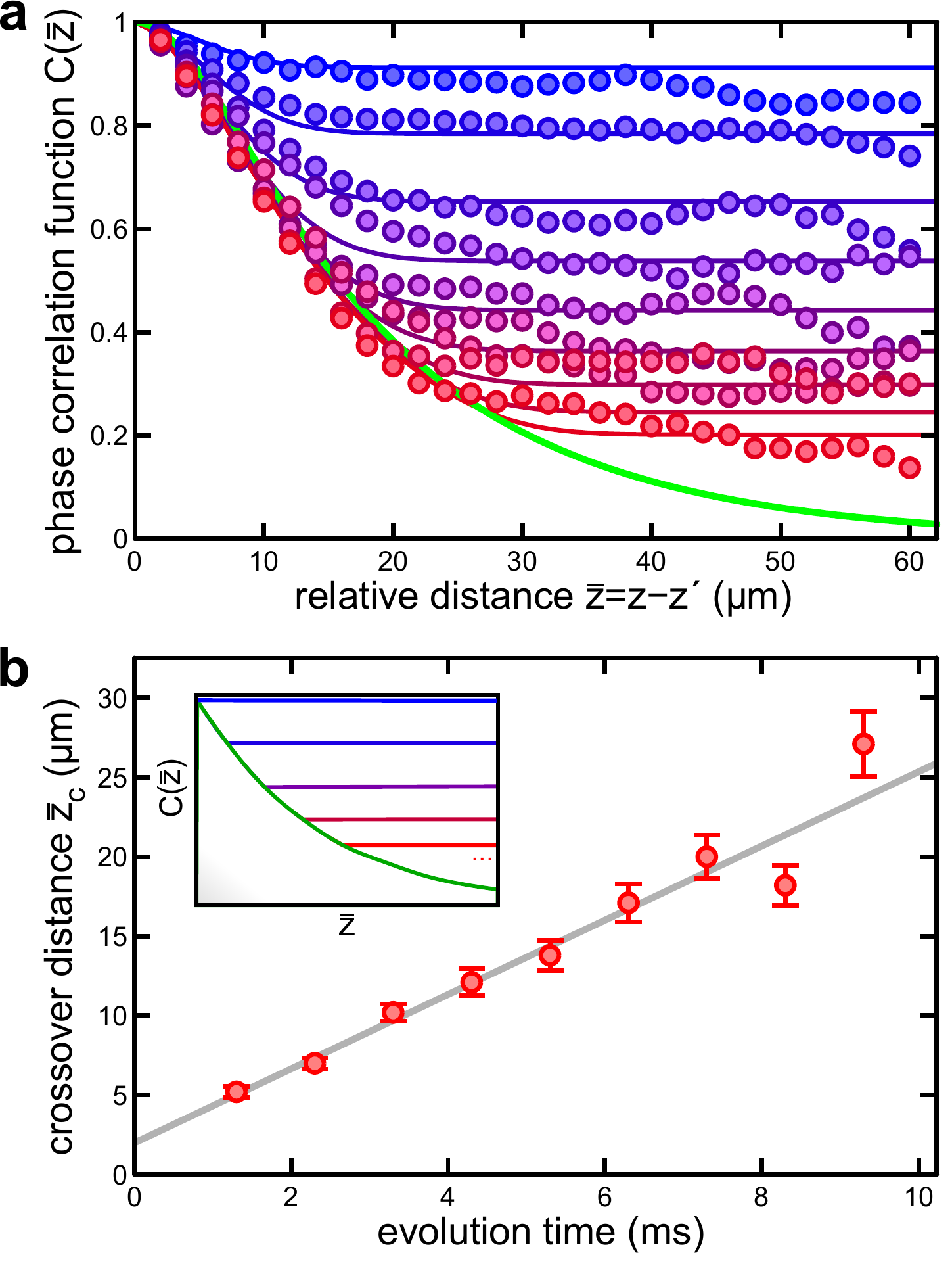} 
\caption{Local emergence of thermal correlations in a light-cone-like evolution. (a) Experimental phase correlation functions $C(\bar z,t)$ (filled circles) compared to theoretical calculations (solid  lines). From top to bottom, the evolution time $t$ increases from $1\,$ms to $9\,$ms in steps of $1\,$ms. The bottom (green) line is the theoretical correlation function of the prethermalized state. For each $t$, the constant values of $C(\bar z,t)$ at large $\bar z$ can be used to determine the crossover distance $\bar z_c(t)$ up to which the system forgets the initial long-range phase coherence (see text for details). (b) Position of the crossover distance $\bar z_c$ as a function of evolution time $t$, revealing the light-cone-like decay of correlations. Error bars denote the uncertainty in $\bar z_c$, following from the standard deviation of the constant values of $C(\bar z,t)$ and the uncertainty in the effective temperature of the prethermalized state (see Supplementary Information). The solid line is a linear fit, the slope of which corresponds to twice the characteristic velocity of correlations. Inset: schematic visualization of the dynamics. The decay of correlations is characterized by a front moving with a finite velocity: for a given time $t$, $C(\bar z,t)$ is exponential (thermal) only up to the characteristic distance $\bar z_c(t)$. Beyond this horizon, long-range phase coherence is retained. Note that in the experimental data shown in (a), the sharp transitions are smeared out by the finite experimental imaging resolution. Figure adapted from~\cite{Langen13b}}
\label{fig:lightcone}
\end{figure}

More insights insights into the emergence of the prethermalized state can be obtained via time-resolved measurements of the two-point phase correlation function. The results of these measurements are shown in figure~\ref{fig:lightcone}. Directly after the quench, the phase correlation function ${C}(\bar z,t)$ is close to unity for any distance $\bar z=z-z'$, where $z$ and $z'$ are arbitrary points along the system. This is a direct manifestation of the long-range phase coherence produced by the splitting process.
After a given evolution time $t$, the phase correlation function decays exponentially up to a characteristic distance $\bar z_c$ and stays nearly constant afterwards: $C(\bar z>\bar z_c,t)=C(\bar z_c,t)$. This means that beyond the distance $\bar z_c$ long-range phase coherence is retained across the system. 
With longer evolution time, the position of $\bar z_c$ shifts to larger distances and the value of ${C}(\bar z> \bar z_c,t)$ gradually decreases. The evolution continues until the system reaches the prethermalized state, where the correlations decay thermal-like throughout the entire system as
\begin{equation}
\mathcal{C}(z,z')=e^{-|z-z'|/\lambda_{Teff}},
\end{equation}
where $\lambda_{Teff}=\hbar^2\nOneD/m k_B T_{eff}$ is the prethermalized correlation length~\cite{Kuhnert13}.

Extracting the crossover points $\bar z_c$ we observe a clear linear scaling of the position $\bar z_c = 2ct$, characterizing the local decay of correlations with time. This observation reveals that an arbitrary point in the gas loses its correlations with other points up to a certain separation $\bar z_c$, while long-range phase coherence persists outside this horizon. The experimental data thus show that the prethermalized state locally emerges in a light-cone-like evolution, where $c$ plays the role of a characteristic velocity for the propagation of correlations in the quantum many-body system. For the data presented in figure~\ref{fig:lightcone}b a linear fit allows to extract a velocity of $c=1.2 \pm 0.1\,$mm/s, which is in good agreement with a Luttinger liquid calculation including the trapping potential~\cite{Geiger13}. This calculation reveals that the underlying mechanism is the spreading of excitations in the system. These excitations can be pictured as quasi-particles which transport information about the quench through the system with their caracteristic velocity $c$~\cite{Lieb72,Calabrese06,Cheneau12}. If the quasi-particles emitted from two points separated by $\bar z_c$ meet after  $t=\bar z_c/2c$
these points establish thermal correlations. This provides evidence for the local relaxation conjecture~\cite{Cramer08}, which provides a general link for the spreading of correlations and the establishment of thermal properties in quantum many-body systems. 

\subsection{Conclusion} 
Matter-wave interferometry with 1D Bose gases offers important insights into the world of quantum many-body systems. For example, our examples demonstrate that non-equilibrium dynamics are far richer than a simple direct relaxation to thermal equilibrium. The pairs of 1D Bose gases introduced in this lecture also allow the realization of many more complex non-equilibrium situations to be explored, for example by introducing a tunnel coupling between the two gases. We expect matter-wave interferometry therefore to have profound implications for our understanding of the emergence of thermal and classical properties in quantum systems,
the study of which is an ongoing theoretical and experimental endeavor.

%



\acknowledgments
We acknowledge fruitful discussions with Tarik Berrada, Thorsten Schumm, Igor Mazets, Remi Geiger, Anatoli Polkovnikov and Eugene Demler. Our work was supported by the Austrian Science Fund (FWF) through the Wittgenstein Prize and the SFB FoQuS project F4010 and the EU through the STREP QIBEC and the European Research Council ERC Advanced Grant QuantumRelax. T.L. thanks the FWF Doctoral Programme CoQuS (W1210), J.-F.S. is supported by the FWF through the Lise Meitner Programme (M 1454-N27). 

\bibliographystyle{unsrt}
\bibliography{bibfiles/references,bibfiles/IFMTrappedBEC,bibfiles/spbs,bibfiles/biblio,bibfiles/refs_for_joerg}

\end{document}